\documentclass[twoside,12pt]{article}
\usepackage{epsfig}
\usepackage{graphicx}
\usepackage{enumitem}
\def\gsimeq{\,\,\raise0.14em\hbox{$>$}\kern-0.76em\lower0.28em\hbox  
{$\sim$}\,\,}  
\def\lsimeq{\,\,\raise0.14em\hbox{$<$}\kern-0.76em\lower0.28em\hbox  
{$\sim$}\,\,} 

\def\gsimeq{\,\,\raise0.14em\hbox{$>$}\kern-0.76em\lower0.28em\hbox
{$\sim$}\,\,}
\def\lsimeq{\,\,\raise0.14em\hbox{$<$}\kern-0.76em\lower0.28em\hbox
{$\sim$}\,\,}
\def\chem#1#2{$\rm{}^{#1}\kern-0.8pt#2$}
\def\reac#1#2#3#4#5#6{$\rm\,{}^{#1}\kern-0.8pt{#2}\,({#3}\,,{#4})\,{}^{#5}\kern-0.8pt{#6}\,$}

\newcommand{\be}{\begin{equation}}
\newcommand{\ee}{\end{equation}}
\newcommand{\bea}{\begin{eqnarray}}
\newcommand{\eea}{\end{eqnarray}}

\begin{document}

\title{ \vspace{1cm} Astronuclear Physics: a Tale of the Atomic Nuclei in the Skies}
\author{M. Arnould and S.\ Goriely\\
\\ 
Institut d'Astronomie et d'Astrophysique, Universit\'e Libre de Bruxelles, Belgium}
\maketitle 
%***************
\begin{abstract}
A century ago, nuclear physics entered astrophysics, giving birth to a new field of science referred to as ``Nuclear Astrophysics''. With time, it developed at an impressive pace into a vastly inter- and multidisciplinary discipline bringing into its wake not only astronomy and cosmology, but also many other sub-fields of physics, especially particle, solid-state and computational physics, as well as chemistry, geology and even biology.  The present Astronuclear Physics review focusses primarily on the facets of nuclear physics that are of relevance to astronomy and astrophysics, the theoretical aspects being of special concern here.

The observational aspects of astronomy and astrophysics that may have some connection to nuclear physics are only broadly reviewed, mainly through the provision of recent relevant references. Multi-messenger astronomy has developed most remarkably during the last decades, with often direct implications for nuclear astrophysics. The electromagnetic view of the components of the Universe has improved dramatically at all wavelengths, from the $\gamma$-ray to the radio domains, providing important new information on the Big Bang and the properties of stars. Neutrino astronomy has made giant steps forward. In particular, the famed ``solar neutrino problem'' is now behind us. The long-sought gravitational waves have at last been detected, with direct relevance namely to the merger of compact stars. The composition of Galactic Cosmic Rays and stellar/solar energetic particles is better known than ever, providing constrains on the GCR physics.

On the stellar modeling side, we broadly brush the progress that has been made based on new observations, and even more so on the spectacular increase in computer capabilities. We briefly outline recent advances regarding the quiescent evolution of stars, as well as the eventual catastrophic supernova explosion of certain classes of them. In spite of significant improvements in the simulations, many long-standing problems still await solid solutions, particularly regarding the details and robustness of explosion simulations. In fact, new questions are continuously emerging, and new facts may endanger old ideas.
  
The lion's share of this review concerns the nuclear physics phenomena that may be at work in astrophysical conditions, with a strong focus on theory. 
Exceptionally large varieties of nuclei have to be dealt with, ranging from the lightest to the heaviest ones, from the valley of nuclear stability all the way to the proton and neutron drip lines. An additional serious difficulty comes from the fact that the nuclei are immersed in highly unusual environments which may have a significant impact on their static properties, the diversity of their transmutation modes, some of which not being observable in the laboratory, and on the probabilities of these modes. The description of nuclei as individual entities has even to be replaced by the construction of an Equation of State at high enough temperatures and/or densities prevailing in the cores of exploding stars and in compact objects (neutron stars). The determination of a huge body of thermonuclear reaction cross sections is an especially challenging task, having to face the ``world of almost no event'' due to the smallness of the relative energies of charged-particle induced reactions relative to the Coulomb barriers, and/or the ``world of exoticism'', as highly unstable nuclei are involved in several nucleosynthesis processes.  

The synthesis of the nuclides heavier than iron is briefly reviewed. Neutron capture mechanisms range from the s-process for the production of the stable nuclides located at the bottom of the valley of stability to the r-process responsible for the synthesis of the neutron-rich isobars. The origin of the neutron-deficient isobars observed in the SoS is attributed to the p-process. Emphasis is put on the astrophysics and nuclear physics uncertainties affecting the modeling of these nucleosynthesis mechanisms.

\end{abstract}

%\pacs{00.00, 20.00, 42.10}
\maketitle
\vfill\eject

\tableofcontents
\newpage

\section{Introduction}
\label{introduction}
%*********************
The year 2019 marks the centenary of the irruption of nuclear physics into astrophysics. The very origin of the long-sought energy needed to make the Sun shine over long enough periods was indeed  hypothesized apparently for the first time by Russell \cite{Russell19}, followed shortly by Perrin
\cite{Perrin20} soon after the first measurements of atomic masses (see \cite{Schatzmann93} for historical 
developments).
 
Following the clarification of the composition of the atomic nuclei in 
1932 with the discovery of the neutron, Gamow, von Weizs\"acker, Bethe 
and others put that idea on a quantitative basis. 
In particular, the energy source of the Sun was ascribed to the 
so-called ``p-p chain'' of reactions, the net effect of which being the 
``burning'' of four protons into a \chem{4}{He} nucleus. 
The energy release is about 6 MeV per proton (meaning about $2 \times 
10^{19}$ kg or $10^{-11} M_{\odot}$ of protons burning per year 
presently in the Sun, where the mass of the Sun $M_\odot \approx 2 \times
10^{30}$ kg).

The Hertzsprung-Russell diagram (HRD) constructed circa 1910 by Ejnar Hertzsprung and Henry Norris Russell and plotting the brightness, classically referred to as the luminosity, of the stars belonging to a given group against their surface temperature or color (see \cite{Babusiaux18} for a modern version of the HRD). The HRD is in fact generally considered as the ``Rosetta stone'' of stellar structure and evolution. Its most remarkable feature is undoubtedly the existence of 
concentrations of stars from a given sample along correlation 
{\it lines} (some spread in these lines results from variations in several characteristics, like the composition of the stars belonging to the selected sample). This key topological feature can be formally demonstrated to be the direct signature of stellar structures in which the energy output can be counter-balanced 
by a nuclear energy production through different chains of reactions (\cite{Cox68} 
for a discussion of this statement). The HRD was thus lending a spectacular support to the ideas that nuclear reactions are indeed key ingredients of the physics of stars.

In the late 1940s, Suess began to argue that certain regularities of the abundances measured in meteorites were reflecting nuclear properties of their isotopes. In particular, the key role of magic numbers in the nuclear shell structure was put forth \cite{Haxel49}, and turned out to be of essential importance in future developments. A myriad of further works have substantiated these early ideas beyond 
doubt, and have put the interplay of nuclear physics and astrophysics on a strong footing into a field commonly
referred to as ``nuclear astrophysics".
The focal role played by the nuclear reactions in the ``alchemy'' of
the Universe also started to be recognized, leading to the development 
of a major chapter of astronuclear physics referred to as the ``theory 
of nucleosynthesis".
Some nucleosynthesis models developed in the late 1940s 
assumed that the nuclides were built in a primordial ``fireball'' at the 
beginning of the Universe \cite{Alpher53}. 
In spite of some attractive features, those models failed to explain 
the mounting evidence that all stars do not exhibit the same surface 
composition. 
They were also unable to explain the presence of the unstable element 
technetium (Tc) discovered at the surface of 
certain giant (``S-type'') stars  \cite{Merrill52}.
(No technetium isotope lives more than a few 
million years.)

The problems encountered by the primordial nucleosynthesis models
put to the forefront the idea expressed by Hoyle in a monumental work
\cite{Hoyle46} that stars are likely to be major nucleosynthesis agents. 
By the late 1950s the stellar nucleosynthesis model, substantiated by 
some seminal works \cite{Burbidge57,Cameron57,Cameron57b}, was 
recognized as being able to explain the origin of the vast majority of 
the naturally-occurring nuclides with mass numbers $A \geq 12$. 
One key theoretical step in the development of these ideas was the 
identification \cite{Opik51,Salpeter52} of the so-called ``$3{\alpha}$''  
nuclear transformation enabling to bridge in stars the gap of 
stable nuclides
at mass number $A = 8$ (see Section~\ref{burn_hesi} for some details). 
In considering the relative abundances of $^4$He, $^{12}$C
and $^{16}$O,   Hoyle went so far as to predict the existence of a
 7.7 MeV
0$^+$   excited state of $^{12}$C as a resonant state in the $^8$Be$ + 
\alpha$ reaction that was soon discovered  experimentally 
\cite{Hoyle54}. 
Despite these early successes, the natural abundances of the light 
nuclides D, \chem{3}{He}, \chem{4}{He}, \chem{6}{Li}, \chem{7}{Li}, 
\chem{9}{Be}, \chem{10}{B}, \chem{11}{B} were difficult to explain 
in terms of stellar thermonuclear processes, and their very  origin has
remained puzzling for some time, until the combined role in their production of a primordial (Big Bang) nucleosynthesis and of the interaction of Galactic Cosmic Rays with the interstellar matter was recognized.

Since those pioneering works, astronuclear physics has advanced at a 
remarkable pace and has achieved an impressive record of success. 
Factors having contributed to the rapid developments include the 
progress in  experimental and theoretical nuclear physics, in ground-based 
or space astronomical observations, and in astrophysical modeling. 
In fact, astronuclear physics has constantly been challenged, and at the 
same time inspired, by new discoveries, many of them marking epochs in 
the history of science. Among them, let us cite (references will be provided later) (1) the discovery in 1965 of the 3K microwave background, which
provided a major support to the Big Bang model and opened a new era of
cosmology, (2) the detection in the late 1960s of neutrinos from the 
Sun, providing the first ``vision'' of the very interior of a star, (3) the mounting evidence that a variety of observed phenomena find an explanation in nuclear radioactivity, largely substantiating what may be the first paper on astronomy with radioactivity \cite{Rutherford29}, (4) the  observation of the supernova SN1987A in the Large Magellanic Cloud that has been a milestone for many fields of astrophysics. In particular, the detection of some neutrinos from this explosion has opened the 
new chapter of the astrophysics of non-solar neutrinos, and (5)  the first detection in August 2017 of a gravitational signal interpreted as originating from the coalescence of two neutron stars (NSs), and opening an additional new chapter of astrophysics, with clear connections to astronuclear physics.
   
In order to take up the continuous challenge from new observational facts,
astronuclear physics concepts and models have to be put on a firmer and 
firmer footing. 
To achieve this goal a deeper and more precise understanding of the  
many nuclear physics processes operating in the astrophysical 
environments 
is crucial, along with improved astrophysical modeling.
Naturally, the acquisition of new nuclear physics data is indispensable 
in the process. 
This quest is, however, easier said than done, given the fact that it is  
generally very difficult, if not impossible, to simulate in the laboratory
the behaviour of a nucleus under relevant astrophysical conditions, or 
even to produce nuclei that might be involved in astrophysical processes. 
Consequently, the development of novel experimental techniques is not 
sufficient, and has to be complemented by progress in the theoretical 
modeling of the nucleus. 
Both experimental and theoretical approaches face great difficulties of 
their own.  
It may be worth noting here that, although initially motivated by  
astrophysics, some experimental and theoretical nuclear physics efforts  
have provided on many occasions unexpected intellectual rewards in 
nuclear physics itself.

This review is meant to be an update and extension of a similar one published twenty years ago \cite{Arnould99}. From the start, we want to make clear to the reader the basic options we have adopted in the formatting of its overall structure and in the choice of its content. We are aware that some of these may lie somewhat away from usual reviews on nuclear astrophysics, and may be somewhat disturbing to some, in particular to astrophysicists who are just nuclear physics data users instead of being full-time nuclear physics practitioners. The main specific features of this review may be summarized as follows:

\begin{enumerate}[label=(\alph*)]
\item  The items of relevance to astronuclear physics just listed above by no means exhaust the questions the field is tackling. They are in reality so varied that it appears impossible to review them all here in any decent way, and we are sorry not to do justice to all the research efforts carried out over the years;

\item We have decided not to review in any extensive way the connections between nuclear physics and astronomical observations or modelings, particularly of elemental or isotopic abundances. This basic choice, which will most likely face criticisms by some and create frustrations, has in fact been made because we  consider that a decent review of the possible relations between nuclear data and astrophysics modeling/observations would require a very lengthy, and what we consider inappropriate, addition of inflationary character to an already quite long review. The connection between nuclear physics and astronomical data is indeed a highly tricky matter that would deserve a very careful discussion of the very many severe uncertainties and shortcomings that pervade astrophysical modelings, if not the analysis of the observations. These uncertainties are very often more or less largely put under the rug, or at least are not properly discussed. Instead, we attempt to provide a (clearly non exhaustive) sample of (preferably recent review) papers dealing with observations or astrophysical modelings that may be of interest to nuclear physicists embarked on astronuclear topics;

\item In contrast, we attempt to cover quite extensively nuclear physics questions that enter astrophysics modeling, even including some phenomena not referenced to in most reviews on astronuclear physics. Emphasis is put on theoretical aspects, with just a series of references focussing more on experimental efforts. Even here, we try to avoid unnecessary repetitions in the form of discussions of topics that are dealt with at length in reviews or textbooks. Note that the literature in the form of published work or preprints is not closely covered after July 1, 2019.
\end{enumerate}

%********************************************************
\section{Observational foundation}
\label{observations}
%********************************************************

The observational foundation of astronuclear physics, and more 
specifically of the theory of nucleosynthesis, rests largely upon the 
determination of elemental and isotopic abundances in the broadest 
possible variety 
of cosmic objects, as well as upon the study of as complete a
set  as possible of observables that help characterizing the objects. 
This knowledge  relies almost entirely on the detailed study of the light 
originating from a large diversity of emitting locations: our Galaxy 
(non-exploding or exploding stars of various types, the interstellar medium (ISM)), external 
galaxies, and perhaps even the early Universe. 
Recent progress in optical astronomy, paralleled by the advent of a 
variety of ``new'' 
(in particular: infrared, UV, X- and $\gamma$-ray) astronomies,
 has led to the unprecedented vision we now have of the sky 
at all wavelengths, ranging from radio-frequencies to $\gamma$-ray 
energies (up to the TeV and PeV domain).  
Dramatic advances in the vision of the electromagnetic Universe have been made possible thanks to ground-based and space borne facilities.
 
The studies of the electromagnetic radiation are complemented with the 
careful analysis of the minute amount of matter of the Universe 
accessible to humankind. 
This matter is comprised for its very largest part in various types of 
solar-system solids. 
The rest is in the form of (extra-)galactic cosmic rays. 
The observation of solar and non-solar neutrinos has also been a major 
step for astronuclear physics as well as for many other fields of 
astrophysics. Very recently, the observational panoply has been broadened further with the detection of gravitational waves.
 
%*************************************************** 
\subsection{Electromagnetic spectra and abundance determinations}
\label{obs_spectra}
%***************************************************************
%
Very roughly speaking, the light from the sky appears to demonstrate some
uniformity of composition of the objects in the Universe, which is most 
strikingly exemplified by the fact that H and He are by far more abundant 
than the heavier species in the whole observable Universe. 
However, they also point to a great diversity of elemental and/or 
isotopic 
abundances that superimposes on that uniformity at all scales ranging 
from stars to galaxies and galaxy clusters.
They imply diverse classes of objects, as well as a diversity of the 
objects belonging to a given class. 
It is impossible here to do justice to the richness of the information
gained by now in this field, and we just limit ourselves to some guiding 
considerations.

%********************************************************
\subsubsection{The spatio-temporal evolution of the composition of our Galaxy}
\label{obs_milky}
%*************************************************
%

In the past decades, spectacular advances have been made in the observations of the stars making up our Galaxy (Milky Way). In particular, large-area sky surveys from state-of-the-art ground-based and space-borne observatories have provided a myriad of multi-wavelength data (spectra), in particular of the surface (photospheric) chemical compositions of stars that are likely not contaminated with nuclear-processed matter from their interiors, and have well preserved the composition of the Galaxy at the place and time of their birth. This is clearly an essential requirement in order for these stars to witness the large-scale time and space variations of the nuclear content of the Galaxy. 
 Data continue to accumulate at a highly impressive pace. These observational efforts are supported by increasingly realistic models and make it clear that the Milky Way is much more complex in its structure and composition than previously imagined. In particular, abundance variations at the level of stellar groups clearly demonstrate again that different elements or groups of elements accumulate at different rates at different locations of the Galaxy. This is certainly the signature of different nucleosynthesis processes acting with unequal spatial and temporal efficiencies. As a very limited selection of the very many papers devoted to the composition of a variety of stars, let us cite {\it e.g.} \cite{Ivezic12,Aguado16,Hollek15,Hansen18,Reddy16,Barbuy18}.

The abundance determinations obtained from the analysis of
photospheric spectra are very usefully complemented with data derived 
from the study of the (molecular) gas and grain components of galactic interstellar 
clouds and circumstellar envelopes. Just to cite a few recent references, see {\it e.g.} \cite{Dishoeck18,Ritchey18}  for a recent analysis of the interstellar content of trans-iron elements, and \cite{Lefloch18} for an account of the Institut de Radio Astonomomie Millim\'etrique (IRAM) program. 
 
%*********************************
\subsubsection{The composition of other galaxies}
\label{obs_galaxies}
%************************************************
%
Information on the composition of other galaxies than the Milky Way is also accumulating very rapidly \cite{Tolstoy09,McConnachie12,Maiolino18}. The analysis of such data forces the conclusion that abundances and their spatial trends may vary quite significantly from galaxy to galaxy, even within a given galaxy class. Substantial local variations are also observed in many instances.  One of the most significant recent advances
in the study of galactic  abundances concerns  high-redshift galaxies referred to as ``Damped  Lyman
$\alpha$ systems (DLAs)'', which provide information on their composition at an early phase in their evolution: the highest the  redshift, the shortest the evolution history of their constituent elements 
from their pristine stage. Among the voluminous literature devoted to the subject, let us just refer to \cite{Maiolino18,Pettini04,Cook15}. The theory of nucleosynthesis and of the formation and nuclidic evolution of galaxies has to cope with this wealth of information.

%****************************************
\subsubsection{Abundances in evolved or exploding stars}
\label{obs_stars}
%*******************************************************
%
On top of the spatio-temporal abundance variations that exist at the scales of stellar groups or sub-groups in galaxies or galaxy clusters, significant abundance differences 
are also well documented at the stellar scale, where individual stars exhibit major abundance differences in a more or less large range of elements, up to the actinides in some cases. 
These differences are of multiple origins. They may be produced in-situ by nuclear processes the products of which are exposed at the stellar surfaces by transport mechanisms from the stellar interiors and/or by the loss of outermost layers through winds, as well as through the contamination from a companion in binary systems. It is impossible to review here the extremely vast literature devoted to the analysis of chemically ``anomalous'' stars, in particular with respect to the Solar System (SoS) standard\footnote{Note that various classes of so-called "chemically peculiar stars" are identified the abundance anomalies of which being likely of non-nuclear origins ({\it e.g.} \cite{Ghazaryan18}).}.
Let us restrict ourselves to some works concerned with different relevant stellar types \cite{Hollek15,Hansen18,Crowther07,Gratton12}.

In addition, exploding objects also exhibit peculiar abundance 
patterns. A very extended literature is devoted to this topic. It largely concerns novae (see {\it e.g.} \cite{Helton12}) and supernovae of different types and sub-types ({\it e.g.}  \cite{Branch98,Filippenko05,Jerkstrand18}). Among the observed supernovae, the explosion referred to as SN1987A that appeared on February 23, 1987 in the Large Magellanic Cloud, a satellite of our Milky Way, occupies a very special place. This wonderful event has been the brightest supernova since 1604 (known as the Kepler supernova), so bright in fact that it could be observed with naked eye. It is also the first such event to have been observed in every band of the electromagnetic spectrum, allowing in particular a detailed compositional information, and also the first to be detected through its initial burst of neutrinos, giving birth to the astrophysics of non-solar neutrinos. As such, it has been a milestone in the observation and theory of supernova explosions. SN1987A has triggered an unprecedented burst of publications
(see {\it e.g.} \cite{McCray93,Matsuura17}; see also the many contributions in \cite{IAU17}).

%****************************** 
\subsubsection{The x-ray and $\gamma$-ray sky}
\label{obs_gamma}
%******************************
%
The electromagnetic view of the sky has been vastly broadened by observations from a large variety of dedicated space-borne x-ray and $\gamma$-ray satellites. High-quality information has been gathered on the properties of compact objects, especially neutron stars, from which the equation of state of nuclear and supra-nuclear extremely neutron-rich matter has been very usefully constrained, complementing the information obtained in the nuclear physics laboratory (see $\it{e.g.}$ \cite{Oertel17,Burgio18,Li19} and Section~\ref{eos}).

At the beginning of the 1980s it was discovered that the electromagnetic 
message from the sky was even richer and more diverse than previously 
thought. 
In fact the ISM was seen to emit a $\gamma$-ray line resulting from the 
de-excitation of the 1.8 MeV level of \chem{26}{Mg} fed by the nuclear
$\beta$-decay of \chem{26}{Al} \cite{Prantzos96}.
This discovery has been followed by the observation that the famed 
supernova SN1987A was emitting $\gamma$-ray lines originating from the 
\chem{56}{Ni}$\rightarrow$\chem{56}{Co}$\rightarrow$\chem{56}{Fe} decay 
chain and from the \chem{57}{Co}$\rightarrow$\chem{57}{Fe} decay.
The decay of \chem{44}{Ti} in the young Cas(siopeia)-A supernova remnant 
has also been observed (see {\it e.g.} \cite{Clayton11,Diehl17}). These observations provide an essential source of information, as well as 
constraints, on the operation of nuclear reactions in 
astrophysical sites. 
In particular the detected emission of $\gamma$-ray lines from supernovae 
was immediately recognized as the clearest demonstration of the 
operation of explosive nucleosynthesis processes (Section~\ref{nucleo_stars}).
 
Gamma-ray lines from nuclear de-exitation are also expected from novae, but remain undiscovered up the now  ({\it e.g.} \cite{Hernanz14,Sieger18}).  

 %***********************************************
\subsection{The composition of the Solar System}
\label{obs_sos}
\begin{figure}[tb]
\hskip-2cm
\begin{center}
\includegraphics[scale=0.4,angle=0]{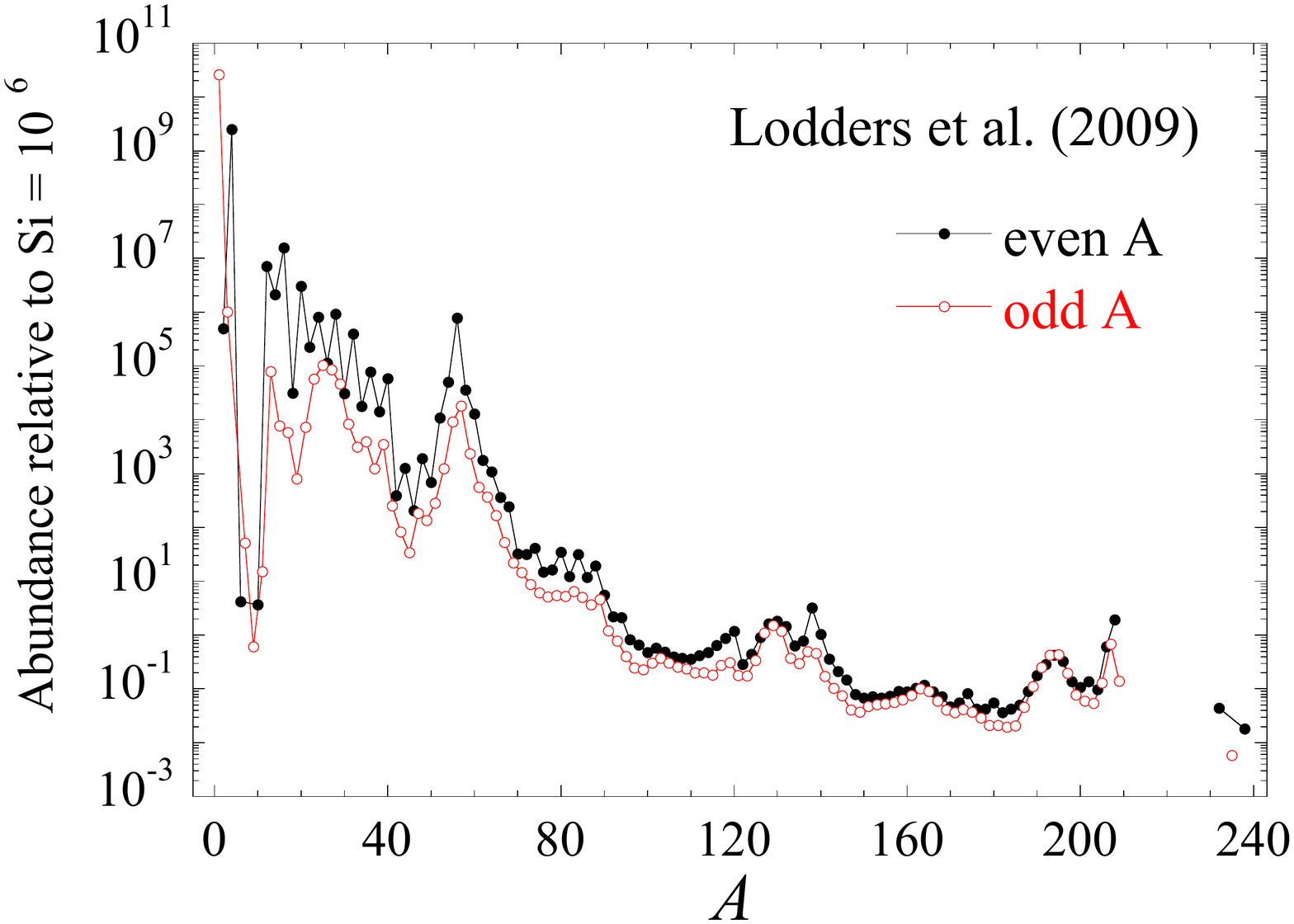}
\vskip-0.6truecm
\caption{Bulk nuclidic composition of the galactic material from which the 
SoS formed about 4.6 Gy ago \cite{Lodders09}. 
The abundances are normalized to $10^6$ Si atoms and are given separately
for even (black full circles) or odd (red open circles) atomic mass numbers $A$. 
They are based on the analysis of the elemental content of meteorites 
of the CI carbonaceous chondrite type, except for ultra-high volatile elements like H, C, N and O obtained from solar spectroscopic data, and from various sources and theory for the noble gases. Terrestrial isotopic composition of the elements is adopted, except for H and the noble gases.}
\label{fig_SoS_A}
\end{center}
\end{figure}

%***********************************************
% 
A milestone in the SoS studies of astrophysical relevance has been
the realization that a meaningful set of abundances likely representative for the composition of the material from
which the SoS formed some 4.6 Gy ago can be derived, in spite of large differences between the 
elemental compositions of constituent members attributable to a large variety  
of secondary physico-chemical and geological processes that have only altered the isotopic compositions in a minor way, except in some specific cases. The SoS abundance distribution is displayed in 
Fig.~\ref{fig_SoS_A} versus atomic mass number.

The understanding of this composition has always held a very special place in astronuclear physics. 
This relates directly to the fact that it provides a body of abundance data whose quantity, quality and coherence remain unmatched, despite the spectacular progress made in astronomical abundance observations. 
This concerns especially isotopic compositions, which are the prime fingerprints of astrophysical nuclear processes.
 
%***************************************************************** **
%***********************************************************************
 
 Even without going into details, some characteristics of the bulk
solar-system composition are worth noticing. 
In particular, H and He are by far the most abundant species, while Li, 
Be, and B are extremely under-abundant with respect to the neighbouring 
light nuclides. 
On the other hand, some abundance peaks are superimposed on a curve which 
is decreasing with increasing mass number $A$. 
Apart from the most important one centered around \chem{56}{Fe} 
(``iron peak''), peaks are found at the locations of 
the ``$\alpha$-elements'' (mass numbers multiple of 4).
In addition a broad peak is observed in the $A \approx 80-90$ region, 
whereas double peaks show up at $A = 130\sim 138$ and $195\sim 208$. 
It has been realized very early that these abundance characteristics provide a clear
demonstration that a close correlation exists between SoS abundances and nuclear properties \cite{Haxel49}.

The composition of the SoS has raised further astrophysical interest
and excitement with the discovery that a minute fraction of its material has an isotopic composition which differs from that of the bulk. 
Such ``isotopic anomalies'' are observed in quite a large suite of 
elements ranging from C to Nd (including the rare gases), and are now 
known to be carried by high-temperature inclusions of primitive 
meteorites, as well as by various types of grains 
found in meteorites (see {\it e.g.} \cite{Clayton04,Nittler16}). The inclusions are formed from SoS material out of equilibrium with the rest of the solar nebula. The grains are considered to be of circumstellar origin and have 
survived the process of incorporation into the SoS.

These anomalies contradict the canonical model of a homogeneous and 
gaseous protosolar nebula, and provide clues to many astrophysical 
problems, like the physics and chemistry of interstellar dust grains, the 
formation and growth of grains in the vicinity of objects with active 
nucleosynthesis, the circumstances under which stars (and in particular 
SoS-type structures) can form, as well as the early history of 
the Sun (in its so-called ``T-Tauri'' phase) and of the SoS solid bodies. 
Last but not least, they raise the question of their nucleosynthesis 
origin and offer the exciting perspective of confronting abundance observations and nucleosynthesis models for a very limited number of stellar sources, even possibly a single one.
This situation is in marked contrast with the one encountered when trying 
to understand the bulk SoS composition, which results from the 
mixture of a large variety of nucleosynthesis events, and consequently 
requires the modeling of the chemical evolution of the Galaxy
(Section~\ref{evol_galaxies}).

Some further excitement has followed the realization that some of the meteoritic anomalies result from the decays within the SoS itself of radioactive nuclides with half-lives in excess of 
about $10^5$~y ({\it e.g.} \cite{Clayton11}). These meteoritic discoveries have been extended further by the detection of deposits of live \chem{60}{Fe} (with a half-life of $2.62 \times 10^6$ y) in various deep-sea sediments. These  deposits are found to be extended in time, and are interpreted as originating from nucleosynthesis events that have taken place during the last ten million years at distances of up to about 325  light-years \cite{Wallner16,Ludwig16}. Small amounts of live interstellar \chem{244}{Pu} (half-life of 81 My) archived in Earth's deep-sea floor during the last 25 My has also been reported \cite{Paul01,Wallner15}.

The possible presence of short-lived radioactive nuclides in live or extinct
form within the early SoS or in deep-sea sediments has far reaching consequences for the 
understanding of the SoS (pre)history and for the astrophysical 
modeling of the stellar origins of these radionuclides. In particular, the contamination by stars of the Wolf-Rayet type (Section~\ref{nucleo_contributors}) in their pre-explosive or explosive phases has been proposed by \cite{Arnould86,Arnould06}. A Wolf-Rayet origin has been made plausible by subsequent studies ({\it e.g.} \cite{Gaidos09,Dwarkadas18} and references therein). Constraints on the production of actinides by the r-process (Sections~\ref{prod_r_site} and \ref{prod_r_mergers}) could also be obtained from deep-sea measurements of \chem{244}{Pu} \cite{Paul01,Wallner15}.
   
%***************************************************
\subsection{The composition of cosmic rays and stellar energetic particle fluxes}
\label{obs_cr}
%***************************************************
 
Cosmic rays are high-energy particles arriving at the Earth. They can be divided into two types, galactic cosmic rays (GCRs) originating from outside the SoS and Solar Energetic Particles (SEPs) (predominantly protons) emitted by the Sun. Energetic particles of the SEP type can also be accelerated at the surface of other stars than the Sun. They are not arriving at the Earth, but can be identified through the products of their interactions with the surrounding stellar material in the form of nucleons or other nuclear species, as well as of $\gamma$-rays. They are referred to as "stellar energetic particles" in the following. Typical SEP (and likely also stellar energetic particle) energies extend from a few MeV into the GeV range in solar (stellar) flares and into the GeV-TeV range for cosmic rays. 

The terminology ``cosmic ray'' is often used to refer only to the GCR flux. Despite the nomenclature ``galactic'', some GCRs may originate from outside the Galaxy. This is the especially the case of the most energetic ones, known as ultra-high-energy cosmic rays (UHECR) with energies above $\sim10^{18}{\rm eV}$ ({\it e.g.}\cite{Allard12}). The precise nature of the galactic GCR sources remains a matter of debate, however, as is the way in which the GCR particles are accelerated up to the GeV-TeV range.  The cosmic ray energies extend in fact largely beyond this range, but those extreme energies belong more to the domain of astroparticle physics, and are not considered here. Excluding these extremely energetic ones, the GCRs are classically considered to be injected into the ISM with a power-law spectrum in energy $E$ somewhat steeper than $E^{-2}$, and than confined possibly through diffusive motion in the magnetized and turbulent galactic disk and halo ISM where they reside for some time before escaping the Galaxy.  It is generally acknowledged that supernovae and supernova remnants likely play an important role in the process of injection and acceleration of GCRs. Many aspects and puzzles of the GCR astrophysics are recently reviewed by \cite{Gabici19}.
 
Much advance has been made in the knowledge of the GCR elemental or isotopic compositions, even if some important information is still missing, particularly regarding isotopic compositions in certain mass and energy ranges ({\it e.g.} \cite{Tatischeff18,Israel18,Genolini18,Evoli18,Lingenfelter19,Gabici19}). It appears now that the observed GCRs are made as some mixture of primary and secondary species, and exhibit both differences and similarities with the SoS composition. They are interpreted in terms of the mixing and interaction of elements from the expanding core-collapse supernova (CCSN) ejecta (see Section~\ref{nucleo_stars}) with those swept up from the ISM.  Locations of choice for this process are groupings of massive stars referred to as OB associations which can be contaminated by the strong winds of massive Wolf-Rayet stars. Type Ia supernovae (SNIa; Section~\ref{nucleo_stars}) are not considered to be significant players in the shaping of the GCR abundance distribution. This is consistent in particular with the absence of these explosions in OB associations.

Some comments are in order here concerning the radionuclide \chem{60}{Fe} synthesized in supernovae. It has been used as a clock to infer an upper limit on the time between nucleosynthesis and GCR acceleration. Its detection implies that the time required for acceleration and transport to Earth does not greatly exceed the \chem{60}{Fe} half-life of 2.6 million years and that the distance of its source does not greatly exceed the distance $\lsimeq 1$ kpc the GCRs can diffuse over this time \cite{Binns16}. This conclusion supports the view that GCRs are much younger than the general galactic material and is in line with their origin in a nearby OB association. It has to be noted that \chem{60}{Fe} has also been detected through a $\gamma$-ray line \cite{Diehl18}, in deep-sea crusts of all majors oceans \cite{Wallner16}, and in lunar samples \cite{Binns16}.

Actinides have also been identified in GCRs \cite{Donelly12}. The data suggest that the cosmic rays currently permeating the solar neighborhood originate from the acceleration of rather standard old (more than $10^8$ years) ISM material that has been contaminated with a few percent of freshly nucleosynthesized material from one or more recent supernovae. Further actinide measurements would be highly valuable in order to constrain further the origin and age of the GCRs.
 
SEPs consist of protons, electrons and ions with energies ranging from a few tens of keV to many GeV.  They originate from a quite large variety of events that translates into a diversity of elemental or isotopic abundances. There is certainly evidence of nuclear reactions in the SEP events from the neutrons and $\gamma$-ray lines observed during solar flares (sudden flashes of increased brightness usually observed near the surface of the Sun). However, the scene is dominated by atomic (especially ionization potentials) and magneto-hydrodynamical phenomena \cite{Reames18}. 

A complementary aspect of GCR and stellar/solar energetic particle astrophysics concerns some features of the SoS composition itself. This question is briefly discussed in Section~\ref{nucleo_gcrs}.

%* ****************************
\subsection{Neutrino astronomy}
\label{obs_neutrino}
%******************************

For years, neutrino astrophysics has built upon experiments designed to detect neutrinos emitted by the Sun, leading to the ``solar neutrino problem'' (see {\it e.g.} \cite{Haxton13}). The origin of the observed solar neutrino deficiency relative to the one predicted by solar models has for years been hotly debated, the main responsibility of the problem being assigned either to particle physics, nuclear physics, or astrophysics. The quarrel is now behind us with the demonstration by the Super-Kamiokande and SNO experiments that the discrepancy between theory and observation can be explained in terms of neutrino oscillations (for a recent review, see {\it e.g.} \cite{Vissani17}). This does not mean that efforts to improve the accuracy of the observations and of the predictions are not pursued (see {\it e.g.} \cite{Agostini17,Pocar18} for experimental developments of the BOREXINO experiment; \cite{Vinyoles17} for improved solar models and some references to recent nuclear reaction rate studies).

The detection of neutrinos from the supernova SN1987A by Kamiokande, IBM, and BAKSAN has opened a new era in neutrino astrophysics ({\it e.g.} \cite{Arnett89}). 
This remarkable observation seems to validate standard models of neutrino 
emission from supernovae, but does not fully pin down at a {\it quantitative} level all the input physics necessary to model supernova explosions, including nuclear physics information. Nowadays, many neutrino detectors are running to detect neutrinos from supernova explosions. Projects also exist to detect the diffuse flux of neutrinos from all past supernovae that pervade space.

As discussed in Section~\ref{nucleo_stars}, neutrinos emerging from the proto-neutron star forming during the explosions of the CCSN type play a key role in this mechanism. They can also be responsible for some changes in the composition of the material ejected by the explosion. In particular, a nucleosynthesis process in which neutrinos are assumed to play a role is referred to as the "neutrino-induced rp-process", or $\nu$p process in short (see {\it e.g.} \cite{Zhang18}, and references therein; see also Section~\ref{burn_rp}).
 
%************************************************
\section{The contributors to nuclei in the Cosmos}
\label{nucleo_contributors}
%*************************************************

The composition of the various constituants of the Universe at the galactic and stellar scales has evolved since the Big Bang. The study of this evolution has been the subject of an enormous amount of observational, experimental and theoretical work. It is just possible to summarize here some the main characteristics of the major nucleosynthesis agents, and we apologize not to do justice to all the many efforts conducted in this extremely wide field.

The changes in the nuclidic composition of the Universe are due to nuclear reactions taking place in conditions of locally established thermodynamic equilibrium. This is the case during the nucleosynthesis epoch of the Big Bang, and later on in stellar interiors. The nucleosynthesis in these conditions is referred to as "thermal nucleosynthesis". Composition changes are also the result of nuclear transformations in too dilute and/or too cold media to establish thermodynamic equilibrium between the reaction partners. This is the case for stellar/solar energetic particles interacting with circumstellar media and for GCRs bombarding the ISM (see Section~\ref{nucleo_gcrs}).This is referred to as "non-thermal nucleosynthesis".

Nucleosynthesis is computed from the solution of nuclear reaction networks linking relevant nuclear species through all possible thermal or non-thermal nuclear reactions that can develop in the considered astrophysical conditions. The extent of these networks is highly variable, ranging from a few light nuclides only in the standard Big Bang conditions (Section~\ref{nucleo_bb}) to thousands of nuclides in the study of the thermonuclear processes responsible for the production of the species heavier than iron 
(Section~\ref{prod_heavy}), or of the non-thermal synthesis produced by energetic particles (Section~\ref{nucleo_gcrs}). For computational ease and feasibility, the common practice in the stellar case is the construction of restricted reaction networks that approximate as best as possible the energy production that is predicted from full networks involving very many reactions with a negligible energy impact. The detailed nucleosynthesis is then calculated in a second step, referred to as a post-processing procedure, decoupled from the detailed stellar model calculations, but making use of the physical conditions (temperatures, densities, timescales) predicted on the basis of model predictions obtained from the use of restricted networks. A few recent exceptions to this post-processing methodology exist, however (see {\it e.g.} \cite{Limongi18}, and references therein).

%****************************************************
\subsection{Thermal nucleosynthesis: the Big Bang contribution}
\label{nucleo_bb}
%********************************************************
%
The standard hot Big Bang (SBB) model provides a very successful and 
economical description of the evolution of the (observable) Universe from
temperatures as high as $T \approx 10^{12}$ K ($t\approx10^{-4}$ s after 
the ``bang'') until the present epoch ($t\approx 10-20$ Gy). 
This model has many far-reaching implications not only 
in cosmology and particle physics, but also in
high-energy nuclear physics (like the relativistic heavy-ion 
physics in relation with the quark-hadron phase transition), low-energy 
nuclear physics (like the physics of thermonuclear reaction rates and weak interaction processes in 
relation to the primordial nucleosynthesis episode), and in  astrophysics 
(through the problems of the formation and initial composition of the 
galaxies). Many of these exciting questions, as well as the details of the
thermodynamics of  the SBB are dealt with at
length in {\it e.g.} \cite{Sarkar96}. 
Here we just limit ourselves to a brief account of some basic 
aspects of the SBB, and in particular of the main features of the
nucleosynthesis epoch of most direct relevance to astronuclear physics.
 
The universal expansion has been discovered by Hubble in 1929: all
galaxies, except those of the Local Group (a gravitationally bound group
of about 20 galaxies to which our Milky Way Galaxy belongs), are receding 
from us (and from each other) with velocities proportional to their 
distances. 
The factor of proportionality is the Hubble parameter $H(t)$, the current 
value of which being the Hubble constant $H_0$. Its precise value has been under active debate for quite long, and is not closed yet, in spite of significant advances (for recent discussions and references, see {\it e.g.} \cite{Aghanim18,Riess18,Shanks18,Rasanen19}). Anyway, the value of the baryon to photon ratio to which the nucleosynthesis during the Big Bang is sensitive is better determined than $H_0$ itself \cite{Kable18}, at least under the assumption that the so-called $\Lambda$CDM cosmology applies.
 
Another pilar of the SBB has been the cosmic microwave background (CMB) radiation discovered in 1965 \cite{Penzias65}, which is a unique laboratory for studying the  initial conditions that gave rise to the observed Universe. Matter and radiation were in equilibrium before  ``de-coupling'' a few 10$^5$ years after the ``bang'', when the temperature had decreased to about 3000 K. After this epoch, radiation cooled during the universal expansion and fits at present with astonishing precision a black-body of temperature $T_0 \approx 2.7$ K with a high angular uniformity ($\Delta T/T<10^{-4}$). 

Nucleons and hydrogen are expected to have been produced in a very early and very hot phase of the Big Bang. This ``baryogenesis'' is not covered here. 
As far as Big Bang nucleosynthesis (BBN) is concerned, it has now entered a high-precision era especially with the recently determined very restricted range of the baryon to photon ratio which has long remained the only free parameter of the SBB model, and due to progress in the evaluation of the "primordial" abundances of D, \chem{3}{He},
 \chem{4}{He} and \chem{7}{Li} that are predicted to have emerged from the SBB  when the temperatures decreased to values in the approximate $10^9 \gsimeq T \gsimeq 10^8$ K range (corresponding to times  
$10^2 \lsimeq t \lsimeq 10^3$ s). 
The confrontation between the BBN expectations and observations largely demonstrates an impressive, and one can dare to say, triumphal success, at least for the first three nuclear species. However, one has to be aware of the fact that this conclusion relies on the necessary consideration of a chain of theoretical and
observational links, each of which bringing its share of uncertainties and
difficulties. Comprehensive discussions of these matters can be found in {\it e.g.} \cite{Aghanim18,Coc17,Pitrou18}.

On the observational side, the determination of the ``primordial'' 
galactic abundances is far from being a trivial matter. It necessitates in particular (model-dependent) abundance
determinations from observations of sites that are expected to have kept at best the pristine SBB abundances. Any disruptive agent of these abundances at a pre-galactic level or at least in the nascent galaxies could more or less largely affect the evaluation of the primordial abundances, and cannot be properly evaluated at present. Additional problems arise for \chem{3}{He}, which can have been destroyed or produced by H burning in very old stars. The most serious difficulty is considered nowadays to come from \chem{7}{Li} the estimated primordial abundance of which is about three times lower than the calculated one \cite{Pitrou18}.

On the theoretical side, very careful evaluations of nuclear reaction and weak interaction rates are made mandatory to make comparison of real relevance between predictions and observations. The main nuclear reactions involve n, p, d and $\alpha$-particle captures on H to \chem{7}{Be} nuclei (see {\it e.g.} \cite{Iliadis15}). A careful evaluation of the remaining uncertainties can be found in {\it e.g.} \cite{Aghanim18,Coc17,Pitrou18}. It is quite remarkable that the lifetime of the neutron may be one of the most important source of uncertainty !  

The success of the standard BBN notwithstanding, many variants to the SBB model have
been developed and studied to a varied extent (see {\it e.g.} \cite{Sarkar96} for a review), some of them stemming in particular from the discrepancies between predicted "primordial" abundances and those derived from observations (largely restricted nowadays to the \chem{7}{Li} one).
 
 %******************************************************
\subsection{Thermal nucleosynthesis: stars}
\label{nucleo_stars}
%*******************************************************************
 
%********* FIGURE  ************************************
\begin{figure}[tb]
\begin{center}
\epsfig{file=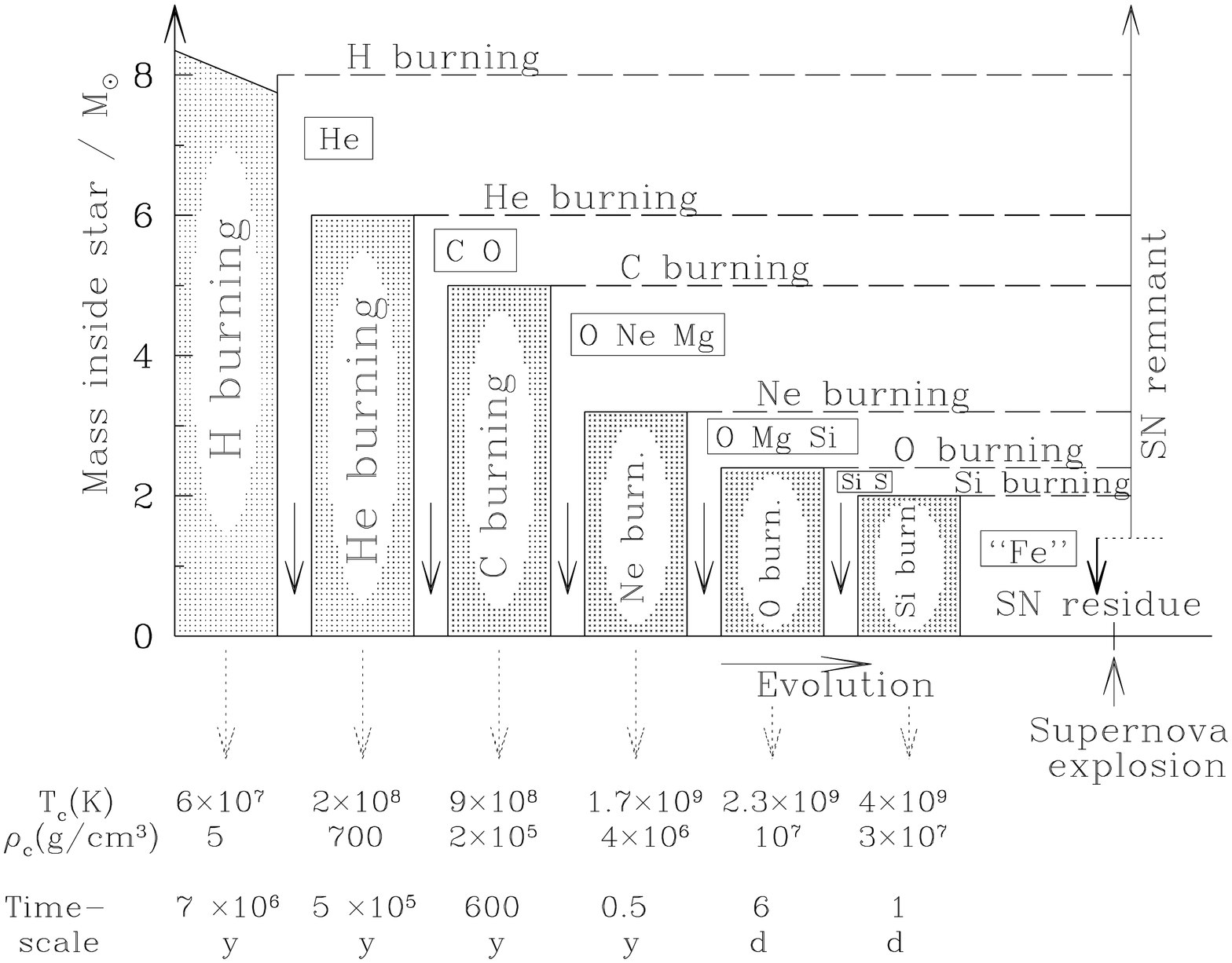,scale=0.5,angle=90.}
\caption{Highly schematic representation in a classical one-dimensional approximation of the evolution of the internal structure of a spherically-symmetric massive star with M $\approx 25$ M$_{\odot}$ (M$_{\odot}$ is the mass of the Sun) with initial composition similar to the solar one.The shaded zones correspond to nuclear burning stages.  A given burning phase starts in the central regions (the central 
temperatures $T_c$ and densities $\rho_c$ are indicated at the bottom of the figure), 
and then migrates into thin peripheral burning shells. In between the central nuclear burning phases are episodes of gravitational contraction (downward arrows). The chemical symbols represent the most abundant nuclear species left after each nuclear-burning mode (``Fe'' symbolizes the iron-peak
nuclei with $50 \lsimeq A \lsimeq 60$). In the depicted illustration, the star eventually explodes as a supernova. The most central parts may leave a ``residue,'' while the rest of the stellar material is ejected into the ISM, where it is observed as a supernova ``remnant''.}
\label{fig_starevol}
\end{center}
\end{figure}
%****************************************************

As pictured very schematically in Fig.~\ref{fig_starevol}, the evolution of the central regions of most stars (for exceptions, see Cat. 1 stars below) is made of successive ``controlled'' 
thermonuclear burning stages where mechanical and energy equilibrium are obtained, and of phases of slow quasi-equilibrium gravitational contraction.\footnote{Mechanical equilibrium or quasi-equilibrium is obtained when the resultant of the forces acting on each element of matter in a large enough fraction of the stellar mass is small enough to ensure at most very slow movements of matter over timescales very long with respect to the free fall timescale. Energy equilibrium results when the nuclear energy production compensates the energy losses through the stellar surface.}
The latter phases are responsible for a temperature increase, while the
former ones produce nuclear energy through charged-particle induced 
reactions. 
Of course, composition changes also result from these very same reactions,
as well as, at some stages at least, from neutron-induced reactions, 
which in contrast do not play any significant role in the stellar energy 
budget. 
The sequence of nuclear burning episodes develops in time with  
nuclear fuels of increasing charge number $Z$ and at temperatures 
increasing from several tens of $10^6$ K to about $4 \times 10^9$ K. 
Concomitantly, the duration of the successive nuclear-burning phases 
decreases in a dramatic way. 
This situation results from the combination of {\it i)} a decreasing energy 
production when going from H burning to the later burning stages and 
{\it ii)} an increasing neutrino production, and the consequent energy losses,
with temperatures exceeding about $5 \times 10^8$ K 
(see \cite{Arnett96} Chap.~10 for details). 
Figure~\ref{fig_starevol} also depicts schematically that a nuclear burning phase, once
completed in the central regions, migrates into a thin peripheral shell. 
As a consequence the deep stellar regions look like an "onion" with  
various ``skins'' of different compositions.
It is important to notice that all stars do not necessarily 
experience all the burning phases displayed in Fig.~\ref{fig_starevol}. Following \cite{Karakas14}, one may distinguish the following classes of stars depending upon their initial masses. They are referred to as "categories" (Cat.) in the following and in Fig.~\ref{fig_nucsite}:
 
(1) Cat.~1: M$\lsimeq0.08$ M$_\odot$ stars that are unable to reach temperatures high enough for leading to any nuclear burning. They make the class of``brown dwarfs'' or even planets for the lightest ones (M $\lsimeq$ 0.01 M$_\odot$), 

(2) Cat.~2: 0.08$\lsimeq$ M$ \lsimeq 0.5$ M$_\odot$ stars burning their hydrogen to leave ``helium white dwarfs'',

(3) Cat.~3: 0.5 $\lsimeq$ M$ \lsimeq 8$ M$_\odot$ stars in which H as well as He can burn, evolving towards ``carbon-oxygen white dwarfs'' (CO-WDs) after a so-called Thermally-Pulsing Asymptotic Giant Branch (TP-AGB) phase involving a complex interplay of peripheral H and He burnings, 

(4) Cat.~4: 8 $\lsimeq$ M$ \lsimeq 10$ M$_\odot$ stars, in which H and He burning is followed by carbon burning, leading to ``oxygen-neon white dwarfs'' (ONe-WDs) after a TP-AGB phase. This phase may end as an ``electron-capture supernova'' (ECSNe) resulting in the formation of a NS, but a thermonuclear explosion of the ``Type Ia Supernova'' (SNIa) (see Cat.~6 below) cannot be excluded \cite{Leung19,Leung19a},
 
(5) Cat.~5: M $\gsimeq 10$ M$_\odot$ stars with of diversity of evolution characteristics and fates that depend more or less severely upon namely their initial mass for a given initial composition, mass loss rate, rotation or magnetic fields. With increasing initial masses, these stars can experience a ``Core Collapse Supernova'' (CCSN) with the formation of a NS or a black hole (BH) (see below for some details and references), ``Pulsational Pair Instability Supernovae (PPISNe)'' which are in fact a special category of CCSNe, ``Pair Instability Supernovae'' (PISNe) (in the approximate M $\gsimeq 100$ M$_\odot$ range), or just collapse to a BH. The reader is referred to
{\it e.g.} \cite{Woosley17,Takahashi16,Takahashi18} for recent works on PPISNe or PISNe. 

Three additional categories relating to binary stars need to be defined:
 
(6) Cat.~6: CO- or ONe-WDs accreting H- or He-rich material from a companion star. Their fate depends on the composition of the accreted material and its accretion rate, as well as on the WD mass. The nuclear burning may affect superficial layers only, leading to explosions of the nova type, or may concern the whole WD structure, eventually triggering SNIa (see {\it e.g.} \cite{Jose15} for details). A specific class of so-called ``Rapidly accreting WDs (RAWD)'' has also been identified, with specific nucleosynthesis signatures \cite{Denissenkov17},

(7) Cat.~7: NSs accreting material from a companion, leading to so-called ``x-ray bursts'' of different types reflecting the explosive burning of H/He-, He- or C-rich fuels (see {\it e.g.} \cite{Jose15} for a review),

(8) Cat.~8: NS-NS or NS-BH mergers, today detected by gravitational wave observations \cite{Abbott17} and potentially a significant source of heavy-elements nucleosynthesis (see {\it e.g.} \cite{Just15}).

An extreme category of stars have masses in excess of an indicative value of about $10^4$ M$_\odot$, and are expected to form with the highest probability at the very early stages of galaxy evolution, when their composition at birth is close to the one emerging from the Big Bang. They are predicted to terminate their evolution as supermassive BHs (see {\it e.g.} \cite{Sun18}). However, the possibility of their explosion as ``general relativistic supernovae'' has also been raised (see {\it e.g.} \cite{Chen14}). 

The stars in Cats. 2 to 4 are classically referred to as low- and intermediate-mass stars, and those in  Category 5 are named massive, very massive or ``supermassive'' ($10^4$ M$_\odot$) stars.
 
It has to be stressed that the mass limits selected above are far from having to be taken at face value. They indeed depend on the physics ingredients adopted for the stellar evolution simulations and their uncertainties (like convective transport of matter in the stellar interiors, referred to as convection), as well as on the initial composition of the stars and their rotational or magnetic field properties. The reader is referred to {\it e.g.} \cite{Portinari98,Karakas14} for an illustration of these uncertainties in the case of low- and intermediate-mass stars, and \cite{Doherty17} for the mass of the progenitors of ECSNe. Note that the first stars to form in galaxies represent extreme cases. They are indeed made of only the BBN products ({\it i.e.} essentially no carbon or heavier elements). They are referred to as Population III (Pop III) stars, while those forming later are classified as Pop II and Pop I stars, the latter ones being the youngest.
 
It has also to be made clear that the true stellar structure is without any doubt 
much more complicated than sketched in Fig.~\ref{fig_starevol}. Additional effects complicate further the stellar structure and evolution picture, like mass loss, rotation and magnetic fields. Among the vast literature devoted to the modeling of stellar evolution in a one-dimensional approximation, let us just cite \cite{Karakas14,Lattanzio16} focussing on low- and intermediate-mass stars, \cite{Limongi18} for the evolution of massive stars, \cite{Clarkson18}  for the evolution of Pop III stars. Very many additional references can be found in these various papers. One cannot emphasize strongly enough that the one-dimensional picture of stellar evolution is far from capturing all the stellar structure and evolution physics, as demonstrated by multi-dimensional simulations made possible by ever increasing computing power (see {\it e.g.} \cite{Cristini17,Mocak18}).

Finally, a large proportion of stars (about two thirds) are not isolated, but belong instead to binary systems. Binarity may have a significant influence on the evolution of the companion stars if they are close enough to each others (``close binaries'') to lead to mass transfer between them ({\it e.g.} \cite{DeMarco17,Boffin17}). Stellar Categories 6 and 7 defined above refer to such systems.

The bottom line of the very many studies devoted to the evolution of single and binary non-exploding stars is that still large uncertainties affect the simulations. The lion's share of these uncertainties is taken by the hydrodynamics modeling, and more specifically by the turbulent transport of materials with different compositions within the stellar interiors, not talking about additional problems raised by rotation, magnetic fields or binarity. Nuclear physics also enters the problem. In general, its accurate treatment does not appear, however, as crucial as the hydrodynamic modeling, being a necessary, but far from a sufficient input in the stellar simulations.

The various non-explosive nuclear burning modes from hydrogen to silicon burning (see Fig.~\ref{fig_starevol}) and their accompanying energy production and nucleosynthesis (see also Fig.~\ref{fig_nucsite}) are discussed extensively in so many places that it is superfluous to review them here thoroughly once more. The reader is referred to {\it e.g.} \cite{Iliadis15} (Chapters 5.1.1-5.3.4) for details. Just a limited overview is proposed in Sections~\ref{burn_h} and \ref{burn_hesi}. 
 
As mentioned above, the quiescent evolution of the stars of Cats. 4 to 7 is eventually followed by an explosive phase. It is a surface phenomenon in the nova or x-ray-burst cases (Cats. 6 or 7), while it deeply affects the whole stellar structure in the case of supernovae of various types (Cats.~4 to 6). ECSNe of Cat~4 and CCSN explosions of Cat. 5 stars have the strongest ties with nuclear physics, and have also been the focus of the largest simulation efforts. These explosions are viewed as the result of the implosion of the Fe core that develops as a result of the various quiescent burning phases depicted in Fig.~\ref{fig_starevol}. In the case of CCSNe of Cat.~5, this implosion follows from the endothermic Fe photodisintegration down to $\alpha$-particles, and even nucleons. In the case of ECSNe of Cat.~4 instead, free electron captures are the main triggering mechanism. Note that the PPISN or PISN triggering results instead from pulsational instabilities that are at work along with a pressure deficit due to the prolific production of electron-positron pairs.

In the ECSN/CCSN case, the implosion is thought to turn into a catastrophic explosion through a very complex chain of physical events. It is out of scope to review here the many efforts that have been devoted to the modeling of the implosion-explosion mechanism. Details about the involved physics and computer simulations can be found in many places ({\it e.g.} \cite{Janka17,Janka12,Connor18,Connor18a,Glas19,Vartanyan19,Burrows19} for the CCSN case and \cite{Leung19} for the ECSN case). In short, some successful  explosions are found, at least for stars in the $M \lsimeq 15  M_\odot$ range, while the situation is less rosy for more massive stars, where some robust explosions are reported, but at the expense of various approximations or ad-hoc perturbations of the pre-supernova models. It is established  today that three-dimensional simulations of the supernova phenomenon are required, and become feasible thanks to ever-increasing computer capabilities. It is also widely recognized that neutrino production and transport through the supernova core material most likely play a key role in the explosion mechanism, but that other processes are at play, like non-radial turbulent fluid motions, various types of instabilities, rotation and magnetic fields. Many uncertainties remain, including quite severe ones in the available pre-supernova models that result from one-dimensional simulations, only some multi-dimensional models for the advanced stages of stellar evolution being constructed at this time (see above). Some of the uncertainties are of nuclear physics origins, like the neutrino-nucleus interactions, free electron capture by very neutron-rich nuclei, or the Equation of State (EoS) of highly neutron-enriched matter at high temperature at nuclear and even beyond nuclear saturation densities. 
 
A ECSN/CCSN is predicted to leave in its central regions a ``residue'' in the form of a NS (observable as a pulsar if it is magnetized and rotating) or a BH, depending in a very complicated manner on the characteristics of the pre-supernova star, like mass and initial composition ({\it e.g.} \cite{Janka17}). The explosive ejection of matter is of course an essential component of the supernova. This ejection results from a shock wave generated in the vicinity of the forming residue and propagating outward through the supernova layers.
This shock wave compresses the various traversed layers, heats them up
before pushing them outward until their ejection into the ISM. This expansion is of course accompanied by a cooling of the material. This heating and cooling process of the layers hit by the supernova shock 
wave allows some nuclear transformations to take place during a quite brief time, modifying more or less significantly the pre-explosion composition of the concerned layers. 
The study of the composition of the ejected material that makes up the 
supernova ``remnant'' is one of the main chapters of the theory of ``explosive nucleosynthesis''.  A voluminous literature is devoted to the study of the outcome of this nucleosynthesis in a large variety of massive stars of different initial compositions, including Pop III stars. It cannot be reviewed extensively here. Most of these studies make use of schematic one-dimentional explosion models, and sometimes not fully state-of-the-art nuclear reaction data that are important input in these studies (see {\it e.g.} \cite{Limongi18,Jose15,Nomoto13,Grimmett18}). Some selected processes are reviewed below (Sections~\ref{burn_hotpp} to \ref{burn_rp} and \ref{prod_r}--\ref{prod_p}).  

Astronomical observations are an essential constraining factor in supernova astrophysics.
Various classes of CCSNe are identified on grounds of their observed properties. One notes in particular H-rich and H-poor events, the latter ones being interpreted as due to a severe pre-explosion loss of the stellar envelope.  Another important piece of observation is the very patchy pattern in the structure and composition of the remnant, which emphasizes the multi-dimensional nature of the supernova phenomenon. Last but not least, the supernova SN1987A has been a cornerstone in this field with the observation of the first neutrinos of non-solar origin. It may also be worth mentioning that the high temperatures and high pressures experienced in the vicinity of the forming residue might be close to be obtained in the laboratory \cite{Blinnikov19}. 

%*******************************************
\subsection{Non-thermal nucleosynthesis: GCRs and solar/stellar energetic particles}
\label{nucleo_gcrs}
%*******************************************

Cosmic rays originate as primary cosmic rays. Their composition of course depends on the precise nature of the source(s), which is not firmly established yet. It is, however, quite safe to claim that they are composed mainly of protons and alpha particles (99\%), with a small amount of heavier nuclei (around 1\%). During their repeated crossings in the galactic disk, non-thermal reactions involving interactions between protons and $\alpha$-particles and heavier nuclei modify the initial composition of the GCRs at injection (the abundance modifications in the galactic halo are limited due to its very low gas density). The composition changes leading to the production of so-called secondary nuclides are followed through propagation models. The basic ingredients of these complex models include the description of the source characteristics (injection spectra and isotopic abundances), a system of transport equations with spatial and momentum diffusion terms (diffusion, convection, reacceleration, energy losses) and their transport coefficients, the 
description of the ISM (gas distribution, radiation and magnetic fields), and, last but not least, particle and  nuclear production and disintegration cross sections shortly discussed in Section ~\ref{reac_nonstat}. These ingredients are discussed in great detail in several recent reviews, among which \cite{Gabici19,Tatischeff18,Genolini18,Evoli18}.

Nuclear interactions with the surrounding circumstellar/circumsolar material also modify the composition of the source stellar/solar energetic particles. These composition alterations have also a direct bearing on the SoS composition itself, including  planetary surfaces, meteorites or cosmic dust, as well as on the surface of certain classes of stars, particularly young and/or magnetically active ones. 

\begin{figure}
\centering
\includegraphics[scale=0.4]{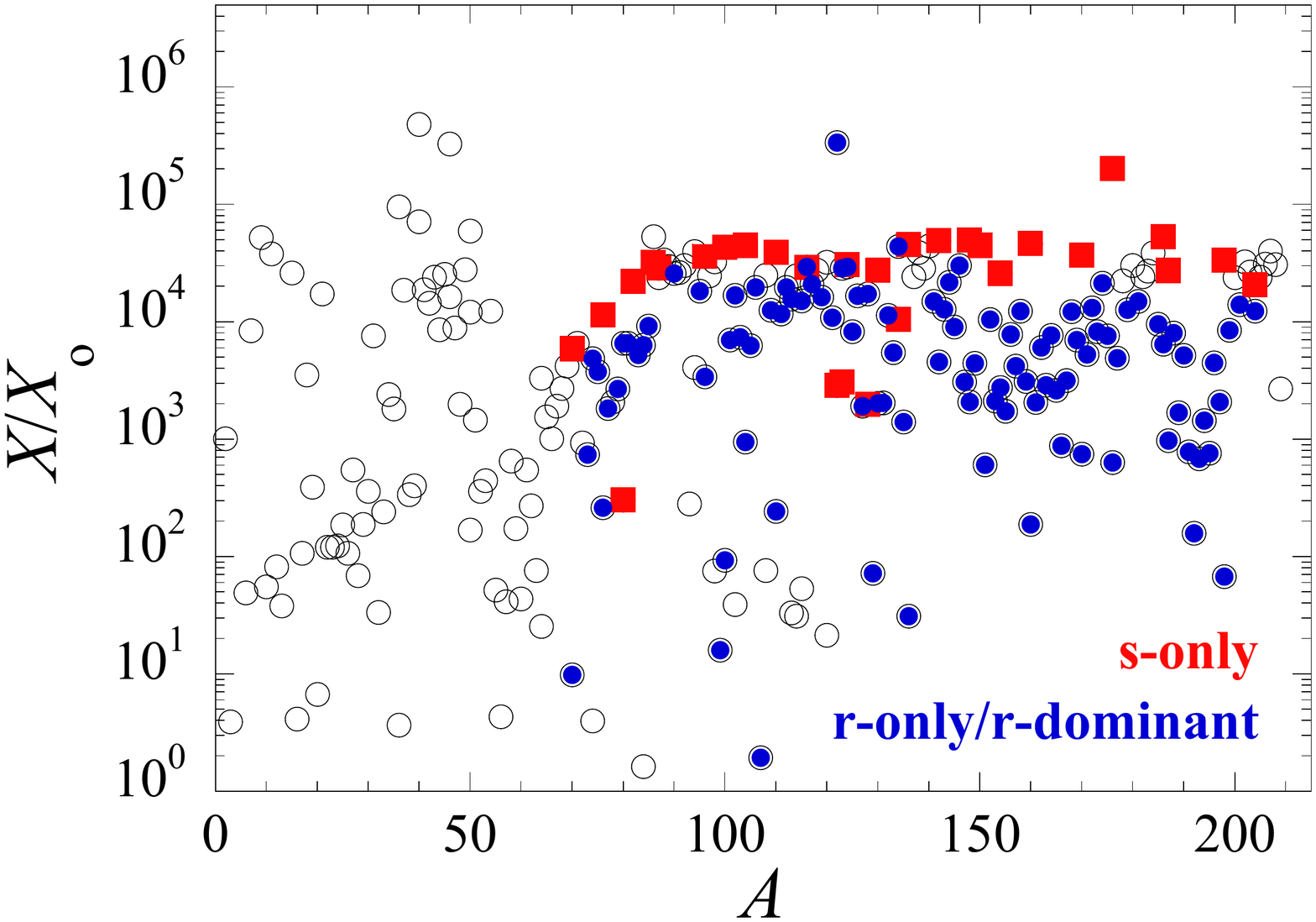}
\vskip -0.7cm
\caption{Overproduction factors $X/X_\odot$ as a function of atomic mass $A$ resulting from the irradiation of a C target by a flux of  $10^{-13}{\rm mb^{-1}s^{-1}}$ of pure $\alpha$-particles of constant energies between 2 and 3~MeV per nucleon during about 3\,800~y. S-only nuclei are shown by red aquares and r-only and r-dominant nuclei, {\it i.e.} nuclei with a SoS r-abundance larger than 50\%, by solid blue circles.  }
\label{fig_sep_spro} 
\end{figure}

The nuclear physics aspects of the non-thermal nucleosynthesis question are touched upon in Section~
\ref{reac_nonstat}. Here, we just briefly overview some non-thermal nucleosynthesis predictions:

\begin{enumerate}
\item  As recently discussed in detail by \cite{Tatischeff18}, it is now classically accepted that the bulk SoS \chem{6}{Li}, \chem{9}{Be}, \chem{10}{B}, most of the \chem{11}{B}, and  at least some  of the \chem{7}{Li} originate from the nuclear interaction of GCRs with the ISM. This mechanism has also been proposed on occasion to explain the synthesis of rare p-nuclides, and more particularly \chem{180}{Ta^m} \cite{Kusakabe18} (see also Section~\ref{prod_p});

\item The origin of relatively short-lived radionuclides, such as $^7$Be, $^{10}$Be, $^{26}$Al, $^{41}$Ca, $^{53}$Mn and $^{92}$Nb and some others that are now extinct, but have been alive in meteorites and lunar rocks at an early stage of the SoS, is also assigned to SEP-induced nuclear transformations at the surface of the early Sun \cite{Leya03,Dauphas06,Liu12}. Although different results may be obtained from such calculations (depending on the adopted parameters (namely the SEP fluence and target and projectile compositions), they agree on the general picture that $^{26}$Al is not produced efficiently enough, while proper amounts of $^{10}$Be, $^{36}$Cl, $^{53}$Mn, and $^{41}$Ca can result, at least locally in a given target. Difficulties may arise, however, when trying to account for the inferred bulk SoS content of these radionuclides \cite{Duprat07};

\item Stellar energetic particles are invoked in order to explain some characteristics of the surface composition of certain stars. This concerns for example the high \chem{6}{Li} abundance observed in metal-poor halo stars, which is very difficult to reconcile with BBN (Section~\ref{nucleo_bb}) \cite{Tatischeff07}. Some spectral features of so-called chemically-peculiar (CP) stars may be another interesting case. Their abundance patterns are usually interpreted on the basis of diffusion processes, that is the diffusive segregation of ionic and isotopic species resulting from the balance between radiative and gravitational forces within the atmosphere and sub-atmospheric regions \cite{Michaud04}. There has been some claim, however, that some short-lived radioactive elements, such as Tc, Pm and $84 \le Z\le 99$ elements, as well as the unusual Co/Fe and $^6$Li/$^7$Li abundance ratios may be present at the surface of the CP roAp star HD~101065, also known as Przybylski's star \cite{Cowley00,Cowley04,Gopka04,Bidelman05}. Such a remarkable observation of short-lived radioactive elements certainly needs to be confirmed, considering the complexity of identifying the spectral lines of ionized radio-elements in warm, abnormal, stratified, and out-of-thermodynamic-equilibrium atmospheres. If firmly established, the origin of these radionuclides (in particular Pm with a 17.7 y half-life for its longest-lived isotope) cannot be found in diffusion processes, but in nuclear processes instead. Another feature of HD~101065 abundances is to exhibit  $35 < Z < 82 $ elements exceeding by 3 to 4 orders of magnitude the SoS values. Again, this quite spectacular feature can hardly be accounted for by diffusion processes, while it can be explained by the action of stellar energetic particles. This has been shown by \cite{Goriely07c} on grounds of a simple approach taking as free parameters the observationally unknown proton and $\alpha$-particle spectra, the time of irradiation, and the possible mixing with nuclearly unprocessed material. High fluences of the order of $10^{27}{\rm cm^{-2}}$ are needed to account for the observations;

\item It may be speculated that a much wider nucleosynthesis than the specific cases mentioned above may result from the action of stellar energetic particles, leading to signatures that fundamentally differ from the ones predicted by thermonuclear processes. Just for illustrative purposes, let us consider the irradiation of a C target by a flux of  $10^{-13}{\rm mb^{-1}s^{-1}}$ of pure $\alpha$-particles of constant energies between 2 and 3~MeV per nucleon during about 3\,800~y. As illustrated in Fig.~\ref{fig_sep_spro}, the production of s-process nuclides (see Section~\ref{prod_s}) in almost solar proportions may result. Such an event cannot be demonstrated to exist, however. It could correspond to the irradiation of a C-rich meteoritic material by an $\alpha$-rich wind or jet ejected at a mean velocity of 22\,000 km/s with a total fluence of about $1.2~10^{25}{\rm cm^{-2}}$ ({\it i.e.} about 120 times larger than the fluence invoked in young stellar objects). In such conditions, the reactions induced by the 2-3~MeV/nucleon projectiles are capable of producing enough neutrons (with a neutron density $N_n$ ranging between $10^7$ and $5~10^8~{\rm cm^{-3}}$ during the 3\,800~y of the irradiation process) that, after thermalization, would be responsible for the neutron capture nucleosynthesis shown in Fig.~\ref{fig_sep_spro}. For high energy projectiles, many production channels are open, so that all the different (p or $\alpha$,$x$n\,$y$p\,z$\alpha$) reactions with the emission of $x$ neutrons, $y$ protons, and $z$ $\alpha$-particles need to be taken into account for each target nucleus. The reaction network typically includes about 5\,000 different species with $0 \le Z \le 102$ and some 250\,000 proton, neutron, and $\alpha$-particle capture reactions. In contrast to thermonuclear nucleosynthesis, cross sections need to be evaluated at stellar particle energies well in excess of the Gamow window. For this reason, such calculations are essentially based on reaction models including the compound nucleus (CN), pre-equilibrium (PE) and direct capture (DC) contributions, as described in Section~\ref{reac_models_general}.  More details on the nucleosynthesis modeling can be found in \cite{Goriely07c}. Of course, the possible contribution of such high-fluence non-thermal processes to the Galactic enrichment still needs to be investigated.
\end{enumerate}

%*****************************************************************
%******************************************************************

 %** ****************************************** 
\section{The nuclidic evolution of galaxies}
\label{evol_galaxies}
%*********************************************

The modeling of the evolution of the nuclidic content of galaxies 
(classically and inappropriately termed the ``chemical'' evolution of galaxies) is without any 
doubt one of the most formidable problems astrophysics has to face. 
This question has been tackled at various levels of sophistication, 
ranging from so-called ``chemo-dynamical models'' to simple `` one-zone'' 
models, the latter ones limiting themselves to the description of the 
evolution of the abundances in the solar neighborhood. 
Because of their complexity, the first types of models are cruder in their 
nucleosynthesis aspects than the latter ones, which do not address any 
galactic thermodynamics- or dynamics-related  issues. 
This immense problem cannot be discussed appropriately here. Let us just mention that one basic ingredient of the models is the stellar creation function, that is the number of stars born per unit area of the galactic disc (in spiral galaxies) per unit mass range and unit time interval.  That question has been discussed at length in 
\cite{Scalo86}. Another main ingredient is of course the mass and composition of the matter ejected by a star with a given initial mass. For details, the reader is referred to the extensive review by \cite{Maiolino18}  (see also {\it e.g.} \cite{Prantzos18,Ojima18} for a more specific focus on the nuclides heavier than iron discussed in Section~\ref{prod_srp_general}).
 
%*********************************************************** **
\begin{figure}
\begin{center}
\includegraphics[scale=0.5]{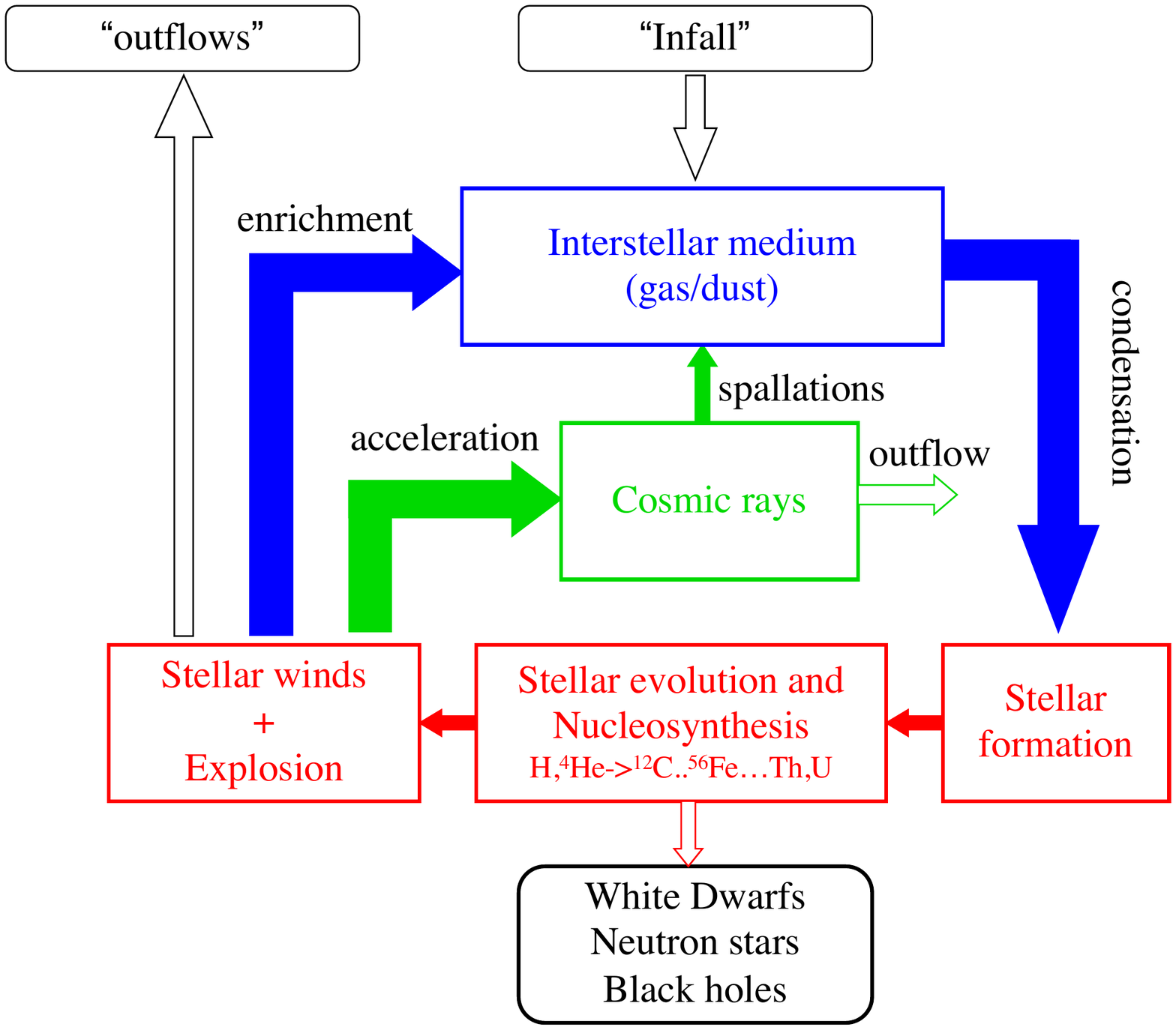}
\vskip-0.8truecm                                   
\caption{A very schematic picture of the galactic ``blender'' (see text). Details of the nucleosynthesis content of the box labelled ``Stellar evolution and Nucleosynthesis'' are provided in Fig.~\ref{fig_nucsite}.}
\label{fig_blender}
\end{center}
\end{figure}
%*************************************************************** 

 %------------------------------------------------
\begin{figure}[tb]
\begin{center}
\includegraphics[scale=0.55,angle=0]{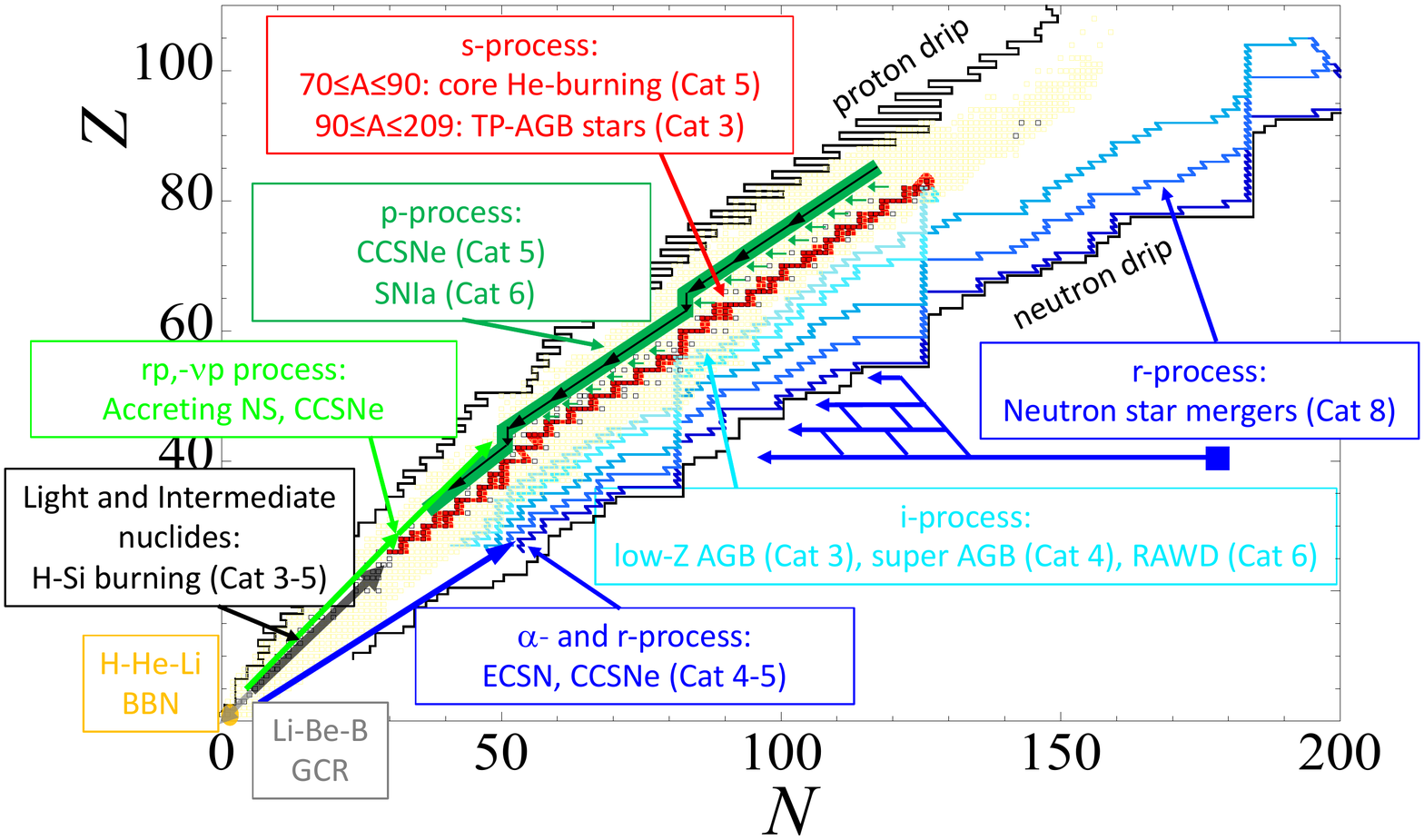}
\vskip -1.5cm
\caption{Schematic representation in the $(N,Z)$ plane of the different astrophysical sites responsible for the synthesis of the stable nuclides. The stellar sites are identified through the different categories (Cat.) defined in the main text. Categories 6 and 8 refer specifically to explosive processes in binary systems. The nucleosynthetic contributions by BBN and by GCR are also displayed. The open black squares correspond to stable or long-lived nuclei and the open yellow squares to the nuclei with experimentally known masses.  Nuclei with neutron or proton separation energies tending to zero define the neutron or proton ``drip lines'' (solid black lines), as predicted from a mass model.}
\label{fig_nucsite}
\end{center}
\end{figure}
%--------------------------------------------------
 
The way the composition of a galaxy evolves is represented in a very sketchy 
manner in  Fig.~\ref{fig_blender}. Let us consider the ISM (made of gas and dust) right after galaxy 
formation. Its composition is assumed to be essentially the one emerging from
the BBN, the standard model of which predicts the presence of 
significant amounts of just H, D, \chem{3}{He}, \chem{4}{He} and 
% 
% ****** footnote ******************
\chem{7}{Li}.\footnote{We neglect here any possibility of 
{\it pre-galactic}
nucleosynthesis of thermonuclear nature \cite{Norgaard76} by still 
putative pre-galactic very massive stars \cite{Ferrara98},
 or of spallation type  by ``cosmological cosmic 
rays'' \cite{Montmerle77}. 
These very early modifications of the Big Bang yields have been advocated 
at several occasions, but are usually not taken into account in galactic 
chemical evolution models.
}
%**end footnote  *****
 
Part of the ISM material is used to form stars which, through a large  
variety of nuclear reactions, transform the composition of their 
constituent material during their evolution.  
At one point or another during that evolution, some material may be 
returned to the ISM through various mechanisms, as already touched upon in Sect.~\ref{nucleo_stars}.

Stars of Cats. 3 and 4 contribute during their quiescent evolutionary phases to the enrichment of the stellar surfaces and/or ISM with H-, He- and C-burning ashes through more or less substantial steady mass losses (stellar winds) (see {\it e.g.} \cite{Karakas14}). This is also the case for stars of Cat. 5 that are massive enough to experience  very strong steady winds, either through strong radiation pressure (referred to as Wolf-Rayet stars; see {\it e.g.} \cite{Limongi18}), or through pulsational instabilities ({\it e.g.} \cite{Woosley17}). In addition, products of other nucleosynthetic processes that are not playing a leading role in the stellar energy budget can also be brought to the stellar surfaces and/or in the ISM. As displayed  in Fig.~\ref{fig_nucsite}, this is the case for stable nuclides heavier than iron produced by the s-process of neutron captures discussed in Section~\ref{prod_s}. The intermediate neutron-capture process, known as the ``i-process'', might also bring its share to some of these nuclides (Section~\ref{prod_i}). 
 
Through their explosive ejection of a more or less large amount of nuclearly processed material, supernovae of various types that terminate the evolution of some stars (especially of Cats. 5 and 6) are essential agents to the evolution of the nuclidic content of galaxies. As depicted schematically in
Fig.~\ref{fig_nucsite}, they are contributors to nuclides up to the vicinity of iron, as well as to heavier neutron-deficient and neutron-rich nuclides through the p-process (Section~\ref{prod_p}) and $\alpha$- and r-processes (Section~\ref{prod_r}).
 
In general, all the stellar material with altered composition is not returned to the ISM. Part is locked up in stellar residues (WDs, NSs, or BHs), and is normally not involved in the compositional evolution of the galaxies, except possibly for stars of Cats. 6 or 8 (explosion of WDs or NSs in binary systems).

 As pointed out in Section~\ref{nucleo_gcrs}, GCRs can also add some limited contribution to the galactic content of some nuclides, particularly Li, Be and B. At least in spiral galaxies like the Milky Way, a fraction of these produced nuclides could escape the galactic disc. Part of the supernova ejecta might also be ejected from the disc (though not sketched in Fig.~\ref{fig_blender}). 
In contrast, some material, possibly of Big Bang composition, might 
fall onto the disc (``infall'') from the galactic halo to 
dilute the stellar-processed material. 

If some rather clear conclusions are emerging from the very many studies of the highly complex problems raised by the observations and modeling of the evolution of galaxies, one has still to live in this field with many open problems and conflicting results, as summarized in {\it e.g.} \cite{Maiolino18}.
  
%******************************************************************************
\section{Nuclear needs for astrophysics} 
\label{nucdata}
%******************************************************************************

\begin{figure}[tb]
\begin{center}
\vskip-1.0truecm
\includegraphics[scale=0.55]{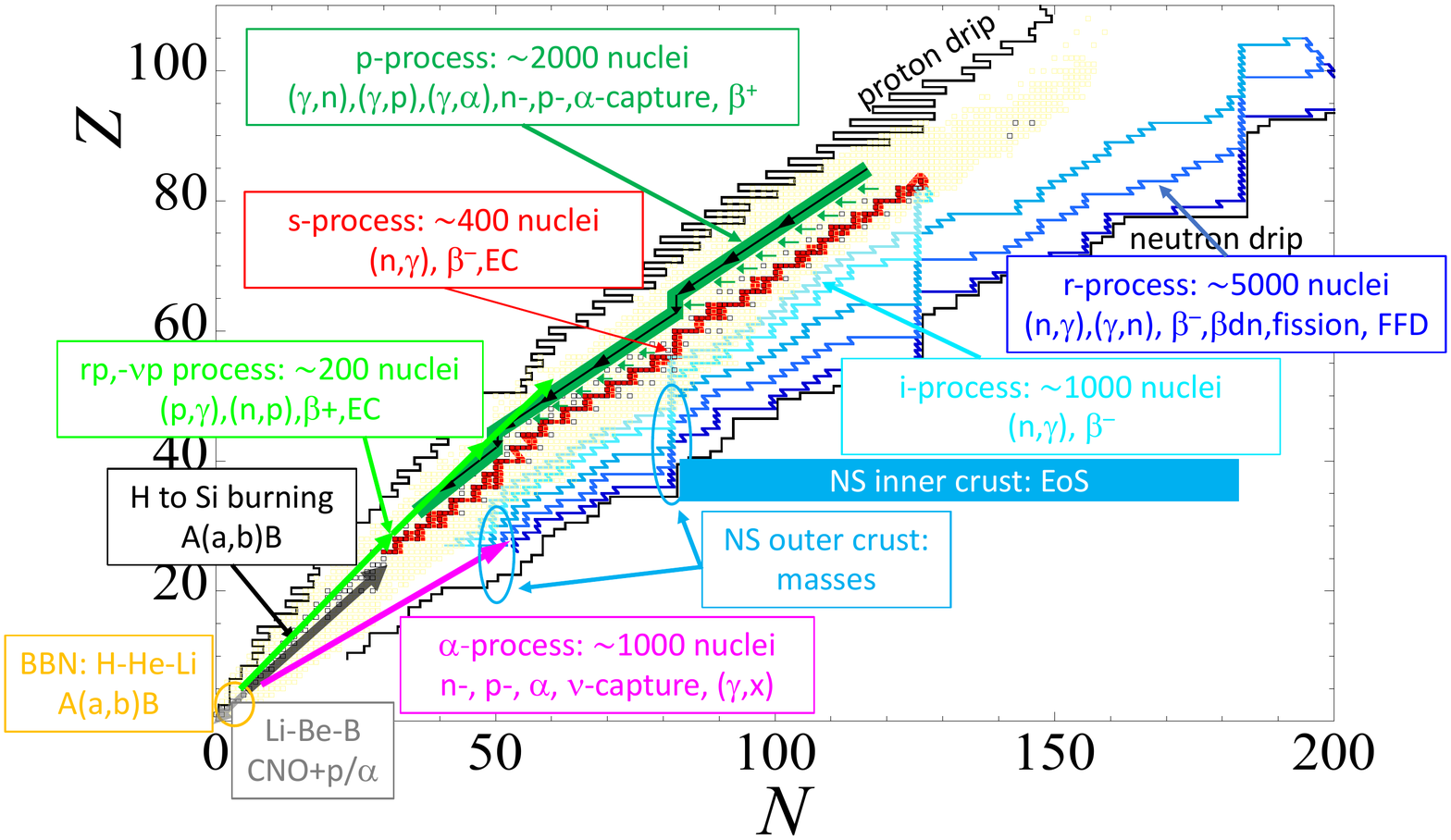}
\vskip -1.0cm
\caption{Schematic representation in the $(N,Z)$ plane of the different astronuclear physics applications, including nucleosynthesis processes, composition and structure properties of NSs. For each process, the nuclear needs are sketched. The open black squares correspond to stable or long-lived nuclei, the yellow squares to the nuclei for which masses have been measured.  Nuclei with a neutron or proton separation energies tending to zero define the neutron or proton ``drip lines'' (solid black lines), as predicted from a mass model.}
\label{fig_nucastro}
\end{center}
\end{figure}

As made clear in the previous sections, the Universe is pervaded with nuclear physics imprints at all scales. 
Figure~\ref{fig_nucastro} illustrates the various nuclear data needs for stellar structure, stellar evolution and nucleosynthesis applications in relation with the different astrophysical sites sketched in Fig.~\ref{fig_nucsite}. The modeling of nucleosynthesis is certainly the most demanding regarding nuclear data, some processes requesting the consideration of as many as thousands of nuclides linked by a huge amount of nuclear reactions (see especially Sections~\ref{prod_r} and \ref{prod_p}).
 
%****************************************************
%--------------------------------------------------

Impressive progress has been made over the last decades in the experimental and theoretical nuclear data hunt of relevance to astrophysics. Still, major problems and puzzles remain. In particular, experimental information only covers a minute fraction of the needs. This situation results namely from the necessary consideration of more or less highly unstable nuclides, most of them being at present impossible to produce in the laboratory. Even when dealing with stable nuclides, their interactions with charged particles operate in most stellar conditions at energies far below the Coulomb barrier (see {\it e.g.} \cite{Iliadis15}). This often implies extremely low cross sections that remain experimentally unreachable in the vast majority of cases. Last but not least, one has to keep in mind that nuclear processes may be more or less deeply affected by the astrophysical very high temperatures and/or densities that cannot be obtained in the laboratory. Theory is consequently badly needed to complement the experimental information.

\begin{figure}[tb]
\begin{center}
\vskip-1.50truecm
\includegraphics[scale=0.55]{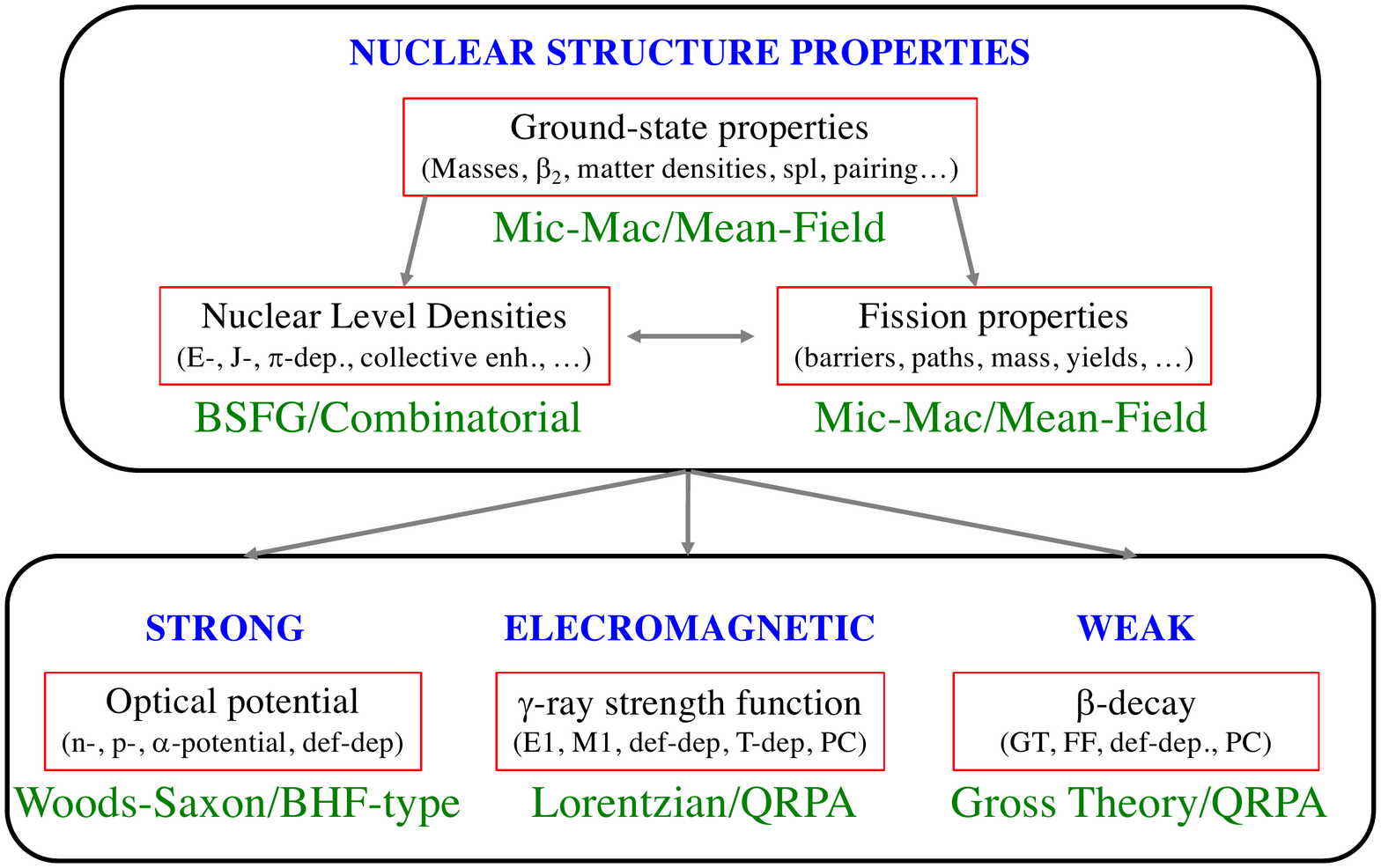}
\vskip -1.cm
\caption{Schematic representation of the nuclear data needs for astrophysics. Nuclear structure information of key importance is displayed. The most often adopted phenomenological (left) / (semi-) microscopic (right) models are given in green (see main text for details).}
\label{fig_nucmod}
\end{center}
\end{figure}

The nuclear models have to meet two major requirements. They have obviously to reproduce available experimental data as {\it accurately} as possible. This is a necessary, but not a sufficient condition. In addition, they have to be of the highest possible {\it reliability}. This requirement is best met by a microscopic, or at least semi-microscopic evaluation of the nuclear properties based as much as possible on first principles. Nowadays, microscopic models can reach the same level of accuracy as phenomenological more or less parametrized  models \cite{Goriely15b,Goriely17a}. The basic nuclear structure properties and interactions of prime astrophysical interest are schematically displayed in Fig.~\ref{fig_nucmod}, along with the phenomenological and (semi-)microscopic models the predictions of which have been most often adopted.

 %------------------------------------------------
%--------------------------------------------------
 
%*******************************************************
\subsection{Masses and nuclear structure properties}
\label{masses}
%******************************************************

Atomic masses pervade all chapters of nuclear physics and astrophysics.   
Figure~\ref{fig_nucastro} (yellow open squares) displays the approximately 2500 nuclides with experimentally known masses. Among them, 286 are naturally occurring, the remaining ones being artificially produced. The worldwide efforts devoted to mass measurements are reviewed in {\it e.g.} \cite{Lunney03,Blaum13,Dilling18,Lunney19,Oliveira19}.
As extended as it is, this  data set does not quite meet the astrophysics requirements. This is especially true when dealing with the r-process (Section~\ref{prod_r}), which involves a large number of neutron-rich 
nuclei unidentified in the laboratory.  
 
Along with experimental masses, many measured nuclear structure properties may help constraining or estimating reaction and decay rates. These include in particular radii, deformations and nuclear level schemes. 
Measured root-mean-square (rms) charge radii between H and Cm are regularly compiled \cite{Angeli13}, as well as known electric quadrupole moments of ground and excited states \cite{Stone16}. Experimental information for nuclei in their ground or excited states are also compiled in various nuclear data libraries and are regularly evaluated for nuclear applications. In particular, the Reference Input Parameter Library (RIPL) of the International Atomic Energy Agency (IAEA) includes updated libraries of measured discrete energy level schemes, deformations, neutron resonance parameters, optical models, photon strength functions and fission properties \cite{Capote09}. 

Modern mass models not only try to reproduce at best experimental masses and mass differences, but also quantities such as charge radii, quadrupole moments, giant resonances, fission barriers, shape isomers, and infinite nuclear matter properties \cite{Lunney03,Goriely17a}. Some of these properties have a direct impact on the predictions of {\it e.g.} reaction rates.

Widely used microscopic-macroscopic mass formulae have extended the early 1935 liquid-drop mass models \cite{Weizsacker35} by introducing microscopic features and by correcting the macroscopic component of the model with the inclusion of infinite and semi-infinite nuclear matter properties and with the introduction of the finite range character of the nuclear forces (for a review, see \cite{Lunney03}). In this framework, the macroscopic and microscopic features are treated independently, both parts being connected exclusively through a parameter fit to
experimental masses. The most advanced form of  the  macroscopic-microscopic models is referred to as the Finite Range Droplet Model (FRDM) \cite{Moeller16}.This model has been widely used in astronuclear physics.  Despite its success in fitting the 2408 $Z\ge8$ measured masses \cite{Wang17} with an rms deviation of 0.60 MeV, it suffers from major shortcomings, such as the incoherent link between the macroscopic part and the microscopic correction, the instability of the mass prediction to different parameter sets, or the instability of the shell corrections. These deficiencies make the model quite unreliable when moving away from the experimental mass region.

With a view to astrophysical applications involving highly unstable nuclides, a series of 32 mass models have been developed recently based on the Hartree-Fock-Bogo\-liubov (HFB) method with Skyrme and contact-pairing forces, together with phenomenological Wigner terms and correction terms for 
the spurious collective energy within the cranking approximation (see \cite{Goriely16a} and references therein).
All the model parameters are fitted to essentially all the experimental mass data. 
While the first model of this series aimed at proving that is was possible to reach a low rms deviation with respect to  all the experimental masses available at the time, most of the subsequent models were developed to further widely explore the parameter space or to take into account additional constraints. These include in particular a sensitivity study of the mass model accuracy and extrapolation predictions to major changes in the description of the pairing interaction, the spin-orbit coupling or  the nuclear matter properties, such as the effective mass, 
the symmetry energy and the stability of the Equation of State (EoS).
 
The 32 HFB models give an rms deviation to the 2408 measured masses \cite{Wang17} ranging between 0.50~MeV for HFB-27 and 0.81~MeV for HFB-1. These rms deviations can be compared to those obtained with other global models, such as the Gogny-HFB mass model with the D1M interaction \cite{Goriely09a} characterized by an rms of 0.80~MeV, the covariant density functional theory with finite-range density-dependent meson-nucleon couplings with more than 1.2~MeV \cite{Pena16}, or the 2012 FRDM version \cite{Moeller16} with 0.60~MeV. 

\begin{figure}
\begin{center}
\includegraphics[scale=0.4]{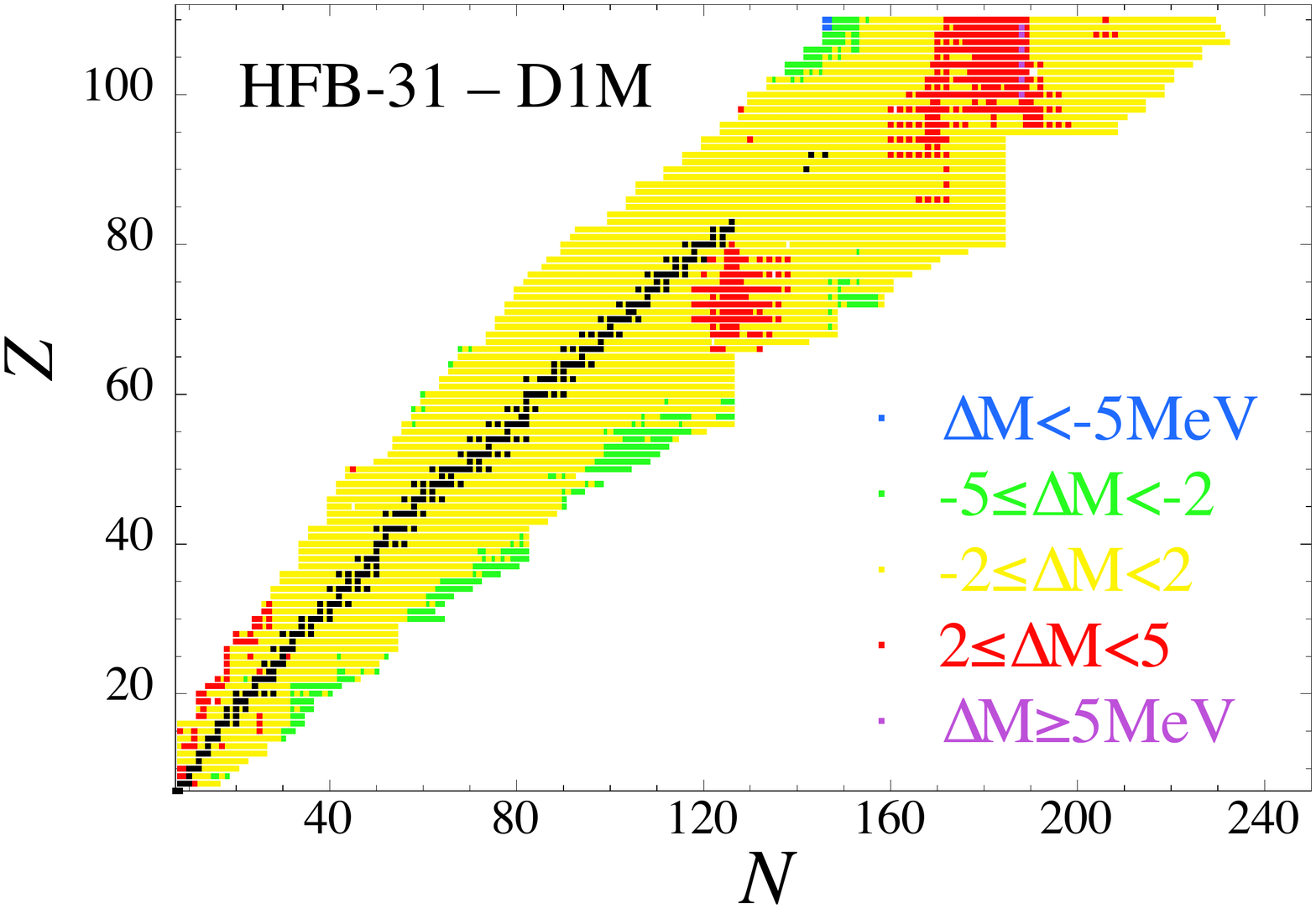}
\vskip -0.5cm
\caption{Mass differences between the HFB-31 \cite{Goriely16a} and D1M \cite{Goriely09a} models for the 8500 nuclei from $Z=8$ up to $Z=110$ located between the HFB-31 proton and neutron driplines.}
\label{fig_hfb-d1m}
\end{center}
\end{figure}

Quite clearly, deviations between the different HFB mass predictions can become significant when moving away from the valley of stability, reaching values of about 3~MeV at the neutron drip line for the heaviest nuclides. Differences are noted not only in the rigidity of the mass parabola, but also in the description of the shell gaps or pairing correlations (see \cite{Goriely14a} for a detailed analysis). A similar situation is found when comparing models of the HFB family to other models. This is illustrated in Fig.~\ref{fig_hfb-d1m} comparing the HFB-31 and D1M predictions. Deviations as high as about 5~MeV are found, especially around the $N=126$ and $N=184$ shell closures. These differences have a disastrous effect on {\it e.g.} neutron capture rates, which can vary by as much as 3 to 5 orders of magnitude.
 
A relativistic or non-relativistic mean-field approach has also been adopted to predict nuclear structure properties \cite{Bender03}. Through a fit to only a limited set of experimental masses, their predictions lead to rms deviations typically larger than 2--3~MeV, increasing to as much as about 5~MeV when the popular SLy4 force is adopted \cite{Stoitsov03}. In such conditions, these mass models are not recommended for astrophysics applications, such as the r-process (Section~\ref{prod_r}).

%------------------------------------------------
%--------------------------------------------------

Further improvements in the mass models are clearly called for. These include development of relativistic and non-relativistic mean field models with state-of-the-art corrections, like the quadrupole or octupole correlations by the Generator Coordinate Method, and a proper treatment of odd-$A$ and odd-odd nuclei with time-reversal symmetry breaking. These models should reproduce not only nuclear masses at best, but also as many experimental observables as possible. These include charge radii and neutron skin thicknesses, fission barriers and shape isomers, spectroscopic data such as the $2^+$ level energies, moments of inertia, but also infinite (neutron and symmetric) nuclear matter properties obtained from realistic calculations, as well as observed or empirical properties of NSs, like their maximum mass or mass-radius relations \cite{Fantina13}. 

%**********************************************************************
\subsection{Nuclei at high temperatures}
\label{masses_hight}
%*******************************************************************

One of the specificities of astronuclear physics applications concerns the 
high temperatures often prevailing in stellar plasmas. In these environments,
nuclei may exist not only in their  ground state, but in excited states as well, which affects more or less deeply their properties.

In hot and dense environments, nuclei can indeed be excited/de-excited namely by (i) photon absorption and emission (both spontaneous and induced), (ii) nuclear excitation by free electron capture and internal conversion, (iii) inelastic and superelastic electron scattering, as well as (iv) nuclear excitation by 
electronic transition and bound internal conversion \cite{Gosselin07}.

In the interiors of non-exploding or exploding stars, these excitation and de-excitation transition rates are fast enough for thermodynamic equilibrium to hold,  at least locally, to a high level of accuracy. In this case, the relative populations of the nuclear excited states follow a statistical Maxwell-Boltzmann distribution ({\it e.g.} \cite{Cox68}). In such conditions, the thermal population of an excited state becomes
significant if its excitation energy is close enough to the thermal energy $kT \approx 8.6 \times (T/10^8$K)~keV at a temperature $T$. The population of low-lying levels in odd-odd nuclei can thus be especially high. The  fraction of nuclei in excited states relative to the ground state population is given by the nuclear partition function defined as
%***************************************
\begin{equation}
G(T)=\sum_\mu \frac{2J^\mu+1}{2J_0+1}\exp\left(\frac{\varepsilon^\mu}{kT}\right) ,
\label{eq_partfunc}
\end{equation}
%****************************************
where $J_0$ is the spin of the ground state and the summation runs over all excited states 
$\mu$ of spin $J^\mu$ and energy $\varepsilon^\mu$. The information required to evaluate $G(T)$
for a large number of stable and unstable nuclei can be found namely in the RIPL database \cite{Capote09}. It may be lacking at high enough temperatures, especially for exotic nuclei. As a consequence, resort to a nuclear level density model is often mandatory (Section~\ref{reac_hf}). 

Low-lying isomeric states may more or less severely hamper the thermal equilibration between ground and excited states due to the slowness of the connecting transitions.  At high enough temperature, however, the thermalization may be obtained through multi-step electromagnetic links involving higher excited states. Cases of astrophysical interest are \chem{26}{Al} \cite{Coc99}, $A=113-115$ isotopes of Ag, Cd and In \cite{Nemeth94}, \chem{176}{Lu}  and \chem{180}{Ta} \cite{Gosselin10}. For example, the ground state and 228 keV isomeric level ($T_{1/2} \sim 6.3$ s) of \chem{26}{Al} are expected to have time to reach thermal equilibrium 
in realistic stellar conditions only at temperatures in excess of about $4 \times 10^8$ K \cite{Coc99}.

The population of excited states not only affects the basic nuclear properties, like masses which turn out to be temperature dependent (as is the difference between atomic and nuclear masses due to the loss of bound atomic electrons by ionization), but also the stability against various channels, like decay rates and reaction cross sections, as discussed in Sections~\ref{decay_beta} and \ref{reac_thermo}.

%***********************************************************************
\subsection{Nuclei at high densities and Equation of State}
\label{eos}
%**********************************************************************

At temperatures exceeding typically $T=10^{10}$~K, a Nuclear Statistical Equilibrium (NSE) involving electromagnetic and nuclear processes generally holds. In order to obtain a full thermodynamic equilibrium, weak interaction processes have to participate as well, implying in particular that the neutrinos involved have a mean free path smaller than the typical dimensions of the considered stellar zones. This can be obtained at densities in excess of about $\rho \gsimeq 10^{-2} \rho_0$, where  $\rho_0 \approx 3 \times 10^{14}$~g/cm$^3$
 is the nuclear saturation density, 
as it is the case namely in the core of CCSNe and in isolated or merging NSs. In fact, the maximum baryon number density in NSs can reach approximately ten times $\rho_0$, as deduced from the observation of NSs with masses estimated to be of the order of 2 M$_{\odot}$ \cite{Fonseca16}. In these conditions of full thermodynamic equilibrium under extreme conditions, the nuclear binding has to be expressed in terms of a nuclear Equation of State (EoS) that describes the energy density and pressure of a system of nucleons and/or nuclei in terms of the average baryon number density \cite{Oertel17,Burgio18,Li19}. The EoS approach replaces the description of nuclei as individual entities and the use of a transport equation simulating the pathways of neutrinos streaming out of the stellar matter.
 
The construction of an EoS relies on methods describing the thermodynamic properties of a strongly interacting  many-body system. Different approaches have been proposed. They include ab-initio many-body models based on realistic few-body interactions, like Brueckner-Hartree-Fock, coupled clusters, quantum Monte Carlo and self-consistent Green's function methods, as well as chiral effective field theory \cite{Baldo12,Carlson15}. More phenomenological mean-field models are also used, based on effective 
interactions such as Skyrme or Gogny forces in non-relativistic models (see {\it e.g.} \cite{Fantina13,Sellahewa14,Pearson18}), or meson-exchange forces in relativistic approaches (see {\it e.g.} \cite{Typel99}). EoS models are constrained by a variety of observables, such as laboratory measurements 
of nuclear properties and reactions, so-called realistic ab-initio calculations, or astronomical observations (see  {\it e.g.}\cite{Fantina13,Oertel17,Burgio18,Lattimer07,Tsang19}). Detailed comparisons of different EoSs and their impact on binary systems containing compact objects as well as the dynamics, neutrino-driven wind properties and nucleosynthesis in CCSNe can be found in {\it e.g.} \cite{Oertel17,Burgio18}.
 
%***************************************
\subsection{Nuclear decays via the weak interaction}
\label{decay_general}
%**************************************

The most familiar forms of weak interaction transformations in the laboratory are the 
$\beta^-$-decay (e$^-$ emission), $\beta^+$-decay (e$^+$ emission), 
and orbital-e$^-$ capture processes. As discussed below, the relative importance and probabilities of these three processes may be quite different in laboratory and astrophysical conditions. Additional mechanisms develop specifically in astrophysical conditions.

%***********************************************
\subsubsection{$\beta$-decays}
\label{decay_beta}
%**********************************************

As demonstrated in particular by the data gathered by \cite{Audi17}, spectacular progress has been made in the measurement of $\beta$-decay half-lives. These experimental advances have in particular made reachable the study of very neutron-rich nuclei with high $Q_{\beta^-}$-values of great importance in nuclear physics, as well as in astrophysical applications, particularly in the modeling of the r-process (see Figs.~\ref{fig_nucsite} and \ref{fig_nucastro}; Section~\ref{prod_r}). Data remain scarce, however, for heavy nuclei in the approximate $Q_{\beta^-} \gsimeq 10$ MeV range \cite{Audi17}.  

As is well known, half-lives decrease rapidly with increasing  $Q_{\beta^-}$-value. One pleasing feature of high $Q_{\beta^-}$ transitions is that the half-lives are close to decrease monotonically, and even to converge to a common value (within perhaps a 
factor of even less than about five) in the range of the highest  $Q_{\beta^-}$ values of 
relevance to the r-process. This convergence even holds when comparing ``light'' and ``heavy'' nuclei, which may be surprising (see {\it e.g.} \cite{Arnould07} for a discussion of this effect).

In order to understand $\beta$-decays in general, and those of nuclei far off the line  of 
stability in particular, it is often much more profitable to work with so-called $\beta$-strength 
functions $S_\Omega$ rather than with matrix elements of individual 
 transitions. A broad overview of the general structure of the $\beta$-strength distributions can be found in {\it e.g.} \cite{Arnould07} (see also below).   
 
\vskip0.2truecm
{\bf Effect of electron degeneracy}. One effect that can contribute to a deviation from the laboratory $\beta^-$-decay half-lives relates to 
the possible reduction of the electron phase-space as the result of the degeneracy of the Fermi-Dirac electron gas that is encountered in a variety of stellar situations.\footnote
%****** footnote *******
{Under local thermodynamic equilibrium conditions that prevail in most stellar interiors \cite{Cox68}, the electron gas is well described by the classical Maxwell-Boltzmann distribution law. 
In various situations, however, use of the Fermi-Dirac distribution is made necessary. The electrons are then
referred to as ``degenerate,'' the degree of this degeneracy increasing with its  ``Fermi energy" ({\it i.e.} with increasing density and decreasing temperature).}
%***** end footnote ********
%
An extreme result of the electron degeneracy is the possible stabilization of $\beta^-$-unstable nuclei in certain astrophysical situations.
 
\vskip0.2truecm
{\bf Contribution of nuclear excited states}. As emphasized earlier (Section~\ref{masses_hight}), the excited levels and ground state of a nucleus are very often populated in thermal equilibrium, possible exceptions being certain isomeric states. These various levels may thus contribute to the decay of a nucleus, so that its effective
$\beta$-decay half-life may strongly depart from the laboratory value. 

A classical example of the importance of the $\beta$-decays from nuclear
excited states concerns \chem{99}{Tc}. Technetium observed at the surface of certain TP-AGB (Cat.~3) stars 
is believed to be \chem{99}{Tc}, which is a product of the s-process of neutron captures 
developing in the He-burning shell of these stars (Fig.~\ref{fig_nucsite} and Section~\ref{prod_s}). 
Under these conditions (typical temperatures in excess of about $10^8$~K), the 141 and 181 keV excited states of \chem{99}{Tc} can contribute significantly to the effective decay rate. In fact, the half-life drops dramatically from the laboratory (ground state) value of $T_{1/2} \approx 2.1 \times 10^5$ y to some 10 y at 
$T \approx 3 \times 10^8$ K. This example illustrates quite vividly that the decay of thermally-populated 
excited states may alter laboratory half-lives most strongly in the 
following conditions: (1) the ground-state decay is slow, as a result 
of selection rules, and (2) (low-lying) excited states can decay through
less-forbidden transitions to the ground and/or excited states of the 
daughter nucleus. Needless to say, the temperatures have to be high enough for the relevant 
excited levels to be significantly populated (see Eq.~\ref{eq_partfunc}).
 
\vskip0.2truecm
{\bf Effect of ionization and bound-state $\beta$-decay}. The loss of bound electrons is common place in
a large variety of stellar conditions, and especially at high temperatures.  Atoms accelerated to relativistic 
cosmic-ray energies are also stripped of their electrons. Highly ionized atoms can also be produced and their half-lives measured in the laboratory, as extensively reviewed by \cite{Litvitov11}.

Ionization may influence the nuclear half-lives in 
several ways. It has first the obvious effect of reducing the probability of capture 
of bound electrons. A less trivial consequence relates to the possible development of
the process of ``bound-state $\beta$-decay,'' for which the emitted electron is captured 
in an atomic orbit previously vacated (in part or in total) by ionization. Note that this process accompanies
$\beta^-$ decays even in the absence of ionization, but its relative contribution is quite insignificant in such conditions. 
%
%*****************************************************************
\begin{figure}[tb]
\hskip-0.2truecm
\includegraphics[scale=0.3,angle=0]{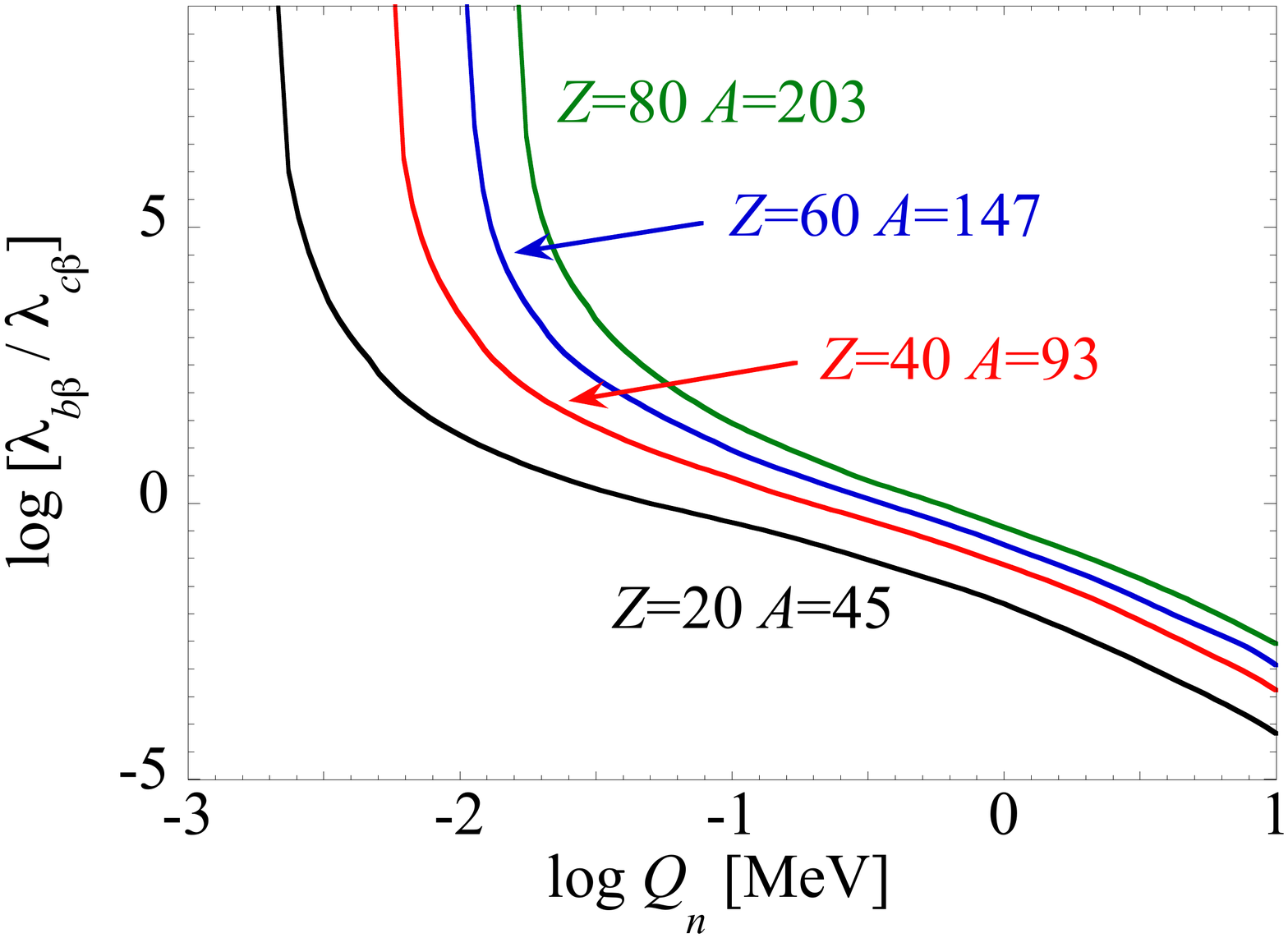}
\vskip-6.3truecm 
\hskip8.5truecm
\includegraphics[scale=0.3]{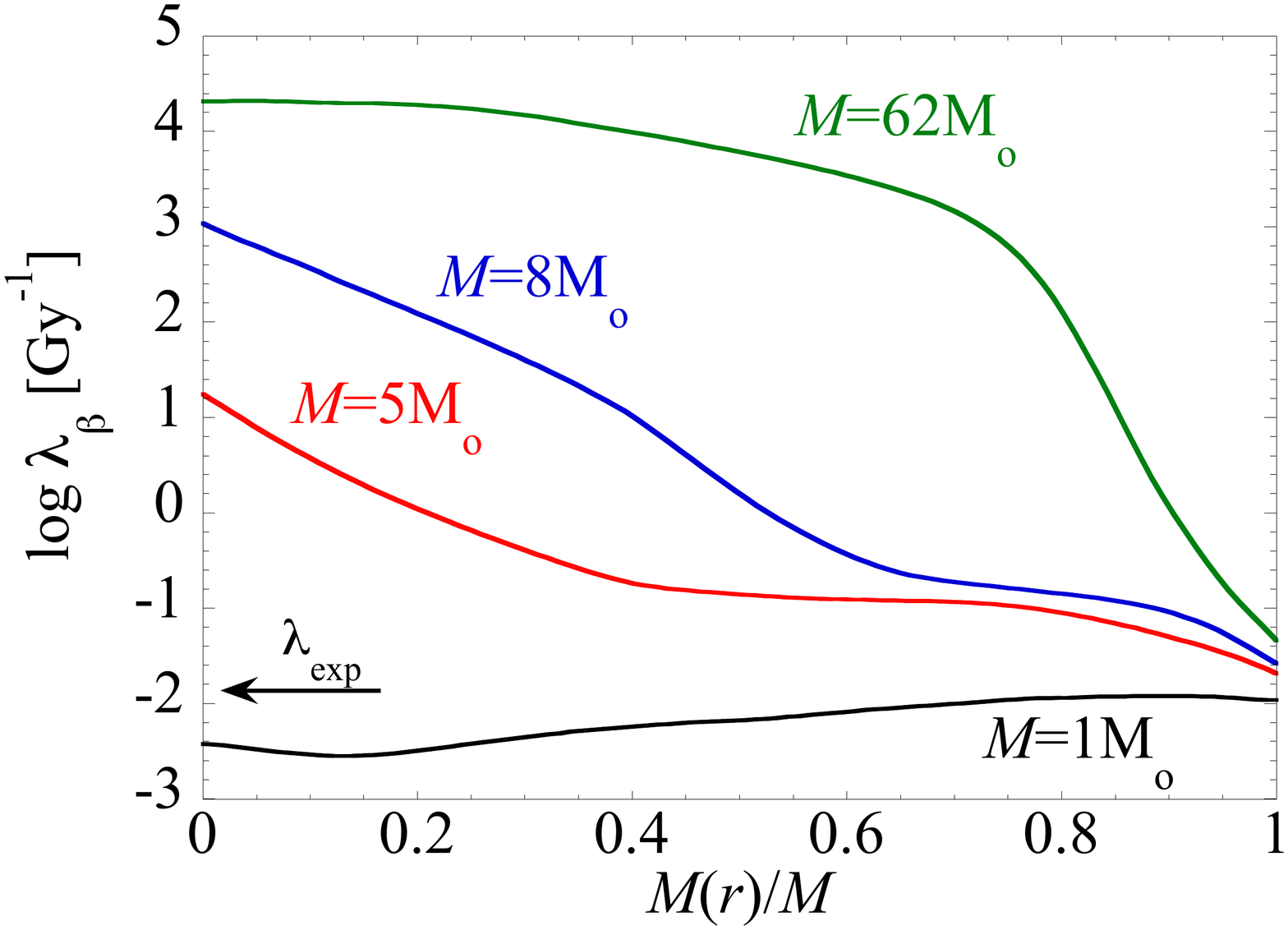}
\caption{{\it Left panel}: Ratio of the bound-state $\beta^-$-decay rates $\lambda_{b\beta}$ to the continuum
 $\beta^-$ decay rates $\lambda_{c\beta}$ versus the neutral atomic mass difference $Q_n$ in fully ionized atoms for different mass and charge numbers. Only allowed and non-unique first-forbidden transitions are taken into account. The creation of an electron in orbitals up to 6$s$ is considered. More details can be found in \cite{Takahashi87};
{\it Right panel}: Effect of ionization on the \chem{187}{Re} $\beta$-decay rate as a function of location inside stars of different masses $M$ (expressed in the mass $M_\odot$ of the Sun) in their phase of central H burning (referred to as Main Sequence stars). The location is expressed in term of the mass fraction $M(r)/M$, where $M(r)$ is the mass inside a sphere of radius $r$. Strong deviations with respect to the laboratory $\beta$-decay rate $\lambda_{\rm exp}$ are observed as a result of ionization, which is the largest in the innermost stellar layers. The stellar rates converge to the laboratory value in the outer layers, where Re becomes neutral. More details can be found in  \cite{Yokoi83}. }
\label{fig_bound}
\end{figure}
%****************************************************
 
 A theoretical conjecture of the existence of bound-state $\beta^-$-decay goes back to 
more than 70 years ago \cite{Daudel47}, but its experimental confirmation had to await until quite recently. 
In fact, it had already been realized in the early 1980s that bound-state $\beta$-decay can be responsible for the transformation of some stable nuclides, like \chem{163}{Dy}, in hot enough stellar interiors \cite{Takahashi83}. 
Subsequently the interest for this process in astrophysics has been 
growing, in particular with regard to specific aspects of the s-process,
and in relation to some cosmochronological studies (Section~\ref{chrono_reos}). An extended set of bound-state $\beta$-decay rates for nuclei mainly involved in the s-process (see Fig.~\ref{fig_nucsite} and Section~\ref{prod_s}) and for wide ranges of astrophysical conditions is found in \cite{Takahashi87}.
Figure~\ref{fig_bound} clearly demonstrates that bound-state $\beta$-decay is of special importance in the case of very low $Q_{\beta^-}$-transitions for which the probability of presence of the emitted electron at the bound electronic orbits is especially high.  
  
The possibility that bound-state $\beta$-decay can lead to the decay of the terrestrially stable nuclide 
\chem{163}{Dy} to which \chem{163}{Ho} decays through electron capture in laboratory conditions with a 
$Q_\beta$ value of 2.6 keV may have consequences for the synthesis of the p-nuclide \chem{164}{Eu}, as first noticed by \cite{Joukoff69}. This bound-state decay has been spectacularly confirmed by a storage-ring experiment at GSI \cite{Jung92}. The measured half-life of the fully ionized \chem{163}{Dy}$^{+66}$ is 47 d, in excellent agreement with the theoretical prediction of 50 d \cite{Takahashi87a}. 

The radionuclide \chem{187}{Re} is also a candidate of choice for the bound-state $\beta$-decay process, in view of its very low neutral atom $Q_{\beta^-}$-value of about 2.5 keV. Note that a value of 2.64 keV is adopted by \cite{Takahashi83}, while \cite{Kienle98} reports 2.66 keV. A recent high-precision measurement leads to 2.492(30$_{stat}$)(15$_{sys})$ keV \cite{Nesterenko14}. These values are quite close to each others, but the bound-state decay rates are very sensitive to the precise $Q_{\beta}$-values in the keV range, as illustrated in Fig.~\ref{fig_bound}. It has been deduced from a GSI storage-ring experiment that the fully-ionized \chem{187}{Re^{+75}} has a half-life of 34 y \cite{Bosch96}, which is more than $10^9$ times shorter than the value for the neutral atoms. This value is in fair agreement with the 12 y prediction made by \cite{Takahashi87a} before the experiment.  

In view of its half-life commensurable with the age of the Galaxy, \chem{187}{Re} can be used for cosmochronological purposes. Bound-state $\beta$-decay complicates substantially the galactic age predictions in view of the effective half-life dependence on the precise location of a \chem{187}{Re} nucleus inside a star, as illustrated in Fig~\ref{fig_bound}. Bound-state $\beta$-decay may also be used as a tool to determine unknown 
$\beta$-decay matrix elements influencing the design of a \chem{205}{Tl} neutrino detector (\cite{Arnould99} for references). Finally, it has to be stressed that the evaluation of bound-state $\beta$-decay rates in stellar plasmas is far from being straightforward, even if the relevant nuclear matrix elements are known. The problems relate in particular  to electron screening effects, which have been estimated so far from a finite-temperature Thomas-Fermi model \cite{Takahashi83}.
  
\vskip0.2truecm
 {\bf $\beta$-decays of radionuclides in cosmic-rays}. The cosmic-ray abundances of some radioactive nuclides can be profitably used for estimating the age of these high-energy particles \cite{Simpson88}, or more precisely the time the cosmic rays have been confined within the disc, and possibly the magnetic halo of the 
%*** footnote ***
Galaxy.\footnote{There is substantial evidence from the observation of
synchrotron emissions that a galaxy containing relativistic cosmic-ray 
electrons in its disc develops a magnetic halo.
The GCRs could spend part of their confinement time 
within such halos.}
%*** end footnote ***
This concerns in particular \chem{54}{Mn}, the neutral atoms of which undergo orbital electron captures 
($T_{1/2} = 312$ d). In high-energy cosmic rays, the orbital electrons are stripped off to 
leave bare \chem{54}{Mn}, which is expected to transform very slowly 
via $\beta^-$ and $\beta^+$-decays.  
The theoretical evaluation of the rates of these laboratory-unknown 
transitions encounters enormous difficulties \cite{Casse73}, and their
measurements are eagerly awaited. Recently, \chem{26}{Al} 
has been resolved from the stable \chem{27}{Al} in the cosmic radiation 
\cite{Simpson98}. Due to the suppression of 
its e$^-$-capture mode resulting from its complete 
ionization, the cosmic-ray  \chem{26}{Al} half-life is increased up to 
$8.7 \times 10^5$ y from the laboratory value of $7.2 \times 10^5$ y.
Another interesting case, in particular for $\gamma$-ray astronomy, 
concerns \chem{44}{Ti}. Its half-life, the laboratory value of which is close to
60 y, may be increased in young supernova remnants because of its possibly substantial ionization
\cite{Mochizuki99}. 
 
 %***************************************
 \subsubsection{Continuum electron captures}
 \label{decay_continuum}
 %**************************************
 
The continuum-e$^-$ capture process is quite common in stellar plasmas, and often
overcomes orbital-e$^-$-captures in highly-ionized stellar material where atoms are immersed in a sea of free  electrons. Continuum-e$^-$ captures display their most spectacular effects in 
situations where the electrons are degenerate. 
Highly stable nuclei in the laboratory may well become $\beta$-unstable 
if indeed the electron Fermi energy is large enough for allowing 
endothermic transitions to take place through the captures of free
electrons with energies exceeding the energy threshold for these 
transitions. Clearly, the higher the electron Fermi energy, the more endothermic the
transitions may be. 

Among the most spectacular endothermic free e$^-$-captures that can 
be encountered in astrophysics, let us mention those on \chem{14}{N}, 
\chem{16}{O}, \chem{20}{Ne} or \chem{24}{Mg} operating in ECSNe (Cat.~4 stars), as discussed in {\it e.g.} \cite{Leung19} (see also \cite{Suzuki19,Kirsebom19} for a recent discussion of the \chem{20}{Ne} case). 
Continuum electron captures have also a remarkable impact on the final stages of the evolution of Cat. 5 stars eventually transforming into CCSN events, which notably require the construction of a high-temperature high-density EoS (Section~\ref{eos}; see {\it e.g.} \cite{Sullivan16,Raduta16,Nagakura18,Titus18,Pascal19}). CCSN models are found to be most sensitive to neutron-rich nuclei in the upper $pf$- and $pfg/sdg$-shells ($A \approx 65 - 120$), and especially around the $N=50$ closed shell just above the doubly-magic nucleus \chem{78}{Ni} \cite{Titus18,Pascal19}.  

A special case of e$^-$-captures is encountered in the so-called Urca-process consisting of alternate e$^-$-captures 
$(Z,A)+e^- \rightarrow (Z-1,A)+\nu_e$ and $\beta^-$-decays $(Z-1,A) \rightarrow (Z,A)+e^- + \bar{\nu}_e$ having the same initial and final nuclides $(Z,A)$. This process has a substantial effect in particular on the cooling of accreting WDs (Cat. 6 stars) and on the eventual ECSN explosions as a result of the release from the stars of 
$\nu - \bar{\nu}$ pairs produced from the Urca-process acting especially on \chem{20}{Ne}, \chem{23}{Na}, \chem{24}{Mg} and \chem{25}{Mg} emerging in rather high quantities from carbon burning (see {\it e.g.} \cite{Schwab17}, and references therein). The effect of the Urca-process on some intermediate-mass nuclei has also been scrutinized \cite{Bravo19}. Neutron star matter may also be the site of the special Urca process 
n $\rightarrow$ p + $l+ \bar{\nu}_l$, p + $ l \rightarrow$ n + $\nu_l$ where $l$ is a lepton, electron or muon, and 
$\nu_l$ is the corresponding neutrino, or the so-called modified Urca process 
n + N $\rightarrow$ p + N + $l + \bar{\nu}_l$, p + N + $l \rightarrow$ n + N + $\nu_l$, 
where N is the additional nucleon which relaxes the momenta restrictions. These processes may have a strong impact on the energetics of supernova explosion and on the cooling of NSs \cite{Shternin18} (and references therein).

The evaluation of the free-e$^-$ capture rates may benefit from the laboratory knowledge of the rate of the inverse $\beta^+$-decays. Advantage of this is taken in \cite{Takahashi87} for continuum electron captures by nuclei not too far from the valley of stability that can be involved in the s-process (Section~\ref{prod_s}). In many instances of astrophysical interest, this information is insufficient, however. Intermediate-energy 
($\gsimeq 100$MeV/u) charge-exchange reactions like ($n,p$), ($d$,\chem{2}{He}), and ($t$,\chem{3}{He})   connecting the same initial and final states as electron captures have been studied experimentally.  Unfortunately, they provide information only on transitions from the ground state of the parent nucleus, so that high-temperature effects on the capture rates cannot be obtained. In addition, a relationship between forbidden transition strengths and charge-exchange cross sections has not been established, making it difficult to extract the relevant strengths.

\begin{figure} [tb]
\vskip-0.5truecm
\center
\includegraphics[scale=0.6]{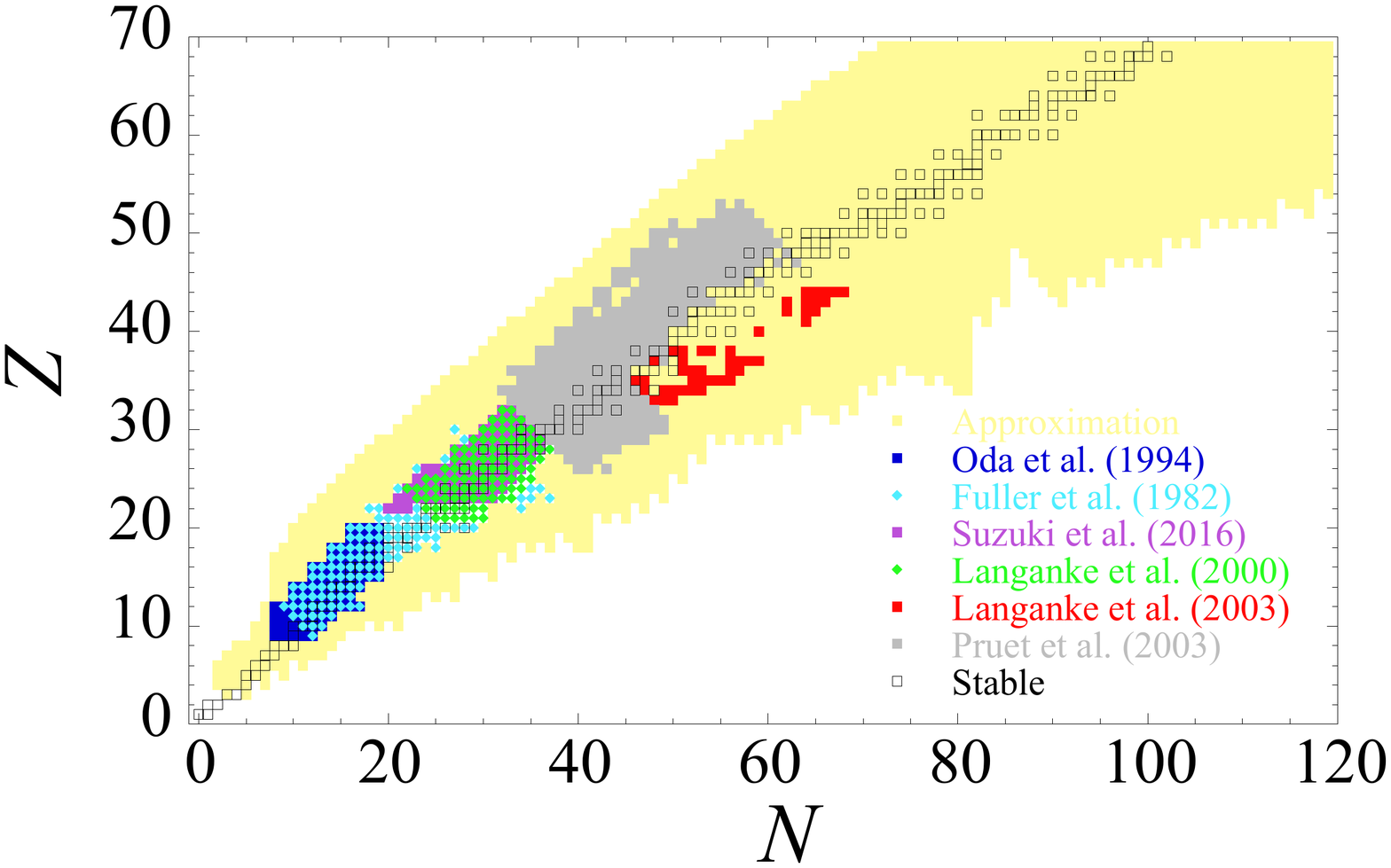}
\vskip-1.0truecm
\caption{Nuclei for which continuum e$^-$-capture rates have been evaluated by Fuller {\it et al.}  \cite{Fuller82}, Oda {\it et al.} \cite{Oda94}, Langanke {\it et al.} \cite{Langanke00,Langanke03}, Pruet {\it et al.} \cite{Pruet03}, Suzuki {\it et al.} \cite{Suzuki16} or approximated (parametrized fit to shell model) \cite{Langanke03}. The black open squares correspond to stable nuclei.}
\label{fig_betalib_NZ}
\end{figure}

Various models have been developed to predict electron capture rates, and in particular (Q)RPA-type and (Monte-Carlo) shell-model calculations. Approximate capture rates expressed in an analytic form have also been developed. The reader is referred to \cite{Sullivan16} for references (see also Fig.~\ref{fig_betalib_NZ}). Shortcomings in the QRPA and shell-model approaches are of similar nature as those affecting the evaluations of the $\beta^-$-rates (Section~\ref{decay_models}). In particular, the shell-model estimates are challenging for astrophysically important nuclei beyond the $pf$-shell, while the QRPA rates appear to systematically overestimate those derived from charge-exchange experiments. Reliable microscopic calculations of capture rates by exotic neutron-rich nuclei are still largely missing.

%****************************************************
%*************************************************** 

In summary, much experimental and theoretical efforts are still required in order to provide more secure estimates of the continuum electron capture rates of astrophysical interest. In particular, investigation of nuclei in the $A \simeq 80$ and $N \simeq 50$ region should take precedence, as changes in their
electron-capture rates are predicted to have a significant impact on the modeling of the pre-SN evolution and CCSN event.
 
Positrons captures are also of importance in certain stellar situations,  and especially in high-temperature 
(typically $T > 10^9$ K) and low-density locations. In such conditions, a rather high concentration of positrons can be reached from an ${\rm e}^- + {\rm e}^+  \leftrightarrow \gamma + \gamma$ equilibrium which favors the ${\rm e}^- {\rm e}^+$ pairs. The competition (and perhaps equilibrium) between positron captures on neutrons and electron captures on protons is an important ingredient of the modeling of CCSNe \cite{Janka17,Janka12}. 

%****************************************************
\subsubsection{Photo-beta disintegrations}
\label{decay_photobeta}
%***************************************************

A competitor to the continuum $e^+$-capture is the so-called photo-beta disintegration, $\gamma + (Z,A) \rightarrow (Z+1,A) + {\rm e^-} + \bar{\nu_e}$, in which a photon virtually dissociates into an electron-positron pair, the positron being absorbed by the initial nucleus $(Z,A)$. This transformation can take place not only in the case of an exothermic $(Z,A) \rightarrow (Z+1,A)$ transition, but also in an endothermic one induced by a high enough photon energy, the distribution of which is given locally by the black body Planck distribution in stellar conditions.  

The photo-beta process and its importance relative to the continuum $e^+$-capture has been studied in the framework of the p-process (see Fig.~\ref{fig_nucsite} and Section~\ref{prod_p}) \cite{Arnould67,Kopytin13}. It is concluded that it could at best be of some efficiency in the synthesis of some specific p-nuclides in a restricted range of stellar conditions.

%****************************************************
\subsubsection{$\beta$-delayed processes}
\label{decay_delayed}
%****************************************************

Beta-decays of highly unstable nuclei may produce daughters that can emit nucleons or are unstable to $\alpha$-decay or fission. These processes are referred to as ``delayed-particle emission'' or ``delayed fission''. The products of these delayed processes may in their turn experience delayed neutron emission and/or fission. The occurrence of these processes is of course conditioned by the energetics of the transformations. In particular, the decay energy window for the emission of $x$ neutrons is defined as $Q_{\beta xn} = Q_\beta - S_{xn}$, where $S_{xn}$ is the separation energy of $x$ neutrons. Note that delayed neutrons are predominantly emitted from isotopes of odd-$Z$ elements, which have a systematically higher $Q_\beta$-value. As a result of these energetic constraints, $\beta$-delayed processes can thus play a significant role in nucleosynthetic mechanisms involving nuclei close enough to the neutron drip line, like the r-process (Section~\ref{prod_r}). it has also been speculated that $\beta$-delayed proton emission could occur along the rp-process path (Section~\ref{burn_rp})

The delayed processes of course add to the complexity of the $\beta$-decay of highly unstable nuclei due to the specific questions raised by the emission of particles and even more so by fission. Much experimental and theoretical effort has been dedicated to the study of these processes. The reader may profitably consult the following works and references therein: {\it i)} for delayed neutron emission, {\it e.g.} \cite{Dillmann17,Tolosa19} for experiments, and \cite{Mumpower16} and \cite{Borzov17} for predictions based on a QRPA+Hauser-Feshbach and on a CQRPA+Density Functional; {\it ii)} for delayed proton emission, {\it e.g.} \cite{Stefanescu18}; {\it iii)} for $\beta$-delayed fission, {\it e.g.} the extensive review of the experimental and theoretical aspects \cite{Schmidt18}, and model predictions based on a QRPA+Hauser-Feshbach model \cite{Mumpower18}.
In spite of all the efforts, the situation is not really rosy. Many experimental data of interest, particularly in the modeling of the r-process, are still missing, and theoretical predictions are often unable to account for the existing ones.

 %--------------------------------------
\subsubsection{$\beta$-decay models}
\label{decay_models}
%-------------------------------------
 Different approaches have been proposed to understand (and wishfully predict)  $\beta$-decays 
of heavy nuclei. Two of them in the extreme can be clearly identified:  on one hand, a macroscopic model 
referred to as the Gross Theory, and on the other the (large-scale) shell model,
 which is fully microscopic. Those in between are global  approaches of various 
kinds, the microscopic character of which is more or less pronounced.  In the first instance,
 they differ  in  their ways of describing the initial ground state and final excited states.

\vskip0.2truecm
{\bf A macroscopic Approach}. The Gross Theory of $\beta$-decay \cite{Yamada65,Takahashi69,Koyama70,Takahashi71} aims at
 describing the general behavior of  the $\beta$-strength distributions $S_\Omega$ in a statistical
manner. With an assumed large number of final states, $S_\Omega$ for allowed and
 first-forbidden transitions are constructed by folding ``one-particle strength functions'' via 
a very simple pairing scheme taking into account the corresponding sum rules and even-odd 
effects. This is in fact the first attempt to predict $\beta$-decay half-lives of nuclei far
 off the line of stability that has taken into account the then-conjectured existence of the Gamow-Teller Giant Resonance (GTGR) \cite{Takahashi73}. 
 The original Gross Theory has been improved in several ways \cite{Tachibana90,Nakata97,Koura17}. Use is made in particular of some empirical and theoretical
 considerations in order to modify the one-particle strength functions and the pairing scheme.
%*******************************************
\begin{figure}
%\vskip-2.7truecm
\center{\includegraphics[scale=0.45]{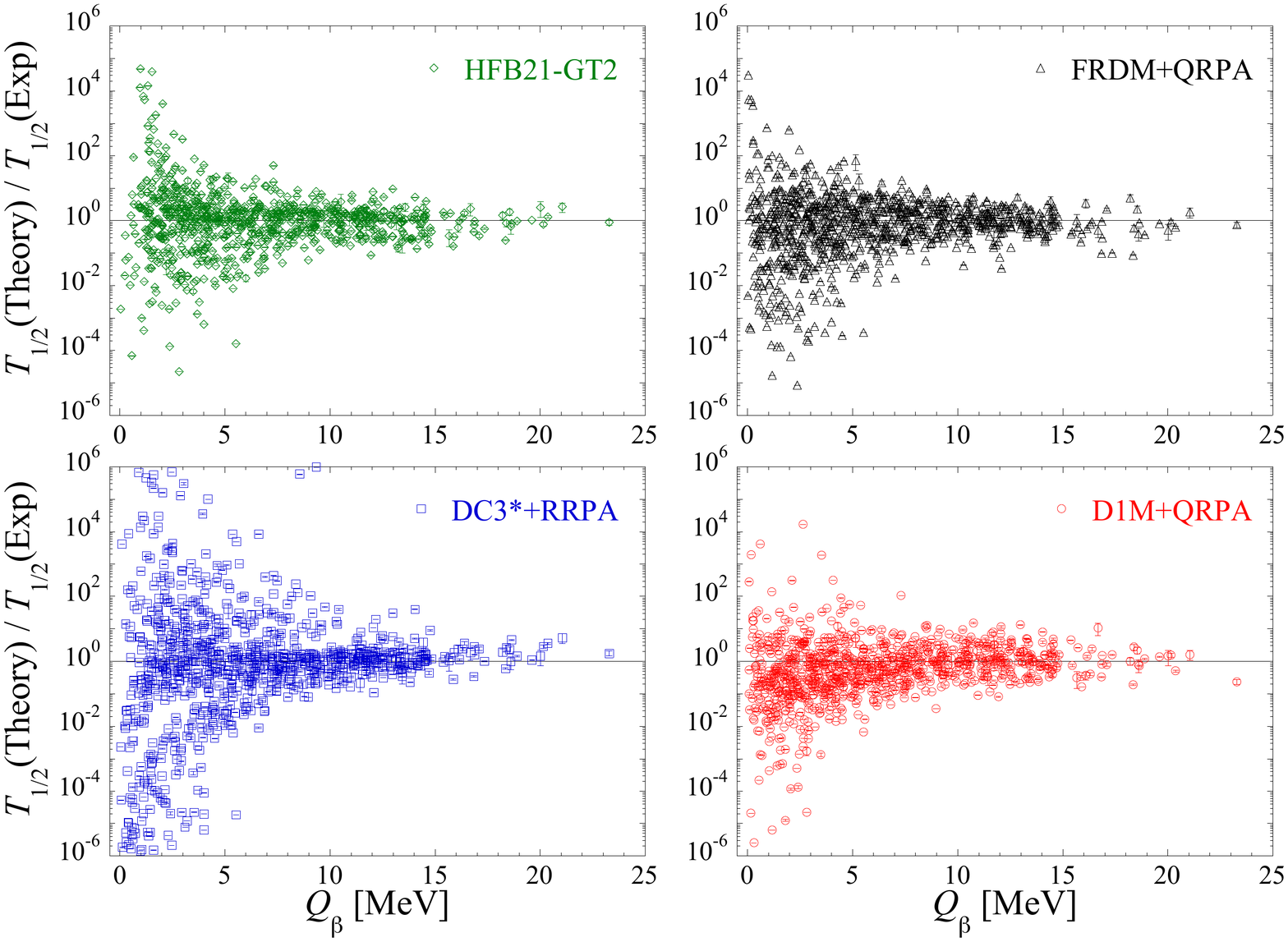}}
\vskip-0.5truecm
\caption{$\beta^-$-decay half-lives predicted by the Gross 
Theory (version 2 based on the HFB-21 masses, see Section~\ref{masses}),  FRDM+RPA \cite{Moeller97},  DC3*+RRPA  \cite{Marketin16} and D1M+QRPA \cite{Martini14,Peru20} relative to the experimental data for all the 950 nuclei with $Z\ge 10$ \cite{Audi17}. The ratios are plotted as a function of the $Q_{\beta^-}$-value obtained from experimental masses \cite{Wang17} and, if not available, from the HFB-31 masses (Section~\ref{masses}).}
\label{fig_beta_exp_mod}
\end{figure}  
%**********************************************

As illustrated in Fig.~\ref{fig_beta_exp_mod}, the Gross Theory, especially in its GT2 version with the adoption of more appropriate $Q_{\beta}$-values than those adopted initially \cite{Takahashi73}, appears to be reasonably accurate in predicting half-lives, at least when high $Q_{\beta^-}$ transitions are involved, as in the case for the decay of very neutron-rich nuclei. This is in particular so near closed-shells  where, normally, microscopic approaches are thought to be superior  ({\it e.g.} \cite{Tachibana95}). The success of the Gross Theory-type of models is especially remarkable in view of their simplicity. The ease with which $\beta$-decay 
probabilities can be re-computed,  in particular with different $Q_{\beta^-}$-values and 
concomitant new parameter fits, is clearly an additional very pleasing feature. The latest GT2 version \cite{Koura17} even cures some problems encountered in the former GT models. 
 
\vskip0.2truecm
{\bf Global (semi-)microscopic approaches}. From the  viewpoint of  microscopic physics, the most efficient way  to get the essential features of the GTGR strength distribution is to embed at the onset of the modeling an effective nucleon-nucleon interaction, notably of the spin-isospin
 ${\bar{\sigma}}  \cdot {\bar{\sigma}} \tau \cdot \tau$  type. The Gamow-Teller force allows 
 particle-hole excitations of the charge-exchange collective mode. Given the residual
 interaction, the final GTGR states can be constructed in an approximate way from the model
 ground state. Essentially all the calculations of relevance for the r-process adopt the 
``random phase  approximation (RPA)'', or its QRPA extension when pairing interactions are included. This model 
 provides the simplest description of the excited states of a nucleus and allows the ground state not to have a purely independent particle character, but may instead contain correlations \cite{Eisenberg72}. 
 
\begin{figure}
%\vskip-2.7truecm
\center{\includegraphics[scale=0.4]{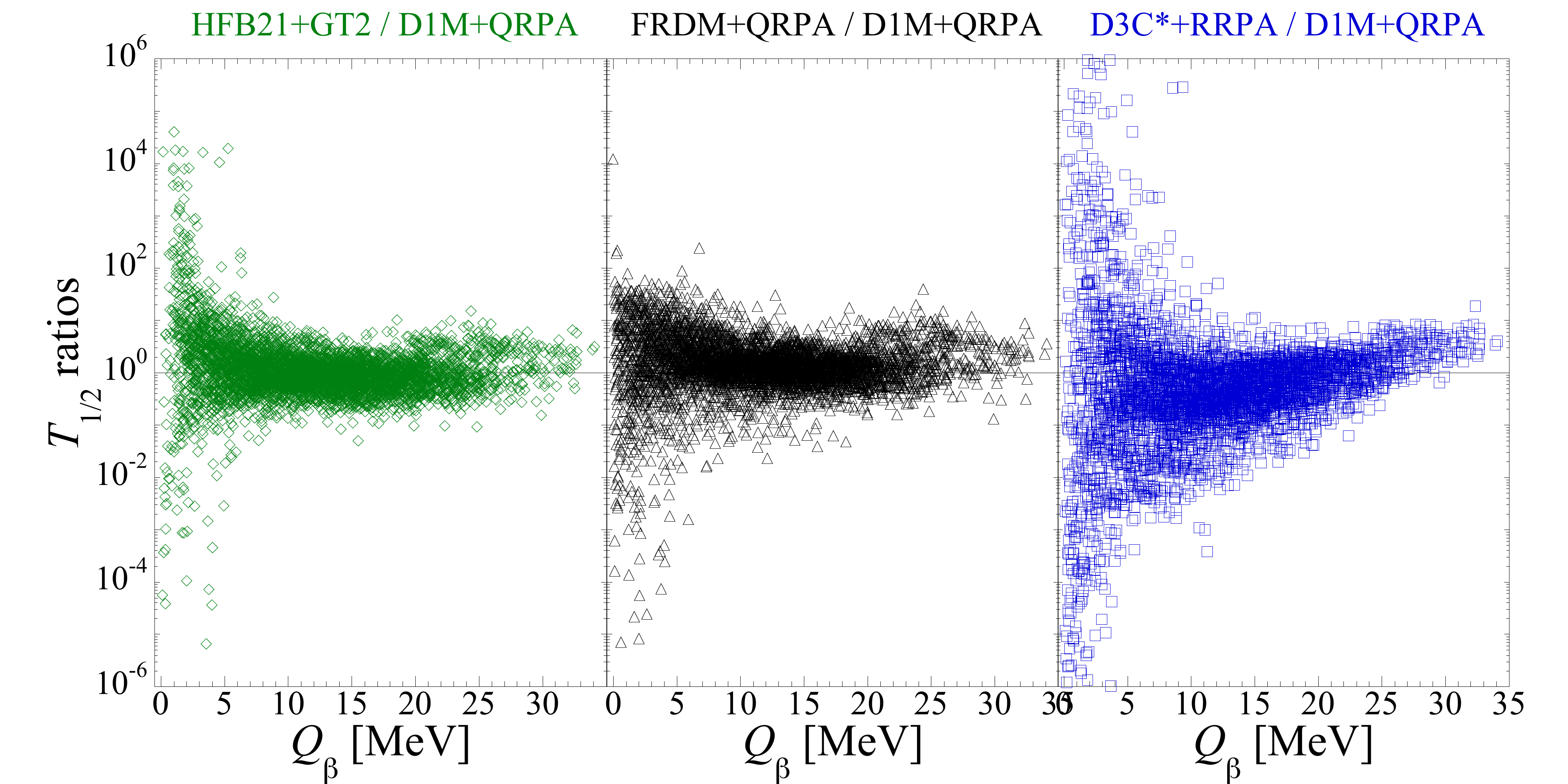}}
%\vskip-2.5cm
\caption{Ratios between calculated $\beta^-$ decay half-lives versus $Q_{\beta}$ for the models identified on top of the figure, and normalized to the D1M+QRPA values of \cite{Martini14,Peru20} for the 5500 nuclei with $10 \le Z \le 100 $ between the valley of $\beta$-stability and the D1M neutron drip line. The HFB-21 masses are adopted for the GT2 model. The FRDM+RPA and DC3*+RRPA values are from \cite{Moeller97} and \cite{Marketin16}, respectively.  The $Q_{\beta^-}$ value are from experimental masses \cite{Wang17} and, if not available, from the D1M masses (Section~\ref{masses}).}
\label{fig_beta_th_mod}
\end{figure}

 A large variety of approaches featuring different levels of approximations and sophistication can be found in the literature. For r-process applications, only a few large-scale calculations exist for all neutron-rich nuclei.  Among them are the FRDM+QRPA model that includes the GT approach for the first-forbidden contributions \cite{Moeller03}, the spherical relativistic QRPA (RRPA) calculation based on the DC3* interaction and including first-forbidden interactions \cite{Marketin16}, and the axially deformed QRPA calculation based on the D1M Gogny interaction \cite{Martini14,Deloncle17,Peru20}. Their capability to reproduce experimental $\beta^-$ half-lives are illustrated in Fig.~\ref{fig_beta_exp_mod}. Considering the 559 experimentally known neutron-rich nuclei with half-lives shorter than 60~s, the rms deviations of the ratios shown in Fig.~\ref{fig_beta_exp_mod} amount to 3.1 for HFB21-GT2, 3.5 for FRDM+QRPA, 8.5 for DC3*+RRPA and 3.3 for D1M+QRPA. Figure~\ref{fig_beta_th_mod} compares the $\beta^-$-decay half-lives for the 5500  $10 \le Z \le 100$ neutron-rich nuclei located between the valley of $\beta$-stability and the neutron drip line predicted by these different models relative to D1M+QRPA. For nuclei with $Q_\beta \gsimeq 10$ MeV of interest for r-process applications, deviations amount typically to a factor of 10. The D3C*+RRPA predictions  show the largest discrepancies relative to D1M+QRPA  in a wide range of  $Q_{\beta}$ values.  
 
%**********************************************
%**********************************************

Some difficulties encountered by the QRPA models are pointed out by \cite{Arnould07}. As far as continuum-QRPA evaluations are concerned, it is not clearly demonstrated yet that this model can successfully reproduce the $\beta$-decay data for a large body of nuclei, including well-deformed ones, if global parameter values are adopted \cite{Borzov05}, which is a very important requirement when dealing with nuclei very far from the valley of stability. One notes also that no consistent, axially deformed, mean-field plus QRPA calculations including first-forbidden transitions exist, in particular for odd number of neutrons and/or protons. In addition, the interplay of quasiparticle-vibration coupling and pairing correlations has been shown to have a non negligible impact on 
$\beta$-decay half-lives predictions \cite{Niu18}. These effects may need to be introduced in future calculations.

\vskip0.2truecm
{\bf Large-Scale Shell-Model Approach}. Through the use of a realistic (two-body) effective interaction, the standard nuclear
 shell-model tactics would  possibly, after much configuration mixing, be  capable of
 successfully describing the $\beta$-decay matrix elements or strength distributions along 
with other nuclear properties, such as the  excitation spectra. This exercise is severely 
hampered, however, by the difficulty of implementing a large enough model configuration space,
 and is next to impossible to perform for heavy nuclei. In practice, therefore, some drastic
 truncation of the model space is often required. With the advent of increased 
computational capabilities, much progress is, however, being made with regard to 
``large-scale'' shell-model calculations, as reviewed by \cite{Caurier05}.
 
A way to ease the basis dimension problem is provided by the shell model Lanczos iteration method ({\it e.g.}
 \cite{Caurier05,Caurier99}). As  shown for Gamow-Teller $\beta$-strength functions, the iterations are rapidly converging ({\it e.g.} \cite{Takahashi86,Caurier05,Mathews83}). The relation between the dimensions of the adopted basis and the actual computation times is discussed by \cite{Caurier05}.

Systematic large-scale shell model calculations may suffer not only from computational limitations, but more fundamentally from the lack of the spectroscopic information that is needed for the proper construction of effective nucleon-nucleon interactions. In particular, the current accuracy of the shell-model  predictions of
 $Q_{\beta^-}$ is insufficient for the half-life computations. Local adjustments of the model parameters is another drawback. The level of Gamow-Teller quenching is an additional embarrassment for this type of approach (see {\it e.g.} \cite{Arnould07} for a brief discussion). All in all, one has to acknowledge that  the seemingly  high-minded philosophy that is claimed to motivate the large-scale shell model calculations is clearly betrayed. Finally, the true predictive power of Bayesian neural network calculations remain to be scrutinized in detail.
 
In conclusion, one may acknowledge that the macroscopic Gross Theory models
 with global parameter values perform remarkably well in their accuracy to reproduce 
experimental data, especially for high $Q_\beta$ values. They also have the important advantage of providing with very limited
 computing efforts all the $\beta$-decay data that are needed in the modeling of the 
r-process, including the contributions of both the allowed and first-forbidden transitions.
 The importance of this has not to be underestimated. Of course, as it is the case for all 
the approximations of the macroscopic type, one may wonder about the reliability of the 
predictions very far from the valley of nuclear stability. As claimed in relation with the
 calculation of nuclear masses (Section~\ref{masses}), the reliability of more microscopic
 models may be expected to be higher. It remains to be seen, however, if this statement 
applies in the field of $\beta$-decay studies. In contrast to the situation encountered for
 mass predictions, the level of accuracy of the currently available microscopic models
 still needs to be improved, even if recent global QRPA calculations come close to compete with the most advanced form of the GT2 model in reproducing experimental data. Microscopic models are on their way 
to producing the $\beta$-decay half-life data needed in the r-process studies, 
including due consideration of allowed and first-forbidden transitions and the proper 
inclusion of nuclear deformation. Much remains, however, to be done in this field.
    
%** ***************
\subsubsection{Neutrino reactions}
\label{decay_neutrino}
%*********************************
%
Various weak-interaction processes involving all sorts of (anti-)neutrinos have an important bearing on a variety of phenomena. They are essential ingredients namely of ECSNe and CCSNe explosions \cite{Janka17}, in the the merging of compact (especially NS) stars, and in the cooling of NSs 
(see Section~\ref{decay_continuum} for an account of the Urca process). Neutrinos emerging from the proto-neutron star formed during a CCSN can also lead to some changes in the composition of the material ejected by the explosion through the so-called $\nu$-process briefly discussed below.

While most of the $\beta$-decay processes of astrophysical interest mentioned above can be dealt with in the classical ``V - A'' theory of the weak interaction \cite{Konopinski66}, the evaluation of neutrino interaction cross sections requires due consideration of both the charged- and neutral-currents of the unified electro-weak 
interaction \cite{Tubbs75,Bruenn85}. This evaluation is done in the framework of the RPA ({\it e.g.} \cite{Sieverding18} for a recent application) or shell models. Many interesting and complicated problems are also raised by various aspects of nucleon correlations and spin fluctuations in neutrino scattering at high 
densities \cite{Raffelt96}. 
  
Some words are in order here concerning the $\nu$-process. Neutrinos can interact with nuclei via inelastic neutral current scattering. This process is dominated by $\mu$ and $\tau$ neutrinos as their average energies are larger than those of the electron neutrinos, and as the neutrino interaction cross sections scale with the square of their energies. Through charged current, electron neutrinos also interact with nuclei, populating excited nuclear levels. These subsequently decay by the emission of nucleons or $\alpha$-particles that interact with nuclei. The associated nucleosynthesis mechanism is referred to as the $\nu$-process (see Fig.~\ref{fig_nucsite} and \cite{Langanke19} for a review).
 
The $\nu$-process is classically considered to contribute to the synthesis of mainly \chem{7}{Li}, \chem{11}{B}, \chem{19}{F}, and of the SoS rare nuclides \chem{138}{La} and \chem{180}{Ta} ({\it e.g.} \cite{Langanke19}). The effectiveness of this production is far from being well established as a result of large uncertainties in the neutrino capture cross sections, as well as in the supernova neutrino spectra. It tends in fact to be lowered based on recent supernova simulations \cite{Langanke19}. In addition, this neutrino-induced nucleosynthesis is not absolutely required, as other production processes have been identified. Certain mass-losing stars or novae (Cats.~5 and 6 stars) may indeed contribute to the galactic \chem{7}{Li} \cite{Cescutti19}. Efficient production of \chem{19}{F} by AGB or even massive mass-losing stars of the Wolf-Rayet type is also demonstrated \cite{Goriely00,Meynet00}. These production channels are supported by observation. Standard p-process in CCSN and SNIa explosions might also be appropriate \chem{180}{Ta} producers \cite{Arnould03} (see also Section~\ref{prod_p_site}). In fact, only \chem{138}{La} is found in a detailed analysis by \cite{Goriely01b} to be a possible product of neutrino nucleosynthesis, but many uncertainties affect this conclusion. Calling for a specific process for explaining the SoS abundance of just one nuclide is of course a rather disturbing situation.
 
Another nucleosynthesis process in which neutrinos are speculated to play a key role is referred to as the ``neutrino-induced rp-process'', or $\nu$p process in short (see {\it e.g.} \cite{Zhang18}, and references therein). In the early neutrino-driven winds of a CCSN, $\bar{\nu_e}$ captures $p(\bar{\nu_e}, e^+)$ on free protons produces a small amount of free neutrons in an otherwise proton-rich matter. These neutrons in their turn produce protons  via $(n,p)$ reactions on neutron-deficient nuclei located on the rp-process path (Section~\ref{burn_rp}), these protons sustaining the rp-process further.

%************************************************
\subsection{Decays via strong interactions}
\label{decay_strong}
%***********************************************

%********************************************
\subsubsection{$\alpha$-decay rates}
\label{decay_alpha}
%********************************************

Nuclear $\alpha$-decays have been studied extensively in the laboratory. Save some exceptional cases for which the thermal population of nuclear excited states can contribute largely to the $\alpha$-decay lifetime in astrophysical conditions \cite{Perrone71}, the available data appear to be sufficient for astrophysical purposes.

Another special case in which $\alpha$-decay is involved concerns the study of the key $3\alpha$-reaction of He burning leading to the stellar production of \chem{12}{C}, which is characterized by a pronounced 3$\alpha$ cluster. From a recent experimental measurement of a very weak $\alpha$-decay channel of the so-called Hoyle state in \chem{12}{C} (see Section~\ref{burn_hesi}), it is concluded by \cite{Aquila18} that the previously proposed reaction rate has to be revised. The $\beta$-delayed $\alpha$-decay of \chem{16}{N} has also been used to constrain the important \reac{12}{C}{\alpha}{\gamma}{16}{O} reaction at energies of astrophysical relevance. A recent experimental study \cite{Kirsebom18} provides new constraints on its rate.

%******************************************************** 
\subsubsection{Fission rates}
\label{decay_fission}
%******************************************************

Fission may play a fundamental role during the r-process nucleosynthesis, especially during the dynamical ejection of matter in NS mergers (see Section~\ref{prod_r_mergers}) due to the significant number of free neutrons per seed nuclei in such environments. More specifically, fission may be responsible for {\it i)} recycling the matter during the neutron irradiation (or, if not, by allowing the possible production of super-heavy long-lived nuclei, if any), {\it ii)} shaping the r-abundance distribution in the $110 \le A \le 170$ mass region at the end of the main neutron irradiation phase, {\it iii)} defining the residual production of some heavy stable nuclei, more specifically Pb and Bi, but also the long-lived cosmochronometers Th  and U (see Section~\ref{chrono_actinides}), and {\it iv)} heating the environment through the energy released. 

As recently reviewed by \cite{Schmidt18}, a vast experimental and theoretical effort has been devoted in recent years to the study of nuclear fission.  This includes the various fission processes of astrophysical relevance, including neutron-induced, spontaneous, $\beta$-delayed and photo-fission and their impact on the r-process nucleosynthesis \cite{Goriely15c,Goriely15d}. Attention has also been paid to the corresponding fission fragment distributions, another ingredients of key importance in the r-process ({\it e.g.} \cite{Goriely13c,Lemaitre19}). It is superfluous to present a new survey of these questions here. We just limit ourselves to a brief comment.
 
It appears quite clearly that the precise impact of the fission processes on the r-process nucleosynthesis remains difficult to evaluate in view of the extreme difficulty to predict reliably the probabilities of the various fission modes. Many complex ingredients indeed enter their calculation. The prediction of the fission path in a multi-dimensional potential landscape, and in particular of the highest, so-called primary, fission barrier is clearly of fundamental importance. Among others, quantities of direct relevance are the collective energy corrections and inertial mass needed to estimate the least-action path and the vibrational zero-point energies in the ground-state configuration \cite{Lemaitre18,Giuliani18}. Additional basic necessary ingredients are the discrete nuclear states and level densities at the saddle points as well as in the isomeric wells along the path. Of course, $\beta$-delayed fission brings its share of complementary difficulties, as $\beta$-decay strength functions enter the evaluation of fission probabilities (see Section~\ref{decay_models}).

%**********************************************
\begin{figure}
%\vskip-3.0truecm
\center{\includegraphics[scale=0.6]{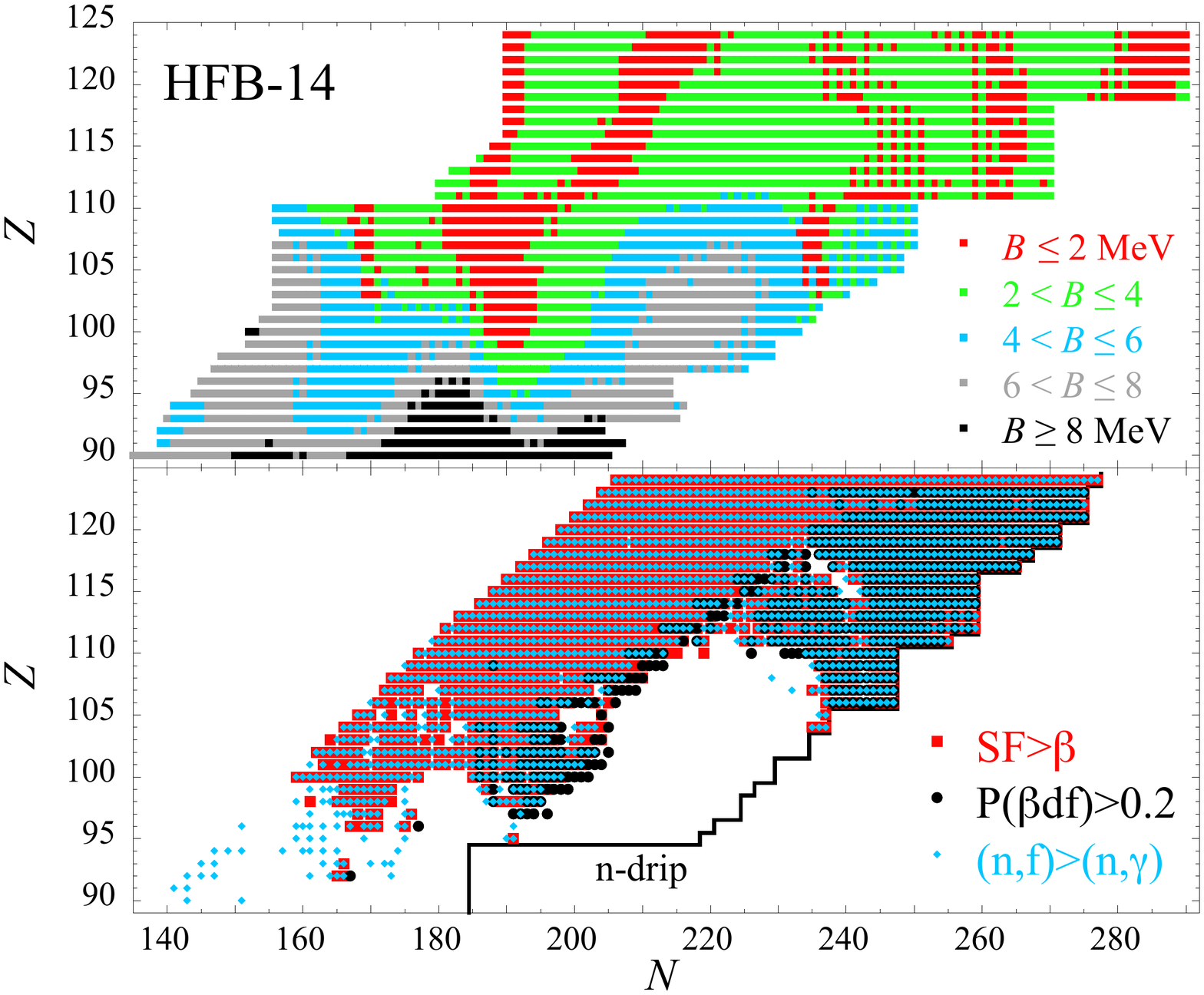}}
\vskip-0.5cm
\caption{(Color online) {\it Upper panel}: Mapping in the $(N, Z)$ plane of the height of the primary fission barrier (in MeV) predicted by the HFB-14 mass model \cite{Goriely07a}.
{\it Lower panel}: Dominant fission regions in the $(N, Z)$ plane. Nuclei for which spontaneous fission is estimated to be faster than $\beta$-decays are shown by red squares, those for which the $\beta$-delayed fission probability (with respect to the total $\beta$-decay) is larger than 20\% by black circles, and those for which neutron-induced fission is faster than radiative neutron capture at $T=10^9$K by blue diamonds. The rates are obtained with the HFB-14 fission paths, as detailed in \cite{Goriely15c,Goriely15d}. The solid line represents the HFB neutron drip line.}
\label{fig_fis_hfb14}
\end{figure}  
%**********************************************

Of particular astrophysical interest are the few calculations of the fission probabilities for the 3000 nuclei with $90 \lsimeq Z \lsimeq 120$ potentially involved in the r-process that rely on microscopic mean-field models based on the Skyrme \cite{Goriely07a}, the Gogny \cite{Lemaitre18} or the Barcelona-Catania-Paris-Madrid \cite{Giuliani18} energy density functionals. The fission channel is predicted to be important in regions of the $(N,Z)$ plane where the primary barrier is typically below 4--5~MeV, as illustrated in Fig.~\ref{fig_fis_hfb14} where fission barriers and rates are estimated on the basis of the Skyrme HFB-14 model \cite{Goriely07a}. While the HFB predictions of the primary barriers show similar trends for exotic neutron-rich super-heavy nuclei  \cite{Lemaitre18,Giuliani18}, significant deviations are found when comparing with macroscopic-microscopic models like the FRDM \cite{Moeller09}, leading to different nucleosynthesis predictions, as discussed in \cite{Goriely15c}. 

It cannot be stressed strongly enough that all the evaluated fission probabilities suffer from large uncertainties. While experimental fission cross sections can be reproduced to within typically a few tens of percents when all the relevant nuclear ingredients are properly tuned,  they can hardly be estimated to within less than a factor of 10 if predicted fission paths and nuclear level densities are adopted, even for experimentally known nuclei. Let us note in particular that the fission half-lives are extremely sensitive to the height of the primary barrier. It is quite remarkable that a modification of the barrier height by about 1~MeV affects spontaneous half-lives by up to 9 orders of magnitude!

The fission fragment distribution also plays a key role in r-process nucleosynthesis simulations since it defines the light species that will be produced by fission recycling. Both the $Z$- and $A$-dependences are needed for all potentially fissioning nuclei. The widely used Gaussian models have little predictive power when dealing with exotic neutron-rich nuclei, so that a growing attention has been paid to global scission-point models. These include in particular the so-called SPY model (for Scission-Point Yields) corresponding to a renewed statistical scission-point model based on HFB ingredients \cite{Lemaitre19} and the so-called General Fission (GEF) model based on global and semi-empirical description of the fissioning system \cite{Schmidt18}. SPY and GEF models are known to predict significantly different fission fragment mass distributions for exotic neutron-rich nuclei, which impacts directly the final abundance distributions of r-nuclei in NS mergers \cite{Goriely15c,Goriely15d}. Differences do not only come from the estimated fragment distributions, but also from the average number of prompt neutrons emitted during the fission process. Such neutrons are recaptured during the nucleosynthesis event, and may affect the  final r-abundances ejected in NS mergers \cite{Goriely15c}.

In summary, despite decades of effort to understand and describe fission processes, the capacity to predict fission probabilities remains rather unsatisfactory, leading to large uncertainties in nucleosynthesis applications where fission plays a key role, such as the dynamical ejecta from NS mergers. Fission properties, in particular fission paths and nuclear level densities at the fission saddle points, but also fission fragment mass distributions and the average number of emitted neutrons need to be determined on the basis of sound and as microscopic as possible nuclear models. Readers are referred to recent works which go beyond the  models  used so far in astrophysics, but still need to be developed and applied to large-scale calculations \cite{Goutte05,Bernard11}.

%********************************************************
\subsection{Nuclear decays and reaction via the electromagnetic interaction}
\label{decay_em}
%*******************************************************

At high temperatures (typically in excess of about 10$^9$K), nuclei may be subject to photodisintegrations 
of the ($\gamma$,n), ($\gamma$,p) or ($\gamma$,$\alpha$) types. More specifically, the importance of these transformations increases as the evolution of a star proceeds beyond C-burning in the non-explosive history of massive stars (see Fig.~\ref{fig_starevol}). The first major stage of photodisintegrations is the so-called Ne-burning episode governed by the \reac{20}{Ne}{\gamma}{\alpha}{16}{O} reaction. Photodisintegrations culminate  at the Si-burning phase terminating with a NSE regime, as discussed in Section~\ref{burn_hesi}. Of course, explosive situations favor photodisintegrations due to the higher temperatures reached in these conditions. In fact, some can already occur during the various hot modes of H-burning (see Sections~\ref{burn_hotpp} - \ref{burn_rp}), and are essential ingredients of the r- and p-processes (see Sections~\ref{prod_r} and \ref{prod_p}).

The photodisintegration rates in a photon bath obeying a Plank distribution law that applies in stellar interiors are briefly discussed in \cite{Iliadis15}. On top of being extremely sensitive to temperature as a direct consequence of  the properties of the Planck spectrum of the photons, the photo-absorption and photo-deexcitation rates strongly depend on the so-called photon strength function describing the average response of the nucleus to an electromagnetic probe. Many experimental and theoretical efforts have been made to improve the determination of the dominant dipole E1 and M1 modes for both the photo-excitation and deexcitation of a nucleus by $\gamma$-ray absorption or emission. Details can be found in \cite{Harakeh01,Goriely19}

In view of the difficulties to derive photodisintegration rates through direct approaches, an indirect method is widely adopted instead. It relies on the detailed balance theorem applied to the reverse radiative captures of nucleons or $\alpha$-particles ({\it e.g.} \cite{Iliadis15}). 

Photon interactions with nuclei are also fundamental for a proper understanding of the UHECR  and more particularly of its composition and energy spectrum as detected on Earth. The UHECR are known to be of extra-galactic origin, so that their initial properties when detected on Earth are affected by interactions with photon backgrounds and cosmic magnetic fields \cite{Allard12}. More specifically, the cut-off in the observed energy spectrum above a few $10^{19}{\rm eV}$ is associated with the interactions of ultra-high energy protons and nuclei with photons from the CMB. While the propagation of UHECR protons has been relatively well studied, the giant dipole resonance and quasi-deuteron interactions of UHECR nuclei with photons still raise questions. This topic has led to a renewed  interest ({\it e.g.} \cite{Khan05,Dedonato09}) both on the experimental and theoretical nuclear physics sides due to the confirmation by the Pierre Auger Observatory of a significant contribution of nuclei to the UHECR spectrum  \cite{Aab14}.  Interactions of heavy nuclei (typically lighter than $^{56}$Fe) with the CMB photons give rise to photodisintegration with nucleon and $\alpha$-particle emission and modify the UHECR composition with respect to the original one in the unknown sources. Photonuclear cross sections including all open channels for break-up and particle emissions up to energies of about 100~MeV are consequently needed for all nuclei (mainly lighter than iron) potentially produced by photoemission. Future experimental and theoretical nuclear physics efforts to provide high-quality data will certainly contribute to a better understanding of the origin of the UHECR.

%************************************************
\section{Thermonuclear reactions}
 \label{reac_thermo}
%***********************************************
 
The 1930's marked the start of a flurry of works on nuclear reactions in astrophysical conditions, with especially the seminal papers by \cite{Bethe38,Weizsacker38,Bethe39} on the energy production in the Sun by the p-p chains and on the hydrogen burning by the CNO cycles. Since then, the fundamental role of nuclear reactions in the nucleosynthesis and energy production in astrophysical situations has been extensively studied, both experimentally and theoretically. 

We do not intended to review here all the very many reactions that have been identified to take place in the various astrophysical situations depicted in Fig.~\ref{fig_starevol} and involved in the processes schematized in 
Fig.~\ref{fig_nucastro}. it is also out of the scope of this survey to discuss the details of the experiments and the theoretical formalisms to be applied in order to calculate the thermonuclear reaction rates of relevance in astrophysical plasmas. The reader is largely referred to {\it e.g.} the textbook of \cite{Iliadis15}, complemented with a recent review of the laboratory works \cite{Oliveira19}. We just restrict ourselves to an overview of some basic distinguishing features of thermonuclear reactions in astrophysical situations.

%*******************
\subsection{Some generalities}
\label{reac_generalities}
%*******************

\noindent (1) In a given astrophysical location, two factors dictate the variety of nuclear reactions that can act as energy producers and/or as nucleosynthetic agents. The abundances of the reactants have obviously to be high enough, and the lifetimes of the reactants against a given nuclear transmutation have to be short enough for this reaction to have time to operate during the evolutionary timescale of the astrophysical site under consideration. These constraints are at the very origin of the well-defined sequence of nuclear phases from H-burning to Si-burning depicted in Fig.~\ref{fig_starevol};
\vskip0.1truecm
\noindent (2) The probability of a thermonuclear reaction in an astrophysical plasma is strongly dependent on some specific properties of this plasma. In this respect, two key guiding features are the distribution of the energies of the reacting partners, and the reaction cross section at a given energy. First, the reacting nuclei are, locally at least, in a state of thermodynamic equilibrium (see {\it e.g.} \cite{Cox68} for a discussion of this property). In such conditions, all nuclear species obey a Maxwell-Boltzmann distribution of energies (some specific exception to this rule are discussed below), from which it is easily inferred that the relative energies of the reaction partners also obey such a distribution. Second, the reaction cross section between charged partners is governed by the probability of penetration of the Coulomb barrier of the interacting nuclei, neglecting for the moment details of the nuclear structure, like resonances. As a result, the effective reaction rate is obtained by integrating the strongly energy-dependent reaction cross sections over the whole Maxwell-Boltzmann energy range. The resulting integrant exhibits a strong maximum, generally referred to as the Gamow peak. It is centered on the ``most effective energy" given by $E_0 = 0.1220(Z_1^2Z_2^2A)^{1/3}T_9^{2/3}$ MeV, where $Z_1$, $Z_2$ and $A$ are the charge numbers and the reduced mass in amu, and $T_9$ the temperature $T$ expressed in $10^9$~K. The Gamow peak is characterized by a  width approximated by $\Delta = 4 (E_0kT/3)^{1/2}$ MeV, where $k$ is the Boltzmann constant  (see {\it e.g.} \cite{Iliadis15}, Eqs. (3.76) and following for the derivation of $E_0$ and $\Delta$). The reactions thus mostly occur in the approximate window from $E_0 - n\Delta$ to $E_0 + n\Delta$ ($n =$ 2 to 3) if indeed the possible role of resonances is neglected. The implications of these considerations are of pivotal importance. It can indeed be concluded that the energy range of astrophysical relevance for reactions between charged particles is largely in excess of the thermal energy $kT$ and much lower to at best somewhat lower than the Coulomb barrier in conditions prevailing during non-explosive phases of the evolution of stars, where thermal energies are found to range between approximately 0.1~keV for H-burning to 350~keV for Si-burning (Fig.~\ref{fig_starevol}). This situation accounts for two facts: {\it i)} the sequence of burning episodes is characterized by a limited number of reactions between nuclei with increasing charges, from H-burning to Si-burning, and {\it ii)} the charged-particle induced thermonuclear reactions of relevance concern mainly the capture of protons or 
$\alpha$-particles which offer the lowest Coulomb barriers.  A limited number of fusion reactions involving heavy ions ($^{12}$C, $^{16}$O) are also of great importance. Explosions occur at higher temperatures, and consequently at higher most effective energies $E_0$ than in non-explosive situations. As a general rule, this is also accompanied with a larger variety of efficient nuclear reactions. 

The considerations above leading to the most effective energy $E_0$ in the case of reactions between charged particles do of course not apply to neutron captures in view of the absence of Coulomb barriers. In this case it can be shown that the most effective energy is of the order of $kT$. It has also to be noted that, in contrast to reactions involving charged reactants, the captures of neutrons do not contribute to the energy budget of a star, but are essential players in the synthesis of nuclides heavier than iron through the so-called s- and r-processes (see Sections~\ref{prod_s} and \ref{prod_r}).
 
\vskip0.1truecm

\noindent (3)  In non-explosive conditions, like in the quiescent 
phases of stellar evolution which take place at relatively low 
temperatures, most of the reactions of interest concern stable nuclides.  
Even so, the experimental determination of their charged-particle induced cross sections  face enormous  
problems, and represent a real challenge ({\it e.g.} \cite{Iliadis15}). This relates directly to the smallness of the cross sections due to the fact that $E_0$ lies well below the Coulomb barrier. Just as an illustration, $E_0$ is about 
1\% of the Coulomb barrier height for the \reac{12}{C}{p}{\gamma}{13}{N} reaction starting the CNO cycle of H burning (see below) at a temperature of $10^7$K. This range of temperatures is typical of H-burning conditions. As a consequence, the cross sections can dive into the nanobarn to picobarn abysses. The situation is somewhat less critical in later evolutionary stages, like Si-burning (Fig.~\ref{fig_starevol}), where temperatures in excess of several times $10^9$K can be reached. In such conditions, $E_0$ values closer to the Coulomb barrier height are obtained, with the concomitant increase in the cross sections.

Nuclear physics is thus often facing the challenge of exploring the ``world of almost no events''. This extreme rarity of events might be thought to imply the inefficiency of the nuclear reactions. This is by far not the case. In contrast to the situation encountered in the laboratory, stars have indeed lifetimes of million to billion years !
\vskip0.1truecm

\noindent (4) In explosive situations, the temperatures are typically higher than in the non-explosive cases. The corresponding increase of the effective energies $E_0$ gives rise to a higher probability of penetration of the Coulomb barriers, and consequently larger cross sections. The price to pay to reach this higher energy domain is huge, however. The nuclear flows indeed depart more or less strongly from the bottom of the valley of nuclear stability, and involve more or less unstable nuclei, sometimes all the way very close to the nucleon drip lines (see Fig.~\ref{fig_nucastro}). The ``world of almost no events'' makes way to the ``world of exoticism'' imposing to nuclear physics the extremely complex task of dealing with nuclei that are often unknown in the laboratory.
A special case concerns photodisintegrations (see Section~\ref{decay_em}) the high temperature dependence of which is largely dictated by the properties of the Planck spectrum of the bath of stellar photons.  
\vskip0.1truecm

\noindent (5) As for $\beta$-decays (Section~\ref{decay_beta}), thermally populated nuclear excited states can contribute to the effective stellar reaction rates, as first demonstrated by \cite{Arnould72}. This effect is especially noticeable in the case of endothermic reactions on targets with low-lying excited states from which the exit particle channels are greatly favored with respect to the ground state due to restrictions imposed by spin conservation selection rules.
\vskip0.1truecm

\noindent (6) As if the evaluation of stellar reaction rates were not complicated enough, a whole new range of problems has opened up with the discovery through a series of remarkable experiments that the reaction cross 
sections measured at the lowest reachable energies are in fact ``polluted'' by atomic or molecular effects induced by the experimental conditions ({\it e.g.} \cite{Iliadis15}). As a result, the situation appears even more intricate than previously imagined, necessitating a multi-step process in order to go from laboratory  data to stellar rates: before applying the usual electron screening corrections relevant to the stellar plasma conditions (see below), it is required first to extract the {\it laboratory} electron screening effects from the experimental cross section data in order to get the reaction probabilities for bare nuclei. In spite of heroic laboratory efforts and complementary theoretical modeling, much obviously remains to be done in order to get reliable estimates of the laboratory electron screening factors.  
\vskip0.1truecm

\noindent (7) In stellar plasmas, a specific electron screening correction has to be applied, and can drastically affect the cross sections for bare nuclei. This correction arises because of the ability of a nucleus to polarize its stellar surroundings. As a result, the Coulomb barrier seen by the reacting nuclei is modified in such a way that the tunneling probability, and consequently the reaction rate, increases over its value in vacuum conditions.  Different formalisms have been developed depending on the ratio of the Coulomb energy of reacting nuclei to the thermal energy. Weak screening applies if this ratio is well below unity, while a strong screening is obtained when this ratio is well in excess of unity. In this case, a very large increase of the reaction rates is predicted. The limiting situation of strong screening is reached when solidification of the stellar plasma leads to the special pycnonuclear regime. In this case, the reactions are not governed by temperature like in the thermonuclear regime, but instead by lattice vibrations in dense Coulomb solids. This limiting regime can be approached {\it e.g.} at the high densities and low temperatures prevailing in WD stars. References to the stellar screening formalism from the weak to the pycnonuclear regime can be found in {\it e.g.} \cite{Yakovlev06,Potekhin13}.

%****************************************
\subsection{Charged-particle induced reaction: a brief overview of experiments}
\label{reac_charged_experim}
%*******************************************

Different energy dependencies of the cross sections between charged particles are identified. By and large, these considerations can be extended to neutron-induced reactions and to photodisintegrations. 
Cases are found where the cross sections exhibit a smooth energy dependence. They are referred to as non-resonant cross sections.  Other cases show strong variations in the vicinity of a particular energy. These are classified as resonant cross sections. One can distinguish between narrow and broad resonances. A special case of interest is encountered when the energy variation of the cross sections at the reaction threshold results from the contribution of the tail of a broad resonance that is located below threshold. These so-called sub-threshold resonance cases play an important role in the very low energy range of astrophysical interest. Another very important limiting case corresponds to resonances that are strongly overlapping as a result of their broadening and of their high enough densities in a small energy interval. In such a situation, referred to as a continuum regime, the cross sections vary smoothly with energy as in the non-resonant case. The various contributions to the total reaction rates mentioned above can be added incoherently if interferences are negligible. To a good approximation, this is indeed the case for narrow resonances, between two broad resonances with different spin-parities, or between a resonance and a non-resonant process of different incoming orbital angular momenta. In other situations, interference effects come into play.

A myriad of experiments dedicated to the measurement of reaction cross sections of relevance to astrophysics have been conducted over the years, with the aim of reaching energy regions as close as possible to the most effective energy $E_0$ introduced above, and thus dealing with extremely small cross sections when charged reactants are involved. Heroic and painstaking efforts have been required in very many cases in order to deliver nuclear data necessary (albeit not sufficient!) to put the astrophysical models on a safer footing.
In many instances, such an experimental activity has been the trigger of
new and exciting technological or physical ideas. The difficulty of providing data in quest and the vast
diversity of the problems to be tackled have always made it unavoidable to 
use the most sophisticated experimental techniques of nuclear physics, or
even to develop novel approaches.  

Many discussions have been devoted to astronuclear physics experiments, and we do not embark here into a review of this vast subject. The reader is referred to textbooks and recent reviews, like \cite{Iliadis15,Oliveira19} and \cite{Rolfs88} for a detailed discussion of sources, accelerators and beam transport systems. Note that unwanted background, in particular of cosmic-ray origin, is one of the main enemy of experimental nuclear physicists in their quest of very small cross sections, as it can substantially contribute to the signal count rates of interest. In order to reduce this background to the lowest possible level, experimental facilities have even been built underground.

Experiments are divided into two broad classes referred to as direct and indirect measurements.
\vskip0.1truecm

 {\bf Direct measurements}. They concern reactions that really take place in 
astrophysical sites. Strictly speaking, they would also have to be conducted in an energy range including the Gamow window centered on $E_0$. This is very seldom the case, especially in non-explosive situations, 
in view of the very low energies involved.
 
Direct methods have been, and still are, widely utilized in the case of stable targets. Typically, use is made of a dedicated accelerator delivering for several weeks low-energy ion beams of high intensity ($\sim 1$~mA) on a target that is able to withstand the heavy beam load (hundreds of watts), and that is also of high chemical and isotopic purity. A few per mil atoms of  impurity can indeed be responsible for a noise exceeding the expected signal.  
In the case of the commonly used inverse kinematics geometry, a heavy-ion  accelerator is often used in conjunction with a windowless gas target of  the static or supersonic jet type. Detectors have generally been the same as those used in classical nuclear  physics, but new detection techniques have also been developed for astronuclear physics experiments.  
 
In the case of unstable targets, two different direct approaches are envisioned, depending upon the lifetimes of the nuclides involved in the entrance channel \cite{Rolfs88}. The radioactive {\it target} technique appears most profitable for radioactive nuclides with lifetimes in excess of about one hour. In contrast, the radioactive {\it beam} method is appropriate for shorter-lived species, and has long been viewed as a new frontier in nuclear physics. It has been vigorously developed over the years in many laboratories around the world. Two basic techniques are used to produce the high-intensity high-purity radioactive beams that are required for the study of the low-energy resonances or non-resonant contributions of astrophysical interest: the ISOLDE post-accelerator scheme, and the projectile fragmentation method. Note that a major breakthrough in the field has been the 1990 first direct measurement in inverse kinematics of the resonant $^{13}$N(p,$\gamma)^{14}$O rate using the ISOLDE post-acceleration scheme at the Belgian Radioactive Ion Beam facility of Louvain-la-Neuve (see {\it e.g.} \cite{Decrock91}). This pioneering experiment has been followed by many others in various laboratories.

Techniques allowing the direct measurement of photon-induced reactions have also been developed ({\it e.g.} \cite{Cwiok18}). As already mentioned above, photodisintegration rates can also be obtained through the application of the reciprocity theorem to the corresponding inverse radiative capture reaction.
\vskip0.1truecm

{\bf Indirect measurements}. The indirect methods are a very important complement, or even an inevitable alternative, to the direct measurements concerning reactions on stable as well as unstable targets. 
This situation relates in particular to the extreme smallness of the cross sections of astrophysical interest, or to the incapability of setting up radioactive beams of the required purity and intensity.
 
Different indirect approaches have been developed and applied to a more or less large extent, like the use of transfer reactions, with {\it e.g.} the special technique of the Trojan horse method ({\it e.g.} \cite{Bertulani18}), the study of  inverse reactions, with the example of the Coulomb break-up technique ({\it e.g.} \cite{Bauer03}), or measurements relating to the decay of radioactive beams, an example of which is an experiment concerning the important reaction  $^{12}$C($\alpha,\gamma$)$^{16}$O ({\it e.g.} \cite{Gai98}; see also the brief reference to the measurement of $\beta$-delayed $\alpha$-decay in Section~\ref{decay_alpha} relevant to this reaction). These techniques are discussed at length in textbooks \cite{Iliadis15,Rolfs88} or review articles ({\it e.g.} \cite{Oliveira19}).
% 
%***********************************
\subsection{Neutron capture reactions: a brief overview of experiments}
\label{reac_neutrons_experim}
%**********************************

Measurement of neutron capture cross sections (especially radiative neutron captures, as well as (n,p) and (n,$\alpha$) reactions, and neutron-induced fission) at energies of astrophysical interest (from a fraction of keV to several hundred keV) have been under very active investigation over the years at a variety of facilities. We do not aim at reviewing these experimental efforts here. The reader is referred to {\it e.g.} the textbook \cite{Iliadis15} or the review articles \cite{Oliveira19,Kaeppeler98,Tessler15,Colonna17} (this list is not meant to be exhaustive).

Let us just notice that various neutron production techniques as well as a diverse cross section measurement methods are adopted. A pleasing feature is that neutrons classically produced by $^{7}$Li(p,n)$^{7}$Be have an energy spectrum that very closely resembles the thermalized (Maxwellian) stellar 
spectra in typical s-process conditions (Section~\ref{prod_s}). The most suitable experimental procedures depend notably on the stability/instability of the targets and reaction products. By and large, the cross sections for the nuclides located at the bottom of the valley of nuclear stability involved in the s-process are known nowadays with a very good accuracy. Experimental data are also available for some nuclides located close enough to the valley. The use of inverse reactions, particularly photodisintegrations, can help obtaining such data. At this point, one has to remember, however, that the experimental accuracy presently reached concerns the capture of neutrons by nuclei in their ground states, and is spoiled by the contribution to the reactions of target excited states, which can be  evaluated by theory only. Some attempts to study the contribution of excited states have started with dedicated measurements of the super-elastic scattering cross sections on long-lived metastable states \cite{Roig06,Roig11}.

%**************************************************
\subsection{Reaction models}
\label{reac_models_general}
%*************************************************
 
Various reaction models have been developed in order to understand or complement the experimental information. Broadly speaking, they are divided into ``non-statistical'' or ``direct" and  ``statistical'' or ``compound" models. 

Direct processes are mostly one-step processes characterized by short reaction times (about $10^{-22}$ s, which  is roughly equal to the time needed for the incident particle to traverse the target nucleus) and by a strong correlation between the initial and final reaction channels. They are encountered mainly when the densities of the participating nuclear levels are relatively low.

In contrast, compound processes involve long reaction times (about $10^{-18}$ s) and proceed by many intra-nuclear collisions leading to the formation of a compound system. Subsequently, the incident energy is shared among the other nucleons, eventually leading to a state of statistical equilibrium. In this case, the compound system is referred to as a compound nucleus (CN).  A sufficient amount of energy may eventually be accumulated for one nucleon (or group of nucleons) to escape. Apart from conservation of total energy and total angular momentum, the incident and exit channels are completely uncorrelated. This is sometimes referred to as Bohr's amnesia hypothesis. These statistical processes are expected to develop when the density of states of the compound systems is sufficiently high. This is considered to be the case mainly for reactions involving targets with mass numbers $A \gsimeq 20$ if they lie close to the line of nuclear stability. This mass limit has to be increased more and more  when moving farther and farther away from stability. Indeed, the nucleon or $\alpha$-particle separation energies decrease to such an extent that the number of available nuclear states becomes insufficient to validate a statistical description of the reaction mechanism. 

As an intermediate between the aforementioned two extremes, a reaction type exists that embodies both direct- and compound-like features. These reactions are referred to as pre-equilibrium, precompound or multi-step processes. Pre-equilibrium emission takes place after the first stage of the reaction, but long before statistical equilibrium of the CN is attained. It is imagined that the incident particle creates step-by-step more complex states of the compound system, and gradually loses the memory of its initial direction.  

The direct, pre-equilibrium and statistical mechanisms all operate in stellar conditions. The first two ones are expected to provide a sizable part of the reaction cross section for incident energies between 10 and (at least) 200 MeV, as it is the case in the interaction of energetic stellar (solar) particles or GCRs with the circumstellar (circumsolar) or ISM material (see Section~\ref{nucleo_gcrs}).

%********************************************
\subsubsection{Statistical Hauser-Feshbach model}
\label{reac_hf}
%******************************************

The statistical Hauser-Feshbach (HF) model \cite{Hauser52} is widely adopted in nuclear physics and astrophysics, in which case the contribution of not only the target ground state, but also of thermally populated excited states has to be taken into account (Section~\ref{masses_hight}). The model is discussed in detail in many places, and need not be reviewed here again extensively. The reader is referred to
{\it e.g.} \cite{Arnould72,Goriely08a,Koning12,Herman07,Aikawa05}.

The HF model relies on the fundamental Bohr hypothesis of formation of a CN the decay of which is uncorrelated with the formation channel, save reaction energy and angular momentum (see above). This hypothesis may not be fully satisfied, however, particularly in cases where a few strongly and many weakly absorbing channels are mixed. As an example, the HF equation is known to be invalid when applied to the elastic channel for which the transmission coefficients for the entrance and exit channels are identical, and hence correlated. These correlations enhance the elastic channel and accordingly decrease the other open channels. To account for these deviations, a width fluctuation correction is introduced into the HF formalism (see {\it e.g.} \cite{Hilaire03}). 

The uncertainties affecting any statistical cross section calculation originate from:
\vskip0.2truecm

\noindent (1) the reaction mechanism itself, {\it i.e.} the CN formation and de-excitation. Compound, pre-equilibrium and direct components may in fact be involved. The last two contributions are often neglected, in particular in astronuclear physics calculations. The impact of this neglect is studied in \cite{Goriely08a,Xu12,Xu14}; 

\vskip0.2truecm
\noindent (2) the evaluation of the nuclear quantities required for the calculation of the transmission coefficients in  the HF equations \cite{Goriely08a}, namely the ground and excited state properties (masses,
deformations, matter densities, excited state energies, spins, parities, \dots), nuclear level densities, $\gamma$-ray strength, optical potential, and fission properties, as depicted in Fig.~\ref{fig_nucmod}.   When not available experimentally, this information has to be derived from nuclear models. For astronuclear physics applications, the various nuclear ingredients should ideally be determined from {\it global}, {\it universal}, and {\it microscopic} models, as discussed in Section~\ref{nucdata}. 
 
A new generation of codes, namely MOST \cite{Aikawa05} and TALYS \cite{Goriely08a,Koning12}, adopt global and coherent microscopic (or at least semi-microscopic) models to the largest possible extent in order to avoid the use of phenomenological highly-parametrized models (Fig.~\ref{fig_nucmod}) \cite{Holmes76,Rauscher01} still of frequent use in astronuclear physics. TALYS offers the possibility to adopt different input parameters and a variety of local or global models for the nuclear ingredients. This allows to evaluate the impact of parameter uncertainties, but also model uncertainties which often dominate parameter uncertainties \cite{Goriely15b,Goriely17a}. 

%------------------------------------------------
\begin{figure*}
\begin{center}
\includegraphics[scale=0.40]{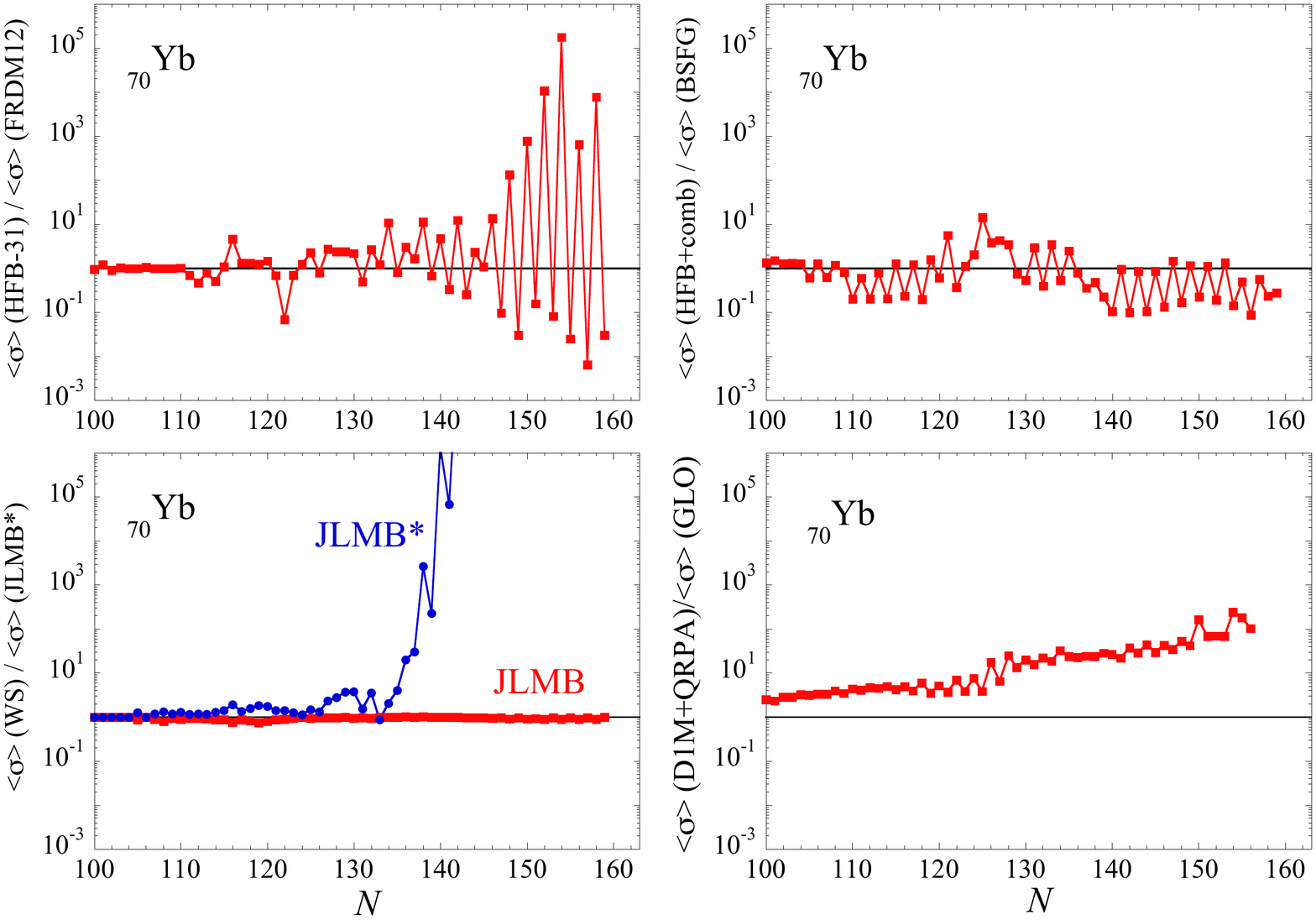}
\vskip -0.5cm
\caption{(Color online) Illustration of some uncertainties affecting the prediction of the radiative neutron capture rates at $T=10^9$~K by the Yb isotopes ($Z=70$) between the valley of $\beta$-stability and the neutron drip line. Considered are the sensitivity to the mass model (upper left), the nuclear level densities (upper right), the optical potential (lower left) and the $\gamma$-ray strength function (lower right). See main text for more details.}
\label{fig_HF_ng}
\end{center}
\end{figure*}
%-------------------------------------------------- 

Figure~\ref{fig_HF_ng} illustrates the uncertainties affecting the prediction of the radiative neutron capture rates in the Yb isotopic chain from the stable isotopes up to the neutron drip line. The Maxwellian-averaged cross section (MACS) at a temperature $T=10^9$~K (characteristic of the r-process nucleosynthesis) is estimated using either phenomenological or (semi-)microscopic models for the various nuclear ingredients of the radiative neutron capture calculations (Fig.~\ref{fig_nucmod}). 

Deviations due to the use of the HFB-31 \cite{Goriely16a} or FRDM \cite{Moeller16} mass models are found to be small close to the valley of nuclear stability, but significantly increase for exotic neutron-rich nuclei. In this case, neutron capture rates may deviate by up to 5 orders of magnitude, as illustrated in Fig.~\ref{fig_HF_ng} (upper left panel). The impact of the nuclear level density model on the capture rates is illustrated in the upper right panel of Fig.~\ref{fig_HF_ng}. The calculations are performed either with the HFB plus combinatorial model \cite{Goriely08b} or with the phenomenological Back-Shifted Fermi Gas (BSFG) model \cite{Koning08}. Deviations up to a factor of 10 with a clear odd-even effect are noticed.

As far as optical potentials are concerned, the phenomenological Woods-Saxon (WS) form ({\it e.g.} \cite{Koning03}) has long been replaced by the nucleon-nucleus optical potential of \cite{Jeukenne77} derived from the Br\"uckner--Hartree--Fock approximation using a Reid's hard core nucleon--nucleon interaction.  This so-called JLM potential has been updated by \cite{Bauge01} through an empirically renormalization of the energy dependence of the potential depth in order to reproduce scattering and reaction observables for spherical and quasi-spherical nuclei between $^{40}$Ca and $^{209}$Bi in a large energy range from the keV region up to 200~MeV.  As seen in Fig.~\ref{fig_HF_ng}, the resulting JLMB potential gives similar radiative neutron capture MACS as the WS potential, a well-known result related to the dominance of the strong interaction over the electromagnetic one in the HF approach.
However, the JLMB renormalization factors are rather well constrained by experimental data except for the isovector contribution to the imaginary part of the potential. This drawback is cured by an adjustment on experimental s-wave neutron strength function data between 1 and 100~keV \cite{Goriely07b}. As illustrated in Fig.~\ref{fig_HF_ng} (lower left panel), the resulting modified JLMB* potential has a drastic impact on the Yb neutron capture rates.  At large neutron excesses, the imaginary component is indeed reduced, lowering the neutron absorption channel, and consequently the radiative neutron capture cross section. In particular,  the rates predicted with JLMB* rapidly drop for the  $N \gsimeq 135$ Yb isotopes, leading to a negligible resonant neutron capture. We stress that a JLMB*-type renormalisation of the isovector component of the imaginary potential in the keV region clearly needs to be further constrained by additional theoretical and experimental works.  

Finally, Fig.~\ref{fig_HF_ng} (lower right panel) illustrates the impact of the photon strength function on the MACS radiative neutron captures. The D1M+QRPA models \cite{Goriely19,Goriely18a} predict an extra low-lying strength with respect to the generalized Lorentzian model \cite{Capote09}. An increase of the neutron capture rates by 1 to 2 orders of magnitude results.

%********************************************
\subsubsection{Non-statistical models}
\label{reac_nonstat}
%*****************************************

\noindent {\bf Direct captures}.  Models for the direct capture (DC) mechanism have been discussed in many places, and are not reviewed in detail here. The reader is referred to
{\it e.g.} \cite{Iliadis15,Descouvemont16,Bertulani16,Mukhamedzhanov17,Fiscic19} for recent works (this list is not exhaustive). In short, they are divided into {\it i)} those involving adjustable parameters, such as the $R$- or $K$-matrix methods, and {\it ii)} ``ab initio'' descriptions, like the potential model, the Distorted Wave Born Approximation (DWBA), and other microscopic models. The first family of models is applicable only when enough cross section data are available above the Gamow window for a reliable extrapolation to lower energies. In contrast, the second one does not require such an information, but only an experimentally-based nucleus-nucleus or nucleon-nucleon interaction.  We restrict ourselves to a description of the direct neutron capture by exotic neutron-rich nuclei of interest in r-process simulations (Section~\ref{prod_r}) for which few, or even no resonant states are available. In these conditions, the neutron DC rate can be 2 to 3 orders of magnitude larger than the one derived from the Hauser-Feshbach approach. 
As already noted above, this DC proceeds via the excitation of only a few degrees of freedom on timescales of the order of the typical time necessary for the projectile to travel across the target, which is much shorter than the one required to form a CN. The DC can be satisfactorily described by the perturbative approach known as the potential model \cite{Xu12,Xu14,Lynn68,Satchler80,Satchler83,Descouvemont08}.  
 
Significant uncertainties still affect the DC predictions. These are related to the determination of the nuclear structure ingredients of relevance, {\it i.e.}  the nuclear mass, spectroscopic factor, neutron-nucleus interaction potential and excited level scheme. Special emphasis has to be put on the determination of the low-energy excitation spectrum including all the spin and parity characteristics. This can be deduced from a nuclear level density model. A combinatorial evaluation of this quantity ({\it e.g.} \cite{Goriely08b,Hilaire12}) has clearly to be preferred over the widely used statistical approach of the BSFG type. An important effort still needs to be put in order to improve the prediction of the necessary nuclear inputs within reliable microscopic models. The transition from the CN to the DC mechanism  when only a few resonant states are available also needs to be scrutinized further in detail, for example within the Breit-Wigner approach or the ``High Fidelity Resonance'' method \cite{Rochman17}.
\vskip0.1truecm

\noindent {\bf Pre-equilibrium mechanism}. Since the basic work of \cite{Griffin66}, a variety of pre-equilibrium (PE) models have been developed. They are extensively described in \cite{Gadioli92,Koning04}, and are just broadly brushed here. 

The PE theories are basically of the (semi-)classical or quantum-mechanical types. Widely used classical PE descriptions are the exciton and hybrid models. They involve the formation of a pre-compound system the decay of which is described in terms of an intra-nuclear cascade. The classical trajectories of the particles inside the nucleus are followed in coordinate space by means of Monte Carlo methods.

Quantum mechanical theories have also been developed to describe the PE mechanism \cite{Tamura82,Bonetti91,Bonetti94,Dupuis11,Dupuis17}, and the increase in computer capabilities has enabled to evaluate cross sections in these frameworks. Although some controversies regarding the underlying quantum statistics in multistep reactions have arisen, quantum mechanical PE theories tend to account for experimental angle-integrated emission spectra with an accuracy comparable to that found in the semiclassical models, and with a higher accuracy for angular distributions, competing with phenomenological experiment-based systematics \cite{Kalbach88}.

%------------------------------------------------
\begin{figure}
\begin{center}
\includegraphics[scale=0.40]{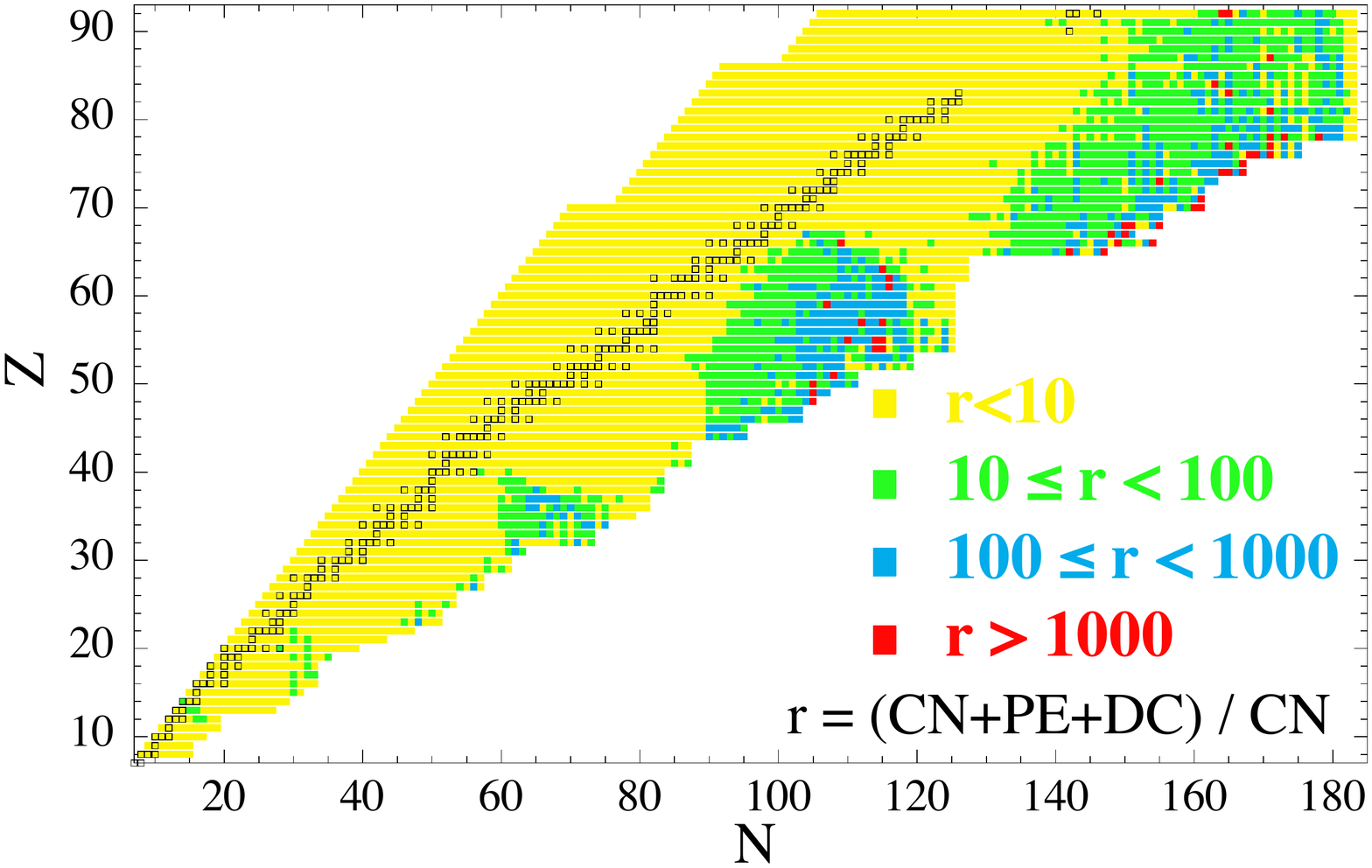}
\vskip -1.3cm
\caption{(Color online) Ratio $r$ of the total (CN + PE + DC) to the CN  neutron capture rates at $T=10^9$~K for all nuclei with $8 \le Z \le 92$ lying between the $N=Z$  and  neutron drip lines. Black squares represent stable and long-lived nuclei.}
\label{fig_HF_dc}
\end{center}
\end{figure}
%-----------------------------------------------

The PE contribution to the reaction cross sections determined within the  two-component exciton  model \cite{Koning04} has been studied systematically along with the DC and CN ones by using the TALYS reaction code \cite{Goriely08a,Koning12,Xu14}. Of particular importance is the fact that the three contributions are calculated consistently with the same nuclear inputs. In particular, the adoption of a single nucleon-nucleus optical potential ensures that the three components are calculated on the same footing, and represent partial fluxes of the same total reaction cross section. 

In stellar conditions, the PE contribution is found to affect the radiative capture rates as well as photoneutron rates of neutron-rich nuclei only at  temperatures $T \gsimeq 3~10^9$~K \cite{Goriely08a}. As shown in 
Fig.~\ref{fig_HF_dc}, the DC increases the radiative neutron capture rates by a factor up to $10^3$ for drip line nuclei, and dominates in fact the capture process by most of the neutron-rich nuclei located between closed neutron shells. However, for some neutron-rich nuclei, DC transitions are found to be forbidden by selection rules \cite{Xu14,Goriely97a}.

\begin{figure}[tb]
\begin{center}
\epsfig{file=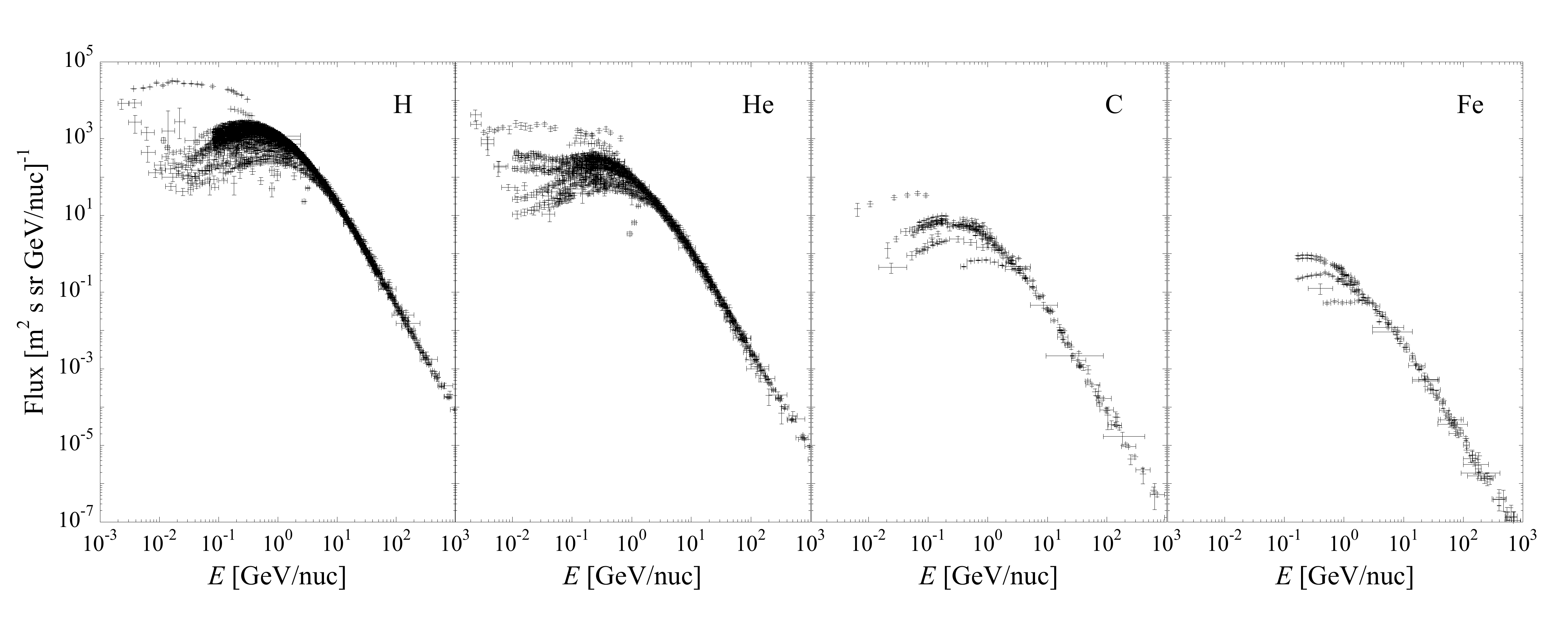,scale=0.35,angle=0.}
\caption{Differential fluxes of GCR protons, Helium, Carbon and Iron measured by different spacecrafts near Earth as a function of their kinetic energy per nucleon.The references to all data and publications can be found in the database \cite{Maurin14} of charged GCRs.}
\label{fig_gcrspectrum}
\end{center}
\end{figure}

A remarkable application of non-statistical cross section models concerns the interactions of the GCRs with the ISM and of stellar/solar energetic particles with their surrounding material, leading to the production of a variety of stable and radioactive residuals from the primary high-energy particles (see Sections~\ref{obs_cr} and \ref{nucleo_gcrs}). They have been at the focus of a wealth of nuclear physics experimental and theoretical studies which are close to impossible to review here more or less exhaustively. The reader is just referred to some recent publications. The GCR nuclear physics aspects are recently reviewed by \cite{Genolini18,Evoli18}, from which it appears that a large variety of experimental cross section data are still missing at the energies of relevance that can be inferred from Fig.~\ref{fig_gcrspectrum}. The interpretation of the experiments and the prediction of unmeasured cross sections are done with the use of a variety of models, ranging from widely used parametrized and semi-phenomenological approaches to more sophisticated ones including cascade-exciton and hybrid Monte Carlo simulation models. Note that some simplification arises from the fact that the collisions can be treated as quasi-free scattering processes when incident nucleon wavelengths are short relative to inter-nucleon distances ({\it i.e.} at energies in excess of 100 MeV), as is especially the case in the interaction of GCRs with the ISM. Still, it is widely acknowledged that there is at present no accurate theory that predicts the necessary cross sections for all collision pairs and energies of interest in GCR physics.

Various nuclear data bases of relevance to GCR studies are available (see Section~\ref{burn_compil}). From their inspection, it appears that nuclear physics lacks behind the high accuracy requested to take full advantage of the very high-quality data provided by the current generation of cosmic-ray experiments.

The next step in the evaluation of the GCR transmutation rates is the integration of the energy-dependent cross sections on an energy spectrum that varies as the GCRs propagate in the Galaxy. Some such spectra near the Earth are illustrated in Fig.~\ref{fig_gcrspectrum}.
 
 %********* FIGURE  ************************************
%****************************************************

As far as SEP transmutations are concerned, much work has been been devoted to the experimental and theoretical determination of cross section data relevant to the production of stable and radioactive residual nuclides by medium-energy protons and neutrons. This interest relates to the fact that this nuclear information is of importance not only in astrophysics, but also in many different fields of basic and applied sciences including space and environmental sciences, medicine, accelerator technology, space and aviation technology, accelerator driven transmutation of nuclear waste and energy amplification. The situation remains less satisfactory for He-induced cross sections. A vast literature is devoted to the study of the very many cross sections required to follow the abundance changes induced by the SEP transmutations. It is not possible to review it here in detail. We just direct the reader to a selection of papers published after 2005 where very many relevant references can be found: \cite{Michel07} for an overview, \cite{Leya11,Michel15} for neutron-induced reactions, \cite{Issa13,Michel05} for proton-induced reactions. Nuclear models are used for comparison with experimental data and for predicting missing laboratory information. They are classically of the HF type (see Section~\ref{reac_hf}). Of wide use is the computer code TALYS already viewed as a state-of-the-art tool in the calculation of thermonuclear reaction rates. A comparison between the predictions of different models can be found in \cite{Leray11}.

%*********************************************
\section{The important chains of charged-particle induced thermonuclear reactions in astrophysical plasmas}\label{burn_general}
%********************************************
  
%*** Sect.5.2.1   **************
\subsection{Hydrogen burning}
\label{burn_h}
%*******************************

Hydrogen burning modes have been studied in substantial details in many places ({\it e.g.} \cite{Iliadis15}), so that it is quite superfluous to review them here again. Let us limit ourselves to some brief comments.

Various non-explosive (``cold") burning modes are identified, starting with the seminal works of \cite{Bethe38,Weizsacker38,Bethe39,Atkinson36}. They develop either in the central regions or in peripheral layers of a large variety of stars (see Fig.~\ref{fig_starevol}), at temperatures in the approximate range of $10^6$ to $10^8$ K. They are characterized by the dominance of $\beta$-decays over proton captures by the unstable nuclei involved, and come to an end when the proton concentration reaches a sufficiently low level.
The dominance of one mode over the others depend not only on temperature, but also on the initial composition. 
In addition to the p-p chains that operate in solar-type stars,
energy production by H burning  can also occur non-explosively through
 the cold CNO cycles, at least if the star contains C, N or O from the start (this excludes Pop~III stars; Section~\ref{nucleo_stars}). The main product of the p-p chains and CNO cycles is \chem{4}{He}. The C, N, and O act just as catalysts in the CNO cycles, and are largely transformed into \chem{14}{N}.
 
Some hydrogen can also be consumed by the NeNa and MgAl chains. These two burning modes only play a minor role in the stellar energy budget, but are of significance in the production of the Na
to Al  isotopes, especially in massive stars. Most important, the MgAl chain may synthesize \chem{26}{Al}, which is a very interesting radioactive nuclide for $\gamma$-ray astronomy and cosmochemistry  (Section~\ref{obs_gamma} and \ref{obs_sos}).

Much experimental and theoretical effort has been devoted to the
reactions involved in the H-burning modes. They are presented in various recent compilations (see Section~\ref{burn_compil}, as well as \cite{Coc15} for some experimental updates up to 2015) which also provide typical uncertainties still affecting the relevant reactions. The yields from the CNO, NeNa and MgAl burning modes based on the adopted rates of the Nuclear Astrophysics Compilation of REactions (NACRE) library  and their uncertainties \cite{Angulo99} have been analyzed by \cite{Arnould99b}\footnote{These results should be updated based on more recent determinations of some of the NACRE reaction rates}. 

Variants of the cold H-burning modes have been identified in explosive situations, namely novae and x-ray bursts. As already noted above, these situations are characterized by burning timescales that are much shorter (of the order of seconds to hours) and by higher temperatures than in quiescent conditions. This is why they are referred to as ``hot H-burning modes".  The consequences of these conditions are threefold. First, proton captures (the rates of which are dramatically increasing with temperature) start competing with $\beta$-decays (which are largely temperature-independent in view of the paucity of thermally populated excited states of the nuclei involved), so that the nuclear flows are driven to the neutron-deficient side of the valley of nuclear stability, in contrast to the situation encountered in the cold burning modes. Second, photodisintegrations can enter the picture (in view of their strong temperature-dependence), which may limit to a more or less large extent the propagation of the flow to neutron-deficient nuclides. Third, the transient nature of the high temperature regimes implies that the material becomes rapidly cool enough for the lifetimes of the nuclei against charged-particle captures to become longer than the typical timescales of the explosions well before H-exhaustion. In such conditions, the charged-particle captures become in fact inefficient, this situation being referred to as the ``freeze-out'' of the reactions.

The nuclear physics consequences of the hot burning modes is that the precise description of the relevant flows requires the knowledge of the rates of captures of protons, but possibly also of  $\alpha$-particles and neutrons by more or less highly $\beta$-unstable nuclei. Except in some specific cases, these reactions have not lent themselves yet to an experimental scrutiny, so that theory must enter the evaluation of their rates. This is even more so as the target excited states entering the astrophysical rates have populations that increase with temperature, so that their contribution to the rates may be increased. The associated nuclear physics uncertainties add to those affecting the modeling of explosions of the nova or x-ray burst types.
 
%******** ***************
\subsubsection{The hot p-p mode}
\label{burn_hotpp}
%*******************************

This mode, as first recognized by \cite{Arnould75}, could in particular develop in nova explosions resulting from  the accretion on a WD of material from a companion star in a  binary system. A variety of reactions of importance have been identified. One of the keys of this type of burning is the \reac{8}{B}{\gamma}{p}{7}{Be} photodisintegration. This reaction impedes the transformation of \chem{7}{Be} into \chem{4}{He}
which characterizes the cold p-p chain, and may thus be responsible for 
some \chem{7}{Li} production (through the \chem{7}{Be} decay) in nova situations. The expectations of \cite{Starrfield78} based on detailed nova models have been substantiated by recent observations of novae and of CRs possibly polluted by nova ejecta ({\it e.g.} \cite{Kawanaka18,Izzo19}).

Several of the hot p-p reaction rates have been scrutinized both 
theoretically and with the help of indirect experimental techniques. This is namely the case for the reactions \reac{7}{Be}{\alpha}{\gamma}{11}{C}, \reac{11}{C}{p}{\gamma}{12}{N} and
 \reac{8}{B}{p}{\gamma}{9}{C}, which may also have some importance in the nova scenario \cite{Boffin93}.
 
%*******************************
\subsubsection{The hot CNO and NeNa-MgAl chains}
\label{burn_hotcno}
%***********************************************
 
As first explored by \cite{Audouze73,Arnould74} on grounds of schematic astrophysical models, the cold CNO switches to the hot CNO mode when \reac{13}{N}{p}{\gamma}{14}{O} becomes faster than the \chem{13}{N} $\beta$-decay.  This occurs typically at temperatures in excess of $10^8$ K. The cold NeNa and MgAl chains evolve into a hot mode at temperatures that are quite similar to the operating  conditions for the
hot CNO. In fact, novae could be favourable sites for the development of both
the  hot CNO and NeNa-MgAl chains.

In the considered hot H burning modes, many reaction rates on unstable nuclei come into play. 
Much theoretical and experimental effort has been devoted to a reliable determination of the rates of some of the reactions that have been identified as keys in the development of those processes. 
In general, these rates have not been measured directly, and are rather evaluated indirectly.
There are, however, some noticeable exceptions to this situation. In particular, \reac{13}{N}{p}{\gamma}{14}{O} has been the first directly measured reaction on an unstable target.
While its rate is now known well enough for practical astrophysical purposes \cite{Arnould92}, several others still require further experimental and theoretical efforts. This question cannot be reviewed in detail here. The reader is referred to {\it e.g.} \cite{Iliadis15,Parikh14} for a discussion of the impact of reaction rate uncertainties on nova model predictions. Let us just note the special case of the stellar destruction by proton capture of the important radionuclide \chem{26}{Al}. This destruction can indeed occur through the proton capture on the \chem{26}{Al} ground state, but also on its isomeric state, which can be thermally populated in the H-burning sites. The direct experimental determination of this rate would require the development of a \chem{26}{Al^m} beam, which obviously represents an interesting technological challenge.

%*******************************
\subsubsection{The rp- and $\alpha$p-processes}
\label{burn_rp}
%**********************************************

With increasing temperatures, and as first explored by \cite{Wallace81}, the hot CNO and NeNa-MgAl modes can transform into the so-called rp- or $\alpha$p-processes when \reac{15}{O}{\alpha}{\gamma}{19}{Ne} or
\reac{14}{O}{\alpha}{p}{17}{F} become more rapid than the corresponding
$\beta$-decays. 
Alternatively, \reac{18}{Ne}{\alpha}{p}{21}{Na} could play this role if 
its still-uncertain rate can indeed become faster than the \chem{18}{Ne} 
$\beta$-decay in appropriate astrophysical  conditions, which could be
encountered in particular in certain x-ray bursters resulting 
from the accretion of matter on a NS.
This chain of transformations is termed the rp-process, and is considered to be the main energy provider of the explosion process.
It could transform into an $\alpha$p-process at temperatures that are 
high enough for $\rm (\alpha\,,p)$ reactions to play a leading role in
 bypassing the $\rm{proton~capture}+ \beta$-decay chains. The $\alpha$p-process is thus not a pure H-burning process per se, but is more properly a combined H- and He-burning mechanism. The nuclear flow associated to the rp- and $\alpha$p-processes could go all the  
way from the C-N-O region up to, or even beyond the iron region (possibly up to mass numbers in the vicinity of $A \approx 100$). 

The rp- and $\alpha$p-processes have been described in many places. The reader is referred to {\it e.g.} \cite{Iliadis15} for a detailed discussion. Let us just stress that a host of reactions on unstable neutron-deficient nuclei, some of them being close to the proton drip line, are involved in these processes. As a result, the experimental study of a significant fraction of the potentially important reactions is difficult to conceive. As a necessary complement to laboratory efforts, use is made of 
a HF model which appears to be adequate for reactions involving at least medium mass nuclei, and except for reactions with low $Q$-values implying relatively low nuclear level densities. The reaction rate data have to be complemented with the evaluation of $\beta^+$-decay and continuum-e$^-$ capture rates.

A discussion of the impact of the nuclear physics uncertainties on the predictions of the rp- and $\alpha$p-process paths, including their termination points in the chart of the nuclides, can be found in {\it e.g.} \cite{Iliadis15,Parikh14}.  The consequences of these uncertainties on the modeling of the x-ray bursts and on their observable properties has been discussed recently \cite{Meisel19}. If indeed the burst energetics of the rp- and $\alpha$p-processes is acknowledged, their precise nucleosynthetic role at the galactic scale remains a matter of debate in view of the possibly very limited efficiency of mass ejection by x-ray bursts. Finally, one has to realize that various variants of the rp- and $\alpha$p-processes are found in the literature, associated to different classes of x-ray bursts (see {\it e.g.} \cite{Jose15}).

We want to conclude this short account of the rp- and $\alpha$p-processes with a word of caution. The reliable identification of key reactions which would have to be  painstakingly measured in the laboratory is by far not a trivial matter. In particular, no truly realistic (three-dimensional) simulations have ever been used to 
analyze in a careful and detailed way the impact of reaction rate uncertainties on the observable properties of the highly complex x-ray bursts. This situation is clearly in danger to lead to some misleading evaluations of the true 
importance of some nuclear data. Nuclear physicists may find it desirable to be fully aware of this situation before embarking on very difficult experiments. 

%*******************************
\subsection{The He to Si burnings}
\label{burn_hesi}
%******************************************

The quiescent burning phases from He to Si occur in the core or in peripheral shells of stars, as depicted schematically in Fig.~\ref{fig_starevol}. This whole series of nuclear episodes occurs in massive enough stars of Cat.~5 only (Section~\ref{nucleo_stars}). In the case of less massive ones (Cat.~3), the nuclear sequence stops at He-burning already, like in solar-mass stars, or at least at C-burning. The explosive modes of He- to Si-burnings develop in different supernova events, explosive He-burning possibly operating in the material accreted at the surface of a neutron star, leading to a variety of x-ray bursts, as already noted above, and even in the merging of WDs in binary systems. The Si-burning episode terminates with the establishment of a NSE in which the composition of the relevant stellar zones is governed by the laws of classical statistical mechanics. The net result of these various burning modes is the progressive transformation of the He-rich material emerging in particular from the H-burning episodes into nuclides forming the Fe-group. 

The astrophysics and nuclear physics of the quiescent and explosive He to Si burning stages have been discussed at length in many places, and do not need to be reviewed again. The reader is referred to {\it e.g.} \cite{Iliadis15} for details, and to {\it e.g.} \cite{Limongi18} for a description of these burning episodes in the framework of a one-dimensional approximation of the non-explosive evolution of massive stars and of an artificially triggered supernova explosion. We restrict ourselves here to some specific comments.

Due to the absence of a stable nuclide with mass number $A=8$, it has taken more than one decade to discover how stars were managing to burn their helium mainly emerging from H-burning (on top of a minor abundance present at the stellar birth). The solution came from \cite{Salpeter52} who imagined a sequential two step process referred to as the 3$\alpha$ reaction. In a first step, two $\alpha$-particles fuse into the extremely short-lived \chem{8}{Be} (half-life of $8.2 \times 10^{-17}$ s). With time, an equilibrium builds up between the production and decay of \chem{8}{Be}, which allows the production of \chem{12}{C} through
\reac{8}{Be}{\alpha}{\gamma}{12}{C}. The probability of this transformation becomes high enough for ensuring an energy equilibrium to be established in stellar zones with temperatures in excess of about $10^8$ K. This success story did not stop here! Remarkably enough, it was soon realized by \cite{Hoyle54} that this energy production process would not be efficient enough unless the second step proceeds via an excited state of \chem{12}{C} with an energy close to 7.7 MeV. The embarrassment was that no such level was known experimentally ! The relief came from \cite{Dunbar53} who triumphantly confirmed in the laboratory the existence of this level predicted on purely astrophysics grounds, and commonly referred to as the "Hoyle state". The reader is referred to \cite{Aquila18} for recent experimental studies.

Another reaction of importance in the He-burning process is \reac{12}{C}{\alpha}{\gamma}{16}{O} which has been at the focus of a flurry of experimental and theoretical efforts. The reader is referred to \cite{deBoer17} for a review of experiments and theory, and to {\it e.g.} \cite{Kirsebom18,Shen19} for experimental approaches (see also Section~\ref{decay_alpha}).
A comment is in order here on the importance of this reaction, which is often claimed as being a key process in astronuclear physics. It is true that a precise knowledge of the $\alpha$-particle capture rate by \chem{12}{C} is a {\it necessary} condition to predict the \chem{12}{C}/\chem{16}{O} abundance ratio emerging from He-burning, which has observational implications, as well as an impact of the evolution of the stars possiby following 
He-burning. However, it is in general not recognized clearly enough that it is not a {\it sufficient} information in these respects. The very many uncertainties and shortcomings affecting the stellar evolution simulations still significantly blur the whole picture that could be provided by the nuclear physics laboratory.

Other reactions are involved in quiescent He-burning ({\it e.g.} \cite{Iliadis15}), and may have an impact on abundance variations in the approximate $A = 12 - 25$ range.  Let us single out the transformation of \chem{14}{N} (the main product of the CNO cycles of H-burning) into variable amounts of \chem{18}{O} or \chem{22}{Ne}, depending on the level of completion of the \chem{14}{N} burning by $\alpha$-particle captures. While \chem{18}{O} may be at the origin of the production of \chem{19}{F} \cite{Goriely00}, the \reac{22}{Ne}{\alpha}{n}{25}{Mg} reaction may produce neutrons leading to an s-process (Section~\ref{prod_s}). The impact of nuclear uncertainties on abundances affected by He-burning is discussed by \cite{Arnould99} on grounds of a very schematic astrophysics model and of the reactions compiled by \cite{Angulo99}, and by \cite{West13} for the 3$\alpha$ and \reac{12}{C}{\alpha}{\gamma}{16}{O}\footnote{Constraining nuclear reaction rates based on astrophysics considerations, as too often attempted in the literature, including \cite{West13}, is a highly risky exercise considering the severe uncertainties and limitations of stellar models.}.

In explosive situations, $\alpha$-particle captures take place through the $\alpha$p-process discussed in Section~\ref{burn_rp}. Some reactions involving unstable targets also occur in addition to those making up the quiescent mode of He burning. A limited r-process could also develop in the peripheral He-burning shell of massive stars during a CCSN event, as mentioned in Section~\ref{prod_r_site}.

Non-explosive carbon burning occurs in the core or in a shell of stars that are massive enough to reach temperatures in excess of about $5 \times 10^8$ K in the C-rich regions (Section~\ref{nucleo_stars}). This burning mode is discussed in detail in many places ({\it e.g.} \cite{Iliadis15}), and need not be reviewed here in detail. We limit ourselves to some comments.

The \chem{12}{C} + \chem{12}{C} is the first reaction between light heavy ions encountered in the course of stellar evolution. The main reaction channels are the production of \chem{20}{Ne} and \chem{23}{Na} through the emission of an $\alpha$-particle and of a proton. The additional endothermic $^{12}$C($^{12}$C,n)$^{23}$Mg reaction is a weak neutron producing channel. Secondary reactions involve the captures of the liberated nucleons and $\alpha$-particles by the He-burning ashes composing the C-burning locations. The main ashes of C-burning are \chem{16}{O}, \chem{20}{Ne}, \chem{23}{Na} and  \chem{24}{Mg}, with trace amounts of various nuclides in the $A < 20$ and $A\geq28$ ranges. Some s-processing may also accompany C-burning (Section~\ref{prod_s}), the neutrons being predominantly produced by $\alpha$-captures on the \chem{22}{Ne} left over from He-burning.

Explosive deflagration/detonation of C may be at the origin of certain events, particularly of the SNIa type (Section~\ref{nucleo_stars}).  The corresponding nuclear pattern is much more extended than in the non-explosive conditions. This is in particular the case in the detonation regime that leads to the production of nuclides up to the iron region. Many reactions are involved that have not been studied in the literature. HF predictions are very often a necessary substitute. Of very special interest is the sub-Coulomb fusion of two \chem{12}{C} ions at energies of astrophysical relevance. It raises quite complex questions related in particular to the energy structure of its sub-Coulomb cross section. It has been at the focus of many experimental and theoretical efforts. The reader is referred to {\it e.g.} \cite{Tumino18,Chien18,Zickefoose18,Torres18,Mori19,Tang19} for recent studies
\footnote{Note that the analysis by \cite{Tumino18} has been hotly debated recently \cite{Tumino18a,Beck19}}. There are prospects to measure the \chem{12}{C} + \chem{12}{C} fusion cross section down to energies close to the effective stellar energy $E_0$ in underground experiments.
 
Following C-burning, the neon/oxygen burnings develop in the central regions or in peripheral shells of stars. These burnings have been discussed in many places ({\it e.g.} \cite{Iliadis15}), and need not be discussed here. Some selected comments follow.

In many instances, Ne burning develops before O starts burning, at temperatures in excess of about $10^9$ K in quiescent burning conditions, and at somewhat higher temperatures ($T \gsimeq 3 \times 10^9$ K) in explosive situations. It is important to note that the fusion of two Ne ions does not take place in stars due their too high Coulomb repulsion. What is called Ne burning is in fact the photodisintegration of \chem{20}{Ne} into \chem{16}{O} through the emission of an $\alpha$-particle. The release of these particles triggers a flurry of reactions starting with the radiative $\alpha$-particle captures on \chem{16}{O}, \chem{20}{Ne} and on the produced \chem{24}{Mg}, leading to some accumulation of \chem{28}{Si}, but also to a smaller, although significant, pattern of ($\alpha$,p) and ($\alpha$,n) transformations, leading to some production of nuclides in the approximate Mg to S range. Note that this phase leads to a net  energy production, although the primary photodisintegration of \chem{20}{Ne} is endothermic. At the typical Ne-burning temperatures, many of the main reaction rates have been measured at energies of astrophysical relevance.  Some uncertainties remain, however, and some secondary rates still have to be evaluated theoretically. Note that the key \chem{20}{Ne} photodisintegration rate is obtained from \reac{16}{O}{\alpha}{\gamma}{20}{Ne} through the application of the reciprocity theorem.  

Oxygen burning builds on the \chem{16}{O} accumulated in the previous evolutionary phases.  It takes place at temperatures in the approximate (1.5 - 2.5)$\times 10^9$ K and (3 - 4)$\times 10^9$ K ranges in quiescent and explosive situations. The main primary \chem{16}{O} + \chem{16}{O} fusion channels produce $\alpha$-particles and protons, as well as some amount of neutrons. Their captures lead to the build-up of a large variety of nuclides in the approximate Si to Ca range. Oxygen burning shows some important new features compared to the previous burning stages. First, photodisintegrations gain importance, especially in explosive conditions. The combined effect of these reactions and of the capture of the released light particles leads to a first step towards the build-up of so-called quasi-equilibrium clusters that will develop further during the next Si-burning phase. The abundance distributions in these clusters are governed by the laws of statistical mechanics, and not by cross section data. Second, the O-burning conditions are prone to the development of continuum electron captures 
(Section~\ref{decay_continuum}). In contrast to nuclear reactions, these weak interaction processes modify the  excess of neutrons per nucleon in the stellar plasma (see {\it e.g.} \cite{Iliadis15}, Eq. 1.36, for a definition of the neutron excess parameter $\eta$. The value of this parameter is very important in the subsequent fate of the stars).

The \chem{16}{O} + \chem{16}{O} fusion near or below the Coulomb barrier has been the subject of many experimental and theoretical studies. The main problems that are raised are of a different nature than in the \chem{12}{C} + \chem{12}{C} case. They relate in fact to the many possible decay channels of the compound nucleus, with the lack of fully reliable data on the competition between two- and three-particle exit channels that produce the same types of light particles. Some prospects of experimental improvements are discussed by {\it e.g.} \cite{Hayakawa16,Courtin17}. Other uncertainties of nuclear origins concern reactions on unstable nuclei, the variety of which is somewhat limited, however, by phtodisintegrations, as well as continuum electron capture rates. Also note that the \chem{12}{C} + \chem{16}{O} fusion may also play some role. The status of this reaction has been recently discussed by \cite{Fang17}.
 
The last stellar energy producing nuclear process is referred to as Silicon burning developing at temperatures around (3 - 4)$\times 10^9$ K or (4 - 5)$\times 10^9$ K in quiescent or explosive environments. Silicon burning has not to be viewed as \chem{28}{Si} + \chem{28}{Si} or \chem{28}{Si} + \chem{32}{S} fusions between ions emerging from the O/Ne burning stages, as the Coulomb barriers are too high to be efficiently overcome at the temperatures reachable in stars. Instead, it corresponds to a complex process of photodisintegration of Si  that is much more extended than encountered in Ne burning. A flow descending from Si down to \chem{12}{C} or even 
$\alpha$-particles develops first as a result of nucleon and $\alpha$-particle emitting photodisintgrations. It is somewhat slowed down by reverse reactions. The capture of the released light particles lead to the build-up of an ascending flow eventually reaching the iron-group region. These captures are counterbalanced by photodisintegrations with the result that a NSE involving nuclear and electromagnetic interactions is obtained between groups of nuclei, leading to the formation of quasi-equilibrium clusters. These are linked by reactions that are for some time too slow to drop into equilibrium. This situation is encountered in particular in the Ca region. Details of the reactions involved in the Si photodisintegration process can be found in {\it e.g.} \cite{Chieffi98}. Several, but not all, of the reactions of interest are known experimentally in the energy range of astrophysical interest. The others, some of which involving unstable targets, are evaluated within the HF model. As already noted above, the abundances within the quasi-equilibrium clusters obey classical statistical physics, and thus depend essentially on binding energies that are known experimentally as the nuclei involved are close enough to the valley of nuclear stability. In addition, continuum electron captures also have to be evaluated, which introduces some extra uncertainties. As initiated during Ne/O burning, these captures modify the neutron excess of the material, with important consequences for the subsequent fate of the stars.
 
With the temperature increase accompanying the disappearance of \chem{28}{Si} at the end of the Si-burning phase, equilibrium involving the previously non-equilibrated reactions is eventually reached, merging the different quasi-equilibrium clusters into a single one involving nuclei from nucleons to the iron-group nuclei. The last link achieving equilibrium is in fact $3\alpha \leftrightarrow ^{12}$C. In such a situation, the composition of the material is essentially determined by the binding energies entering the classical statistical mechanics equations and by the neutron excess of the matter that evolves with the physical conditions, as the weak interaction processes remain out of equilibrium due to the insufficient coupling of (anti)neutrinos with matter. The nuclides of the iron group dominate the composition, with variations in their relative abundances dictated by the changing neutron excess. Note that densities in excess of $10^{13}$ g/cm$^3$ have to be reached in supernova conditions for the weak interactions to reach equilibrium in their turn, leading to a so-called ``$\beta$-equilibrium''. In this situation, an EoS dictates the nuclear characteristics of the stellar material. (Section~\ref{eos}).
  
%******************************************
\subsection{Compilations of nuclear data and thermonuclear reaction rates for astrophysics} 
\label{burn_compil}
 %*****************************************
 
As made clear in many of the previous sections, astronuclear physics is in high demand of a huge variety of nuclear data including the static and decay properties of nuclei, as well as their interactions with a variety of particles and nuclei. In spite of significant experimental efforts, massive recourse to theoretical predictions is mandatory in many applications.

The availability of compilations providing an easy access to evaluated and well documented nuclear data is an essential tool for astrophysics modeling, as first recognized by \cite{Fowler67} more than 50 years ago. An important step in the compilation effort has been taken by the Nuclear Astrophysics Compilation of REactions referred to as NACRE \cite{Angulo99}. It concerns laboratory data for charged-particle induced reactions involved in Big Bang nucleosynthesis and in the quiescent H- and He-burning modes collected up to 1999. It includes several new important features with respect to the former compilations. Among them, let us stress that proposed adopted rates are complemented with lower and upper limits based on published experimental data. These rates are presented in tabular form that is very advantageous in many respects compared to the purely analytical presentation adopted in all previous compilations. Note that the contribution of target excited states to the effective stellar rates is obtained through a HF treatment. An update of NACRE, referred to as NACRE II, is also available \cite{Xu13}. For targets with mass numbers $A < 16$, it revises 34 NACRE rates based on experimental data published after the NACRE release, and on a new theoretical handling of these data. NACRE II complements another compilation of charged-particle induced reactions based on a Monte Carlo approach, and referred to as STARLIB. It involves about 70 reactions in the 14 - 40 mass range ({\it e.g.} \cite{Iliadis15a}) complemented with neutron capture reaction and $\beta$-decay rates based on experiments and theory (uncertainties are not evaluated for theory-based rates). Solid as it may look mathematically, the Monte Carlo methodology adopted to evaluate the reaction rate uncertainties is not free from some problems of physical nature, as briefly acknowledged by \cite{Iliadis15}. Another compilation, referred to as KADoNiS (Karlsruhe Astrophysical Database of Nucleosynthesis in Stars) \cite{Dillmann14} is dedicated to experimentally-based neutron capture cross sections between \chem{1}{H} and \chem{210}{Bi} of relevance to the s-process, and to proton and $\alpha$-particle captures between \chem{70}{Ge} and \chem{209}{Bi} pertinent to the p-process (Section~\ref{prod_p}).  

As a major extension of NACRE and NACRE II, a library of nuclear data referred to as BRUSLIB (BRUSsels LIBrary) has been constructed \cite{Xu13b}, and its regular updates are available electronically from the nuclear database at http://www.astro.ulb.ac.be/bruslib. In addition to the NACRE and NACRE II data, BRUSLIB contains the latest predictions of a wide variety of nuclear data, including nuclear masses, radii, spin/parities, deformations, single-particle schemes, matter densities, nuclear level densities, E1 strength functions, fission properties, and partition functions. These are provided for all nuclei lying between the proton and neutron drip lines over the $8 < Z < 110$ range. The evaluation of these data is based on a single microscopic model 
(Section~\ref{nucdata}) that ensures a good compromise between accuracy, reliability, and feasibility. In addition, these various ingredients are used to calculate about 100 000 HF neutron-, proton-, 
$\alpha$ and $\gamma$-induced reaction rates estimated by the TALYS reaction code. Note that this same code is used for the prediction of reaction rates in STARLIB.

A nuclear NETwork GENerator NETGEN complements the BRUSLIB package \cite{Aikawa05}. It is an interactive, web-based tool to help astrophysicists in building up a nuclear reaction network as defined by each user. It makes use of the NACRE, NACRE II and other sources for experimentally-based reaction rates and of the BRUSLIB theoretical rates. It also includes experimental and several calculated (temperature-dependent) $\beta$-decay rates. The reader is referred to \cite{Xu13b} for full details.  

Finally, a recent compilation of nuclear properties for astrophysical and radioactive-ion-beam applications has been published \cite{Moeller19}. Apart from several nuclear structure data, it tabulates quantities related to 
$\beta$-delayed one- and two-neutron emission probabilities, average energy and average number of emitted neutrons, $\beta$-decay half-live with respect to GT decay with a phenomenological treatment of first-forbidden decays, and $\alpha$-decay half-lives for 9318 nuclei ranging from \chem{16}{O} to  \chem{339}{136} and extending from the proton to the neutron drip lines. The reported data are based on the use of a FRDM model adopting a folded-Yukawa single-particle potential, and of a QRPA approximation with a separable interaction for the calculation of the Gamow-Teller $\beta$-decay rates and $\beta$-delayed neutron-emission probabilities.

The calculation of abundances is obtained by solving nuclear networks that involve very often a large number of reactions. Evaluating the impact on the predicted abundances of reaction rates uncertainties is an important task in order to guide experimental and theoretical efforts, but is far from being a trivial exercise. This is even more so as the uncertainties affecting a given reaction sometimes influence in a quite indirect way the abundance of a given nuclide. This is the reason why misleading conclusions can be drawn from just sequential and unrelated variations of some reaction rates within some prescribed range of uncertainties. An alternative approach is to estimate abundances through a Monte Carlo approach which allows a random sampling over reaction rates. Illustrations of this approach can be found in {\it e.g.} \cite{Iliadis15a}. However, this Monte Carlo makes the fundamental approximations that the reactions rates are uncorrelated. This approximation is usually not correct, especially when dealing with theoretical rates which are clearly correlated through the HF formalism and corresponding nuclear ingredients. Both parameter and model uncertainties of nuclear inputs need, therefore, to be propagated into reaction rates estimates and consistently taken into account in reaction network simulations to study the impact of correlated rates on energetics and nucleosynthesis. Some application of this uncertainty analysis has been performed, for example, for the p-process nucleosynthesis and can be found in \cite{Arnould03} (see their Fig. 35, in particular).

Various nuclear data bases of relevance to the non-thermal regime relevant to GCR studies are also available, as well as a recent attempt to characterize the uncertainties in the calculation of the isotopic production cross sections (ISOPROCS project). The reader is referred to \cite{Genolini18,Evoli18} for references. As already noted in Section~\ref{reac_nonstat}, nuclear physics has not reached yet the level of accuracy that would allow to take full advantage of the very high-quality data provided by the current generation of cosmic-ray experiments.

%
 %***********************************************************
\section{Nucleosynthesis of the heavy nuclides}
\label{prod_heavy}
%**********************************************************
 
%---------------------------------------------------------- 
\subsection{The s-, r- and p-nuclides in the Solar System: generalities}
\label{prod_srp_general}
%----------------------------------------------------------
 
Since the very beginning of the development of the theory of nucleosynthesis 
({\it e.g.} \cite{Burbidge57,Cameron57,Cameron57b}), it has become a common practice to split the SoS abundance distribution of the
nuclides heavier than iron (Fig.~\ref{fig_SoS_A}) into three separate distributions giving the image of the SoS
content of the so-called p-, s- and r-nuclides. These are defined as the stable nuclides located in the chart of the nuclides on the neutron-deficient side of the valley of nuclear stability (p-nuclides), at the bottom of the valley (s-nuclides) and on its neutron-rich side (r-nuclides). 

This splitting has greatly helped clarifying the very nature of the processes responsible for the synthesis of the three classes of nuclides. It has been soon realized that the capture of charged particles by nuclei heavier than the iron peak was in general inefficient in stellar conditions. The Coulomb barriers between interacting partners indeed turn out to be high enough for the lifetimes against charged particle captures of the nuclei under consideration to be substantially longer than the typical stellar evolutionary lifetimes. Neutron captures were consequently considered as the nucleosynthetic mechanism of choice, at least for the s- and r-nuclides, and are referred to as the ``s-process'' (Section~\ref{prod_s}) and the ``r-process'' (Section~\ref{prod_r}). The situation has been for long somewhat less clear-cut for the p-nuclides. It is now considered that the ``p-process'' responsible for their production is dominated by photodisintegrations of pre-existing s- and r-nuclides, with some corrections brought by neutron captures (Section~\ref{prod_p}). The contribution of proton captures to some among the lightest p-nuclides has also been envisioned. The corresponding process, referred to as the ``rapid proton capture'' (or rp) process (Section~\ref{burn_rp}) has been suggested to develop as a result of accretion of matter onto NSs, but is not expected to significantly contribute to the Galactic enrichment.

%************** FIG *****************************
\begin{figure}[tb]
\center{\includegraphics[scale=0.35]{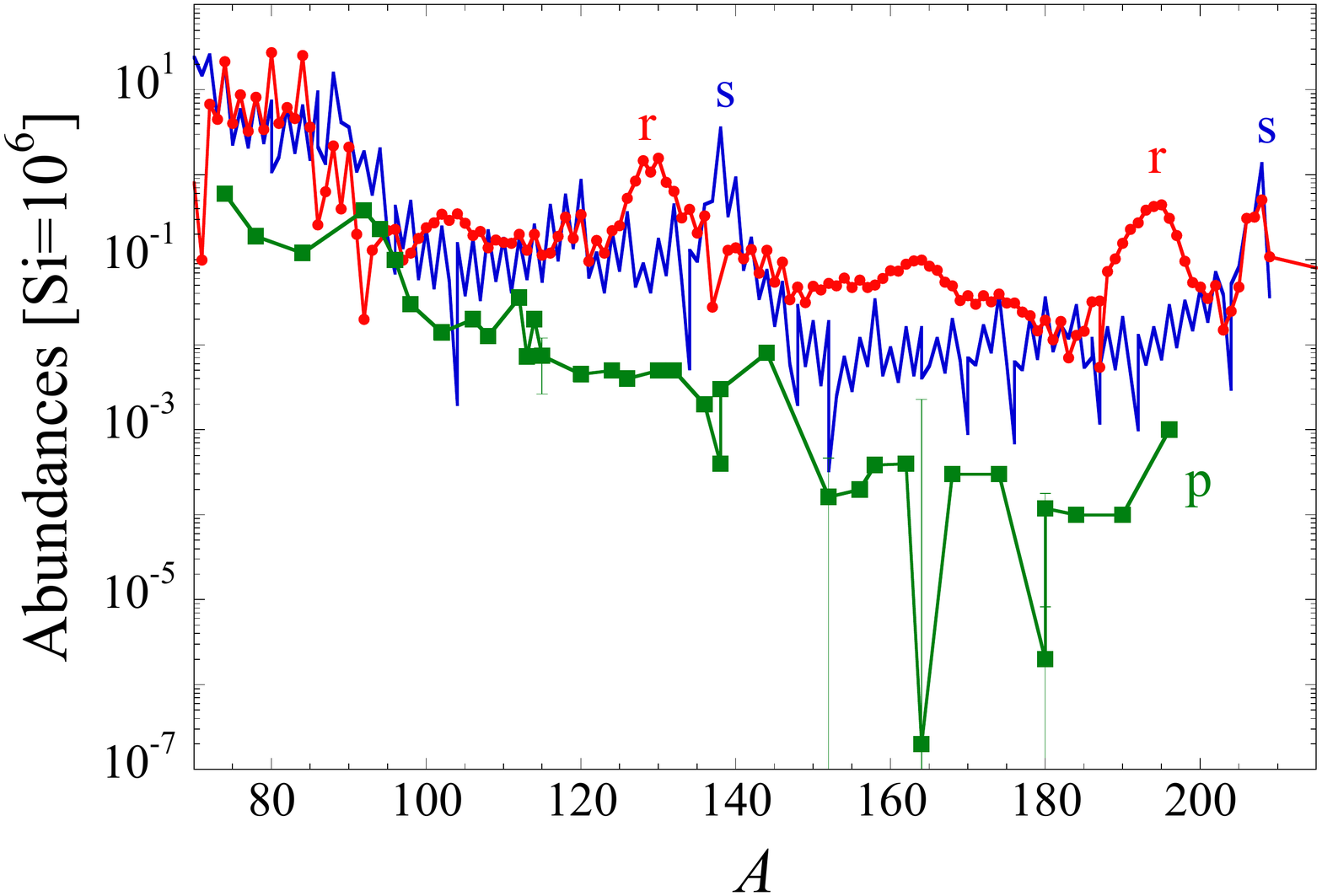}}
\vskip-0.3truecm
\caption{Decomposition of the SoS abundances of heavy nuclides into s-process ({\it solid line}), r-process ({\it dots}) and p-process ({\it squares}) contributions. The
uncertainties on the abundances of some p-nuclides that come from a possible s-process contamination
 are represented by vertical bars. See Fig.~\ref{fig_sos_sr} for the uncertainties on the SoS s- and r-nuclide data.}
\label{fig_sos_srp}
\end{figure}
%-------------------------------------------------------------------------- 

%************** FIG *****************************
\begin{figure}
\center{\includegraphics[scale=0.4]{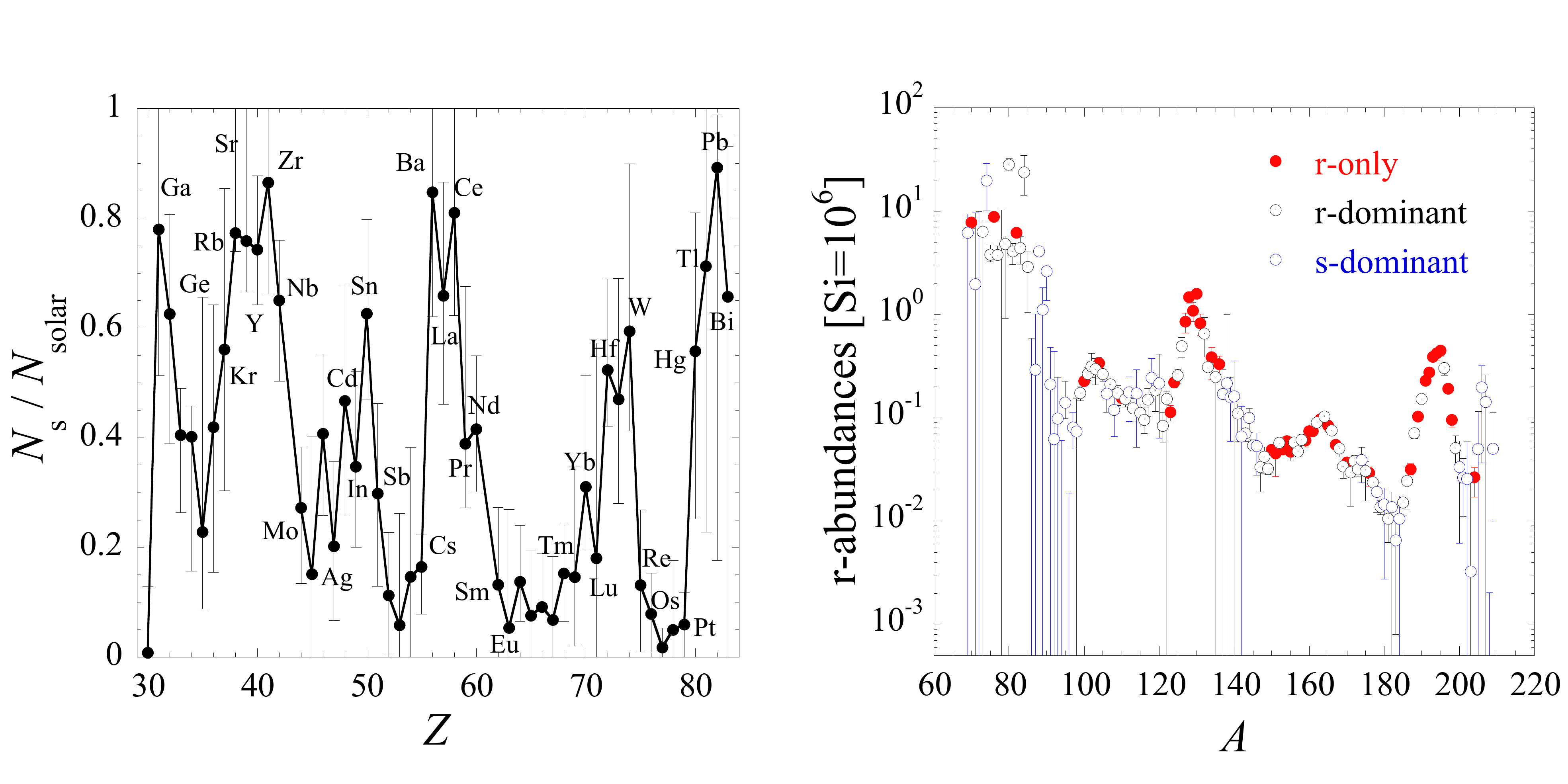}}
\caption{{\it Left panel:} S-process contribution to the SoS abundances $N_{\rm solar}$ of the elements with $Z \geq 30$ from a model discussed in {\it e.g.} \cite{Arnould07}. Uncertainties are represented by vertical bars.
{\it Right panel:} SoS isotopic r-abundances from a model discussed in {\it e.g.} \cite{Arnould07,Goriely99}. Different symbols identify different relative levels of r-process contribution. The s-dominant nuclides are defined as those predicted to have more than 50\% of their abundances  produced by the s-process. The s-process contribution varies between 10 and 50\% in the case of the r-dominant species, and does not exceed 10\% for the r-only nuclides. Uncertainties are represented by vertical bars.}
\label{fig_sos_sr}
\end{figure}
%-------------------------------------------------------------------------- 

A rough representation of the splitting of the SoS abundances above iron into s-, r- and p-nuclides is displayed
in Fig.~\ref{fig_sos_srp}. In its details, the procedure of decomposition is not as obvious as
it might be thought from the very definition of the different types of nuclides, and is to
some extent dependent on the models for the synthesis of the heavy nuclides 
(see Sect.~\ref{prod_srp_general}). These models
predict in particular that the stable nuclides located on the neutron-rich/neutron-deficient side of
the valley of nuclear stability are produced, to a first good approximation, 
by the r-/p-process only, and are referred to as ``r-only'' and ``p-only'' nuclides (see Figs.~\ref{fig_nucsite} and \ref{fig_nucastro}). The
situation is more intricate for the nuclides situated at the bottom of the valley of nuclear
stability. Some of them are produced solely by the s-process, the typical flow of which runs
very close to the bottom of the valley, as illustrated in Fig.~\ref{fig_nucsite}.
They are referred to as ``s-only'' nuclides, and are encountered only when a stable 
r-isobar exists, which ``shields'' the s-isobar from the r-process.  As a result, only even-$Z$ heavy elements possess an s-only isotope. The others can be produced in various relative amounts essentially by the s- and r-processes. These nuclides of mixed origins are called ``sr'' (note that some nuclides may have a small p-component on top of the s-one. They are referred to as ``sp'' nuclides). 

The details of the s- and r-contributions to the SoS depend on the astrophysical models and on the adopted nuclear physics. They are discussed at length in \cite{Arnould07,Arnould03,Goriely99}. The results are summarized in Fig.~\ref{fig_sos_sr}. As stressed in particular in \cite{Arnould07}, the splitting procedure between s-, r- and p-nuclides faces more or less large uncertainties that are unfortunately quite systematically swept under the rug in the literature. The question of the uncertainties is clearly of large enough importance to deserve a careful study, especially in view of the sometimes very detailed and far-reaching considerations that have the SoS s-r splitting as an essential starting point.
 
Figure~\ref{fig_sos_srp} shows that about half of the heavy nuclei in the solar material come
from the s-process, and the other half  from the r-process, whereas the p-process is
responsible for the production of about 0.01 to 0.001 of the abundances of the s-  and
r-isobars. It also appears that some elements have their abundances
dominated by an s- or r-nuclide. They are naturally referred to as s- or r-elements.
For example, Ba is traditionally considered as an s-element, whereas Eu is an
r-element (Fig.~\ref{fig_sos_srp}).  
Clearly, p-elements do not exist. If this naming remains valid in other locations than the 
SoS, which may well not be the case (Section~\ref{obs_stars}), stellar spectroscopy can provide information on the s- or r-  (but not the p-) abundances outside of the SoS. In fact, a wealth of observations demonstrate significant
departures from the  SoS s- or r-element abundances. Such departures exist in the SoS itself in the form of
``isotopic anomalies'' (Section~\ref{obs_sos}), or in stars with different ages, galactic
locations, or evolutionary stages (Section~\ref{obs_stars}). The SoS abundances and their s-, r-
 and p-process contributions  do not have any ``universal'' character.
        
%**********************
\subsection{An overview of the s-process}
\label{prod_s}
%************************************

%*****************************************
\subsubsection{Some generalities}
\label{prod_s_general}
%***********************************

The ``slow'' neutron-capture (s-)process relies on the assumption that 
pre-existing (``seed'') nuclei are exposed to a flux of neutrons that
is weak enough for allowing a $\beta$-unstable nucleus produced by 
an (n,$\gamma$) reaction to decay promptly, except perhaps for some relatively long-lived nuclides which may
instead capture a neutron. The nuclides where $\beta^-$-decays and neutron captures may compete are referred to as ``branching points'' in the s-process path, which constitute an important ingredient  of the s-process ({\it e.g.} \cite{Kaeppeler89} for a review). The handling of these branches necessitates the knowledge of the stellar 
$\beta$-decay rates, as well as of the probabilities of neutron captures by unstable nuclei close to the line of stability.

A so-called ``canonical'' model of the s-process has been initially developed by \cite{Clayton61}, and has received some refinements over the years ({\it e.g.} \cite{Kaeppeler89}). It needs not be reviewed here in detail. Let us just mention that it does not rely on any specific astrophysical model, its outcome depending solely on the nuclear input. This model views the SoS s-abundance pattern as originating from a superposition of two exponential distributions of the time-integrated neutron exposure,
$\tau=\int_0^t N_{\rm n} v_T {\rm d}t$ (where $v_T$ is the most probable relative neutron-nucleus velocity at temperature $T$ and $N_n$ the neutron density). These distributions are traditionally held responsible for the so-called weak ($70 \lsimeq A \lsimeq 90$) and main ($A \gsimeq 90$) components of the s-process. A third
exponential distribution is sometimes added in order to account for the $204 < A \le 209$ s-nuclides.
Through an adequate fitting of the parameters of the $\tau$-distributions, the superposition of the two or three resulting abundance components provides constraints on the temperatures, neutron concentrations  and operation timescales appropriate to the development of the s-process(es) responsible for the successful reproduction of the abundance distribution of the s-only nuclides in the SoS. Note in particular that the abundance peaks observed at $A = 138$ or 208 in the SoS s-process abundance distribution (see Fig.~\ref{fig_sos_srp}) relate directly to the smallness of the neutron capture cross sections at the magic  $N = 82$ and 126 neutron numbers. Small neutron capture cross sections are indeed responsible for a staggering in the s-process flow, and consequently of an accumulation of material at these locations.

Despite the success of the canonical model in fitting the solar s-nuclide distribution, some of its basic assumptions deserve questioning. This concerns in particular a presumed exponential form for the distribution of the neutron exposures $\tau$, which has been introduced by \cite{Clayton61} in view of their mathematical ease in abundance calculations. In addition, the canonical model makes it difficult in the s-nuclide abundance 
predictions to evaluate uncertainties of nuclear or observational nature. As a result, the concomitant uncertainties in the solar s-abundances are traditionally not evaluated. The shortcomings of the canonical model are cured to a large extent by the so-called ``multi-event''  s-process model described in detail in \cite{Goriely99}. Quite importantly, this model does not make any {\it a priori} hypothesis on the distribution of neutron exposures, but still does not rely on any specific stellar model and on the precise neutron producing mechanisms.

%***************************************
\subsubsection{Specific s-process sites}
\label{prod_s_site}
%***************************************

As demonstrated by many observations, the surfaces of a variety of low-mass ($M \lsimeq
3 M_\odot$) AGB stars (Cat.3 stars; Section~\ref{nucleo_contributors}) are enriched  with certain s-elements, implying that they have been synthesized in-situ and transported to the stellar surface by convective-type motions referred to as ``dredge'-up''. The identification of the element Tc at the surface of certain types of AGB stars (see \cite{Merrill52} for this discovery and {\it e.g.} \cite{Karakas14} for more recent references) is the most spectacular proof of the recent operation of this mechanism, as Tc has only unstable isotopes with lifetimes much shorter than typical evolutionary timescales, as already noted in Section~\ref{introduction}. These observations also make it plausible that AGB stars eject part of their synthesized s-nuclides into the ISM 
through their winds, and thus contribute to the galactic, and in particular SoS, s-nuclide enrichment. 
As dust particles are known from astronomical observations to form
in their ejecta, AGB stars could also be the source of certain anomalous
meteoritic grains containing various heavy elements with an s-process isotopic pattern.  

As reviewed in great detail by \cite{Karakas14}, the s-process in AGB stars is thought to occur in their He-burning shell surrounding a nuclearly inert C-O core, either during recurrent and short convective episodes (``thermal pulses''), or in between these pulses. A rather large diversity of s-nuclide abundance distributions are predicted to be produced. A fraction of the synthesized s-nuclides (along with other He-burning products)
could then be dredged-up to the surface shortly after each pulse. 

At least in low-mass AGB stars, it is generally considered that the necessary neutrons for the development of
the s-process are mainly provided by \reac{13}{C}{\alpha}{n}{16}{O}, which can operate at temperatures around 
$(1 \sim 1.5)\times 10^8$ K. The efficiency of this mechanism is predicted to be the highest in stars with metallicities [Fe/H] lower than solar ([Fe/H] $\ll 0$). Note that the $Z=0$ (Pop III star) case does not follow the rule, since seeds for neutron captures (esp. Fe) are absent. An efficient production of s-process elements starting from the neutron captures on the C-Ne isotopes may still be possible, as shown  in \cite{Goriely01c}.

An s-process could also develop in  intermediate-mass ($3 - 8 M_\odot$) AGB stars, as reviewed by \cite{Karakas14}, as well as in massive super-AGB stars \cite{Doherty17}. It differs, however, from the one developing in lower-mass stars, the neutrons being expected to be produced mainly at the base of the thermally pulsing He-burning shell by \reac{22}{Ne}{\alpha}{n}{25}{Mg} following the classical sequence
\chem{14}{N}($\alpha,\gamma$)\chem{18}{F}($\beta$)\chem{18}{O}($\alpha,\gamma$)\chem{22}{Ne} that burns
the \chem{14}{N} produced in the preceding CNO cycle. This comes from the fact that higher temperatures are required to burn \chem{22}{Ne} than \chem{14}{N}, temperatures increasing with increasing stellar mass and decreasing metallicities.

The bottomline of the above-mentioned s-process studies is that low- and intermediate-mass stars are classically considering as the preferred sites for the main s-process. However, it cannot be stressed strongly enough that, in their details, the conclusions drawn above suffer from more or less large uncertainties. The most severe ones certainly come from the predictions of the s-processing in between thermal pulses. The neutron production in these locations indeed depends sensitively on the mechanism of proton ingestion into underlying He-rich layers in amounts and at temperatures that allows the operation of the \reac{12}{C}{p}{\gamma}{13}{N}($\beta^+$)\chem{13}{C}($\alpha$,n)\chem{16}{\rm O}, while the production of \chem{14}{N} by \reac{13}{C}{p}{\gamma}{14}{N} is inefficient enough to avoid the hold-up of neutrons by the \chem{14}{N} neutron poison (see {\it e.g.} \cite{Karakas14,Goriely00} for details about this process). This mixing mechanism is very difficult to model in common one-dimensional models. Limited multi-dimensional efforts have been undertaken that stress the shortcomings of one-dimensional approaches ({\it e.g.} \cite{Herwig05,Herwig06}). Other effects could contribute to modifications in the classical one-dimensional simulations, including magnetic fields and rotation. These multidimensional effects could also impact the production of neutrons by \reac{22}{Ne}{\alpha}{n}{25}{Mg} during thermal pulses.
 
The efficiency of the transport (dredge-up) of the produced s-nuclides to the AGB star surfaces 
(as demanded by the observations) is also far from being well understood. 
This subject is in fact a matter of debate, different recent models or adopted parametrisations leading in
some cases to quite different conclusions concerning the characteristics, and
even the very existence, of this transport episode ({\it e.g.} \cite{Karakas14,Goriely18}). The dredge-up efficiency, as well the efficiency of the neutron production are in fact often treated as adjustable parameters in order to explain the large variety of observational data.

Nuclear uncertainties add their share to the overall s-process modeling. In general, these uncertainties have, however, a more limited impact than those of astrophysical origin. This results from the fact that the s-process path lies close to the valley of nuclear stability, so that most neutron capture cross sections have been measured at supra-thermal energies of astrophysical relevance \cite{Dillmann14}. The situation is somewhat less pleasing at branching points involving unstable nuclides with generally unmeasured neutron capture cross sections and more or less uncertain astrophysical $\beta^-$-decay rates \cite{Takahashi87,Goriely99}. In addition, uncertainties affect also the rate of the neutron producing reaction \reac{22}{Ne}{\alpha}{n}{25}{Mg} \cite{Longland12} and, to a lesser extent, of \reac{13}{C}{\alpha}{n}{16}{O} \cite{Xu13}.

Massive (Cats.~3 and 4) stars, and more specifically their He-burning cores and, to some extent, their C-burning shells, are also predicted
to be s-nuclide producers through the operation of the
\reac{22}{Ne}{\alpha}{n}{25}{Mg}. 
This neutron source can indeed be active in
these locations that are hotter than the
 He shell of AGB
stars. In addition, \chem{22}{Ne} burning can also be activated in the carbon burning shell of massive stars. Many calculations performed in the framework of realistic stellar
models come to the classical conclusion that this site is responsible for a
 substantial production
of the $A \lsimeq 100$ s-nuclides (weak s-process), and can in particular account for the SoS abundances of these species. In short, it appears in particular that massive stars can be responsible for a substantial production of heavy s-nuclides at low enough metallicities. In addition, rotation is responsible for a large increase in the amount of produced light s-nuclides, or even of heavy ones for large enough rotation rates and low-enough metallicities. Uncertainties in the modeling of massive stars (concerning in particular the treatment of convection, the distribution of rotational velocities in the stellar interiors, and mass loss rates) transpire in the  s-process predictions (see \cite{Prantzos18} for references).

Very low-metallicity (Pop III or very low-$Z$ Pop II) massive stars constitute a special case of s-process. They may provide an additional s-process site through the possible ingestion of protons into He-rich burning zones already encountered in low- or intermediate-mass stars. The neutrons produced through \reac{13}{C}{\alpha}{n}{16}{O} could explain the observed abundances of the light, and at least medium-mass (up to about Ba) s-nuclides in certain massive carbon-enhanced metal-poor (CEMP) stars. Unusually, \reac{16}{O}{n}{\gamma}{17}{O} could act in those stars as a neutron poison instead of the classical \chem{14}{N}. However, its efficiency could be drastically limited by the ensuing \reac{17}{O}{\alpha}{n}{20}{Ne} reaction \cite{Banerjee18}. Rotation is considered by some authors as being of key importance in the merging of protons and He, and consequently in the efficiency of the associated s-process \cite{Choplin17}. As in the AGB case, the modeling of the proton injection into He-rich layers is very difficult. On the nuclear physics side, additional uncertainties come in particular from the rate of $\alpha$-captures by \chem{17}{O}.

%*********************************************
\subsection{An overview of the i-process}
\label{prod_i}
%*********************************************

The s- and r-processes introduced very early in the development of the theory of nucleosynthesis have to be considered as the end members of a whole class of neutron capture mechanisms. Supported by some observations that were difficult to reconcile solely with a combination of the s- and r-processes, a process referred to nowadays as an intermediate or i-process has been put forth, with neutron concentrations in the approximate 10$^{12}$ to 10$^{16}$ neutrons/cm$^3$ range. The mechanism envisaged to be responsible for this production was the ingestion of protons in He-rich layers, leading to the operation of the \reac{13}{C}{\alpha}{n}{16}{O}. This is analogous to the mechanism already considered to be active in thermally pulsing AGB stars (Section~\ref{nucleo_contributors}), but the higher neutron concentrations are expected to result from the very low metallicity of the considered stars and the activation of \reac{13}{C}{\alpha}{n}{16}{O} in convective regions at higher temperatures. The conditions under which the neutron concentrations could lead to levels of neutron productions characteristic of the i-process have been analyzed in detail by \cite{Jorissen89} on grounds of a parametric approach allowing to scrutinize the effects of temperature, density, initial \chem{12}{C}/p, 
\chem{12}{C}/\chem{4}{He}, \chem{12}{C}/\chem{16}{O} and \chem{18}{O}/\chem{22}{Ne} abundance ratios, and proton ingestion timescales. This study has been complemented over the years by a number of numerical simulations of proton ingestion during core He flash in very low-metallicity low-mass stars ({\it e.g.} \cite{Campbell10}), during the thermal pulse phase of massive AGB (super-AGB) stars of very low metallicity ({\it e.g.} \cite{Jones16}), during the post-AGB phase (``final thermal pulse'') ({\it e.g.} \cite{Herwig11}), during rapid accretion of H-rich material on WDs ({\it e.g.} \cite{Denissenkov17,Denissenkov18,Cote18}), or during shell He burning in massive very low-metallicity Pop II or Pop III stars ({\it e.g.} \cite{Clarkson18}).
  
 It is suggested that i-process heavy element patterns observed in peculiar stars, like the most iron-poor star currently known, SMSS J031300.36-670839.3, several CEMP-r/s (carbon enhanced metal-poor with simultaneous presence of s elements and Eu) stars, as well as the Sakurai's object (V4334 Sgr). Some isotopic anomalous grains of extrasolar origin found in certain meteorites (Section~\ref{obs_sos}) have also been suggested to have a possible origin in an i-type process developing during the final thermal pulse experienced by a post-AGB star like the Sakurai's object \cite{Arnould93}. This suggestion has received support from the analysis of certain carbide grains showing \chem{32}{S} enrichment \cite{Fujiya14} and of rare graphite grains \cite{Jadhav14}.

The one-dimensional modeling of the i-process faces severe uncertainties, as it is common in all the situations involving mixing of H- and He-rich stellar layers. Clearly, three-dimensional magneto-hydrodynamical simulations are required in order to put the nucleosynthesis conclusions on safer grounds. Only some preliminary results have been obtained in these complex modelings ({\it e.g.} \cite{Clarkson18,Herwig11}). Nuclear physics uncertainties add their share to the problem. This mainly arises from the fact that typical i-process neutron capture paths are located more or less deeply into the neutron-rich side of the valley of nuclear stability (Fig.~\ref{fig_nucastro}), imposing the necessity to predict neutron capture cross sections that cannot be measured in addition to $\beta^-$-rates that are uncertain in astrophysical conditions, due in particular to the contribution from excited states (Section~\ref{decay_beta}) (see {\it e.g.} \cite{Denissenkov16} for some discussion of nuclear uncertainties).
 
%***  Sect.6.4  **********************************  
\subsection{An overview of the $\alpha$-process and the r-process}
\label{prod_r}
%**************************************************
\subsubsection{Some generalities}
\label{prod_r_general}
%************************************************

As mentioned above, a specific mechanism, referred to as the r-process, is called
for in order to explain the origin of the stable neutron-rich nuclides heavier than iron and the actinides observed in the SoS and in a variety of stars. It relies on a chain of captures of neutrons whose concentrations 
are by far higher than in the s-process, and in fact exceed a concentration of 10$^{20}$ neutrons/cm$^3$.
As this requirement is clearly impossible to meet 
in quiescent stellar evolutionary 
phases, deep supernova layers in the vicinity of a forming NS 
residue have been quite naturally envisioned as a possible 
r-process site. 
However, for decades, it has not been possible to substantiate this 
connection on grounds of detailed supernova models.

It is out of the question to present in the following an exhaustive and rapidly growing list of the myriad of observational and theoretical astrophysics works devoted to the r-process. The reader is referred in particular to recent reviews \cite{Cowan19,Kajino19}, to the very many publications of the "R-Process Alliance" (individual references can be found under this consortium name), and to the observational data collected in the framework of the AMBRE project \cite{Guiglion18}. Nuclear physics experimental and theoretical investigations of high relevance to the r-process are already touched upon in various subsections of Section~\ref{nucdata}.

The contribution of the r-process to the SoS is discussed in Section~\ref{prod_srp_general}. From the observations of galactic as well as of extragalactic metal-poor stars, there is mounting evidence that all analyzed such stars show some contamination of trans-iron nuclides. These nuclides are generally considered to be of r-process origin, which points toward an early galactic/extragalactic contamination by this process that pre-dates the s-process, even if some early operation of the s-process cannot be totally excluded. 

Another important piece of information about the r-process comes from the observation of significant star-to-star abundance variations, especially in the $Z \leq 50$ range. Similarities also exist between the patterns of abundances of heavy neutron-capture elements in the range from around Ce ($Z = 58$) to Os ($Z = 76$) in some r-process-rich metal-poor stars, like CS 22892-052 \cite{Sneden96},  and the SoS one. This convergence has led in the literature to a recurrent claim that the r-process is ``universal''. This conclusion has been put in doubt by \cite{Goriely97} who have proposed that the abundance convergence in the $58 \lsimeq Z \lsimeq 76$ range can be the natural signature of nuclear properties, and does not tell much about the astrophysics of the r-process.  Recent observations also indicate that star to star variations in the r-process content of metal-poor globular clusters may be a common, although not ubiquitous, phenomenon \cite{Roederer10,Roederer11}. Stars such as HD 88609 or HD 122563 have been found to be significantly deficient in their heavy elements \cite{Honda07}. CS 22892052 and HD 122563 are now interpreted as two extreme cases representative of a continuous range of r-process nucleosynthesis patterns \cite{Roederer10}.
 
Of special interest is the detection of Th and U in some stars, and in particular in the most actinide-rich star 2MASS J09544277+5246414 presently known \cite{Holmbeck18}. Finally, as already noticed in Section~\ref{obs_cr}, GCRs could also have a r-nuclide component, including actinides \cite{Tatischeff18}.

On the theoretical side, the early works of \cite{Burbidge57} and \cite{Seeger65} have proposed the simplest and most widely used form of the r-process scenario, referred to as the canonical r-process 
 model (CAR). It assumes that pre-existing material made of pure \chem{56}{Fe} is 
driven by neutron captures into a location of the neutron-rich region determined by the
 neutron supplies and by the highly temperature-sensitive reverse photodisintegrations. Details on the other simplifying assumptions adopted in CAR can be found in {\it e.g.} \cite{Arnould07}.
Although this model does not make reference to any specific astrophysics scenario, 
but builds on nuclear properties only, it has 
 greatly helped paving the way to more sophisticated approaches of the r-process. It has in particular pointed out that events characterized by high temperatures ($T \gsimeq 10^9$ K), high neutron concentrations ($N_n \gsimeq 10^{20}$ neutrons/cm$^3$) and short neutron irradiation timescales (less than a couple of seconds) are needed to account for the SoS r-nuclide abundances. These conditions imply nuclear flows very deep inside the neutron-rich side of the valley of nuclear stability (neutron separation energies lower than 2-3~MeV) during neutron irradiation. These requirements have been associated early on to stellar explosions of the supernova type.
 
 A parametric approach of the r-process extending CAR through the relaxation of some of its simplifying assumptions has been developed by \cite{Goriely96,Bouquelle95}. In its formulation, it is identical to
 the model used for the s-process in the decomposition  between the SoS s- and
 r-abundances (Section~\ref{prod_srp_general}). It demonstrates that suitable superpositions of events of the CAR type can successfully reproduce the SoS r-abundance distribution.\footnote
%**************footnote**************
{In order to avoid unnecessary confusions, it may be worth noting that the multiple events considered in the model may not only refer to {\it numerous stars}, but also to a {\it suite of thermodynamic conditions} that can likely be encountered {\it in a single object}. In addressing those aspects statistically,  
a ``multi-event'' approach is likely more realistic than the consideration of a few  events.}
%************************************ 

In associating the r-process with supernova explosions, several attempts to go beyond the above-mentioned
models have been made by taking into account some evolution of the 
characteristics of the sites of the r-process during its development (see {\it e.g.} \cite{Arnould07} for references to early works). These models, referred to as ``dynamical'',  
by and large rely on a generic CCSN scenario in which a so-called ``hot bubble'' region is created by neutrino heating at the periphery of a nascent NS (Section~\ref{nucleo_stars}). The hot bubble consists of a rapidly
expanding matter, referred to as a neutrino (transonic) ``wind'' or (subsonic) ``breeze'' with high entropy and with a more or less significant neutron excess ({\it e.g.} \cite{Arnould07} for some details and references). At short times, or at temperatures of about $(10 \sim 7) \times 10^9$ K, the bubble composition is
determined by NSE favoring $\alpha$-particles and some neutrons.
As the temperature decreases further along with the expansion, the
$\alpha$-particles recombine to form heavier nuclides, starting
with the $\alpha\alpha$n$\rightarrow$$^9$Be($\alpha$,n)$^{12}$C 
reaction. It is followed by a complex sequence of $\alpha$-particle and
nucleon captures [especially of the ($\alpha$,n), (p,n) and (n,$\gamma$) type], and of the inverse transformations, 
synthesizing heavy nuclei even beyond Fe. This is the so-called ``$\alpha$-process'' first investigated by
\cite{Delano71} and analyzed in detail by \cite{Meyer98}. Note that this $\alpha$-process must not be confused 
with the same term used by \cite{Burbidge57}, which is now referred to as ``Ne burning'' (see {\it e.g.} \cite{Arnould07} for a short review and references). It is worth noting that the pivotal role of r-process seed devoted to \chem{56}{Fe} in the CAR-type r-process approaches disappears in the models involving the 
$\alpha$-process trigger.

As reviewed in {\it e.g.} \cite{Arnould07}, the development of the r-process under wind/breeze conditions depends on a proper combination of {\it i)} the expansion timescales that may be influenced by the deposited neutrino energy and on the mass loss rate, these two characteristics having in their turn an impact on the time variation of the wind/breeze temperature, {\it ii)} the ``radiation entropy'' (defined by Eq. 37 of \cite{Arnould07}) and {\it iii)} the electron fraction (net electron number per baryon) $Y_e$. 
 
 Multi-event simulations within the neutrino-driven wind approach have been performed in \cite{Goriely16c} under some simplifying assumptions. These calculations can successfully reproduce the overall SoS r-abundance distributions, but it remains of course to be seen if the favorable conditions leading to such an agreement can be met in realistic explosion simulations.

In addition to the astrophysics questions, the $\alpha$- and r-processes 
raise major nuclear physics problems. They indeed involve a wealth of
very neutron-rich $A \gsimeq 12$  nuclei whose properties are more or less poorly known experimentally, 
a large fraction of them even remaining to be produced in the
laboratory. In particular, masses, $\beta$-decay and neutron
capture rates have in most instances to be estimated theoretically (see various subsections of Section~\ref{nucdata}). The impact of the related uncertainties on abundance predictions has been
discussed in many places. Neutron-induced fission, as well as $\beta$-delayed 
neutron emission or fission may also bring more uncertainties in the
calculated r-process yields, and even in the very energetics of the process.
These additional difficulties relate directly to our poor knowledge of
fission barriers of very neutron-rich actinides (Section~\ref{decay_fission}).   

Early in the development of the theory of nucleosynthesis, an alternative to the high-temperature
 r-process CAR model has been proposed 
\cite{Tsuruta65}. It relies on the fact that very high densities (say
 $\rho > 10^{10}$ g/cm$^{3}$) can lead material deep inside the neutron-rich side of the
 valley of nuclear stability as a result of the operation of endothermic free-electron
 captures (Section~\ref{decay_continuum}), this so-called ``neutronization'' of the 
material being possible even at the $T = 0$ limit. The astrophysical plausibility of this
 scenario in accounting for the production of the r-nuclides has long been questioned, and
 has remained largely unexplored until the study of the composition of the outer and inner 
crusts of NSs \cite{Baym71} and of the decompression of cold neutronised matter 
resulting from tidal effects of a BH on a NS companion \cite{Lattimer77}. 
The decompression of cold NS matter has recently been studied further very actively with the use of a variety of EoSs (Section~\ref{eos}). In view of the renewed interest for a high-density r-process, 
a simple steady-flow model has been developed without reference to a specific astrophysical scenario \cite{Arnould07}.  
 
%**************************************
\subsubsection{Specific r-process sites: CCSNe}
\label{prod_r_site}
%**************************************

Since the early 1990's, a rich nucleosynthesis has been expected to take place in the neutrino-heated ejecta of CCSNe in the close vicinity of the NS. This mechanism has been discussed in so many places recently ({\it e.g.} \cite{Arnould07,Cowan19,Kajino19}) that an additional review is superfluous. 

Despite the initial excitement from CCSN models, it has now become clear that the neutrino-driven wind model faces severe theoretical difficulties in meeting the required conditions for a successful explosion, and a fortiori for a robust r-process. In fact, only the r-nuclides roughly in the $A \lsimeq 100$ range are predicted to be possibly produced in more or less parametrized supernova models. 
Various physical effects have been added to the standard CCSN models in order to examine their possible impacts on the explosive fate and on the production of r-nuclides. Among these, let us cite fluid instabilities, rotation and magnetic fields (so-called "magneto-rotational" effects) possibly leading to jet developments, acoustic waves, and accretion of material on rotating WDs ({\it e.g.} \cite{Janka17,Arnould07} for a review). A quite large variety of more or less parametrized and qualitative models have been proposed on such grounds, leading to the conclusion that the r-process might be aided by some at least of these additional effects, with the possibility of ejection into the ISM.  

Fully self-consistent, non-rotating non-magnetized two-dimensional hydrodynamic ECSNe (Cat. 4 stars; Section~\ref{nucleo_stars}) simulations predict the ejection of neutron-rich nuclides between the Fe group and at least up to, and even possibly slightly beyond the $N = 50$ region, including Pd, Ag and Cd \cite{Wanajo11,Wanajo18}. The production of light neutron-rich material is reduced further in the case of the CCSN explosion of more massive stars in the range of 10 to 27 M$_\odot$ (Cat. 5 stars) explored by \cite{Wanajo18}. No well-developed r-process is identified in the simulated explosions. It has to be stressed, however, that the nucleosynthesis has not been followed up to the freeze-out of the nuclear processing. In addition, firmer conclusions have to await three-dimensional simulations, as artifact effects can be induced by two-dimensional models.

Collapsars are also proposed as possible sources of r-nuclides. They are a rare subclass of CCSNe of rapidly rotating and highly magnetized massive stars, and are generally considered to be at the origin of observed long 
$\gamma$-ray bursts. The production of r-nuclides in these events may be associated to jets predicted to accompany the explosion, or to the accretion disk forming around a newly born central BH.  

The production and ejection of r-nuclides in self-consistent successful three-dimensional models of magneto-hydrodynamic jets has started to be studied, and shows some promising results ({\it e.g.} \cite{Haveli18,Winteler12,Nishimura15}). However, the jet models face some problems. One of them concerns the considered, possibly unrealistic, large pre-collapse magnetic fields of the order of $10^{13}$~G, as well as the adopted high rotation rates. The level of alignment of the magnetic field  on the rotation axis is another source of embarrassment. In addition, the resulting magneto-rotational jets may suffer from instabilities that could very significantly reduce the efficiency of the production of r-nuclides, particularly the heavy ones beyond the second r-process peak \cite{Mosta18}. Further work is clearly needed in order to put the conclusions regarding the r-process efficiency of the jet model on safer grounds.

The production of r-nuclides has been investigated through three-dimensional general-relativistic magneto-hydrodynamical simulations of late-time neutrino-cooled accretion disks around a newly-formed BH created as a result of the collapsar process \cite{Siegel19b}. BH accretion rates ranging between 0.003 and 0.1 M$_\odot{\rm s}^{-1}$ needed to explain the observed energetics and timescales of long $\gamma$-ray bursts give rise to neutron-rich outflows provided charged-current weak interactions are properly taken into account. Such a scenario is very similar to the wind ejecta from BH-torus accretion disks resulting from NS mergers 
(Section~\ref{prod_r_mergers}), except that the total accreted mass is found to be typically one order of magnitude larger in collapsars than in mergers. The result is that masses up to 1 M$_\odot$ may be ejected from collapsars, this large amount of material ejected per event compensating for their low frequency (collapsars are expected to be even rarer than NS mergers with  a few time $10^{-5}$ event per year and per galaxy). Another advantage is the fast evolution of massive stars with respect to the time lag necessary for a binary NS system to merge. This allows to overcome some of the observational challenges met by r-process models based uniquely on the NS merger scenario \cite{Siegel19b,Siegel19}. This mechanism avoids namely the uncertainties associated with the jet model discussed above, and predicts larger ejected mass than the other proposed scenarios. 
 
 %----------------------------------------------------------------
 \begin{figure}
\centering
\includegraphics[scale=0.25,angle=0]{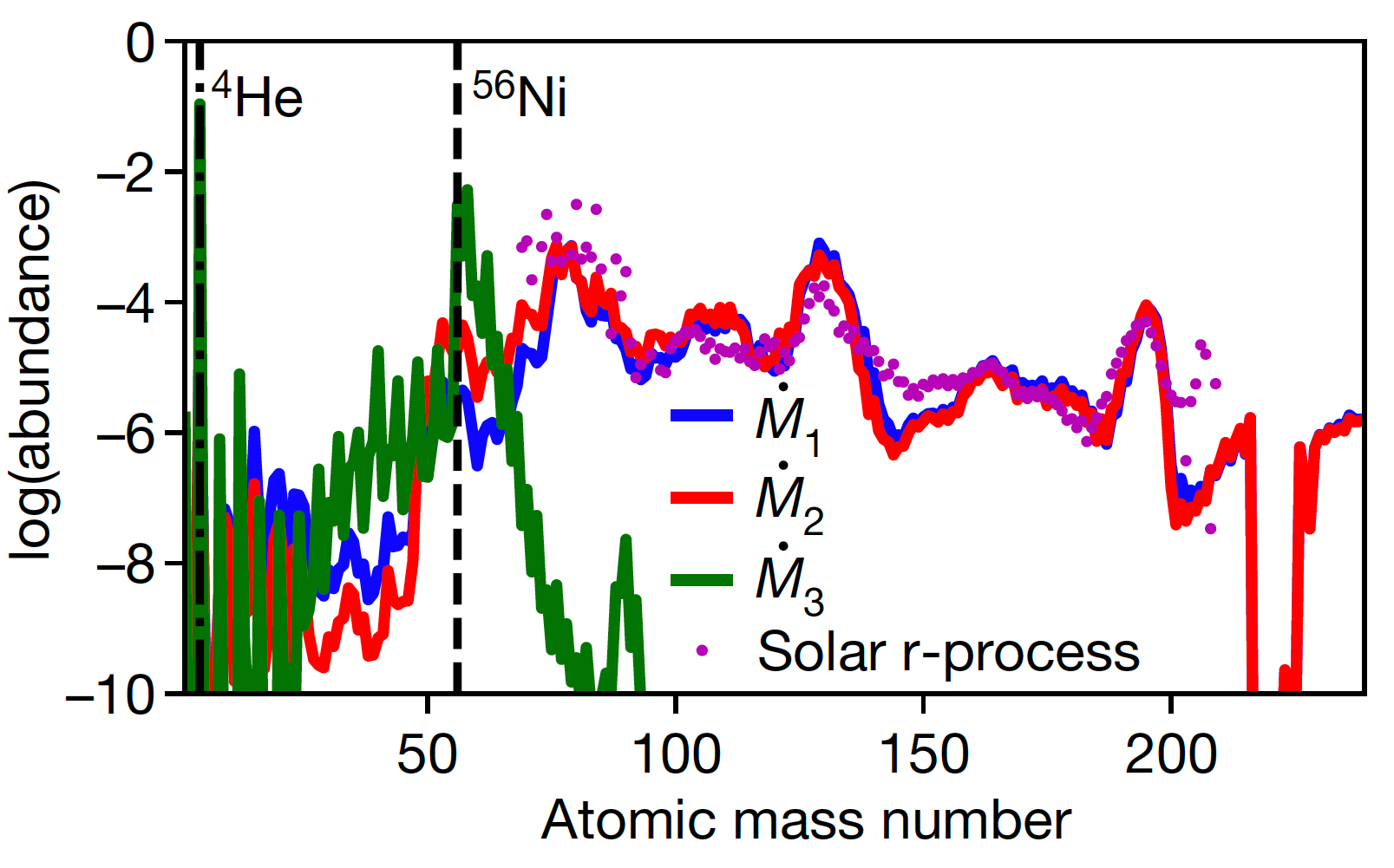}
\caption{Abundance distributions of nuclei synthesized in the disk outflows at three different accretion stages identified by their accretion rates: $\dot{M}_1 > \dot{M}_2$  with  $\dot{M}_2$ close to about 
$10^{-2} M_\odot$ s$^{-1}$ correspond to the $\gamma$-ray burst phase. A lower rate $\dot{M}_3$ characterizes a later post-burst phase at which the outflow is dominated by \chem{56}{Ni}. See Ref.~\cite{Siegel19b} for more details.
\newline
{\it Source}: Reprinted figure with permission from \cite{Siegel19b}.}
\label{fig_collapsar} 
\end{figure}
%-----------------------------------------------------------------------------

%*******************************************
\subsubsection{Specific r-process sites: NS mergers}
\label{prod_r_mergers}
%*******************************************

Since the early 2000s, special attention has been paid to NS mergers  as r-process sites following the confirmation by  hydrodynamic simulations that a non-negligible amount of matter could be ejected from the system.
Newtonian \cite{Ruffert01, Janka99, Korobkin12}, 
conformally flat general relativistic \cite{Oechslin07,Goriely11b,Just15},  as well as fully relativistic \cite{Kyutoku13,Hotokezaka13a, Wanajo14, Foucart14} hydrodynamical simulations of NS-NS and NS-BH mergers with microphysical EoSs have demonstrated that typically some
$10^{-3}\,M_\odot$ up to more than 0.1\,$M_\odot$ can become gravitationally unbound on roughly
dynamical timescales due to shock acceleration and tidal stripping. Also the relic object (a hot, transiently stable hypermassive NS \cite{Baumgarte00} followed by a stable supermassive NS, or a BH-torus system), can lose mass through outflows driven by a variety of mechanisms  \cite{Ruffert99, Dessart09, Fernandez13, Siegel14, Metzger14, Perego14}. 
 
\begin{figure}
\centering
\includegraphics[scale=0.50]{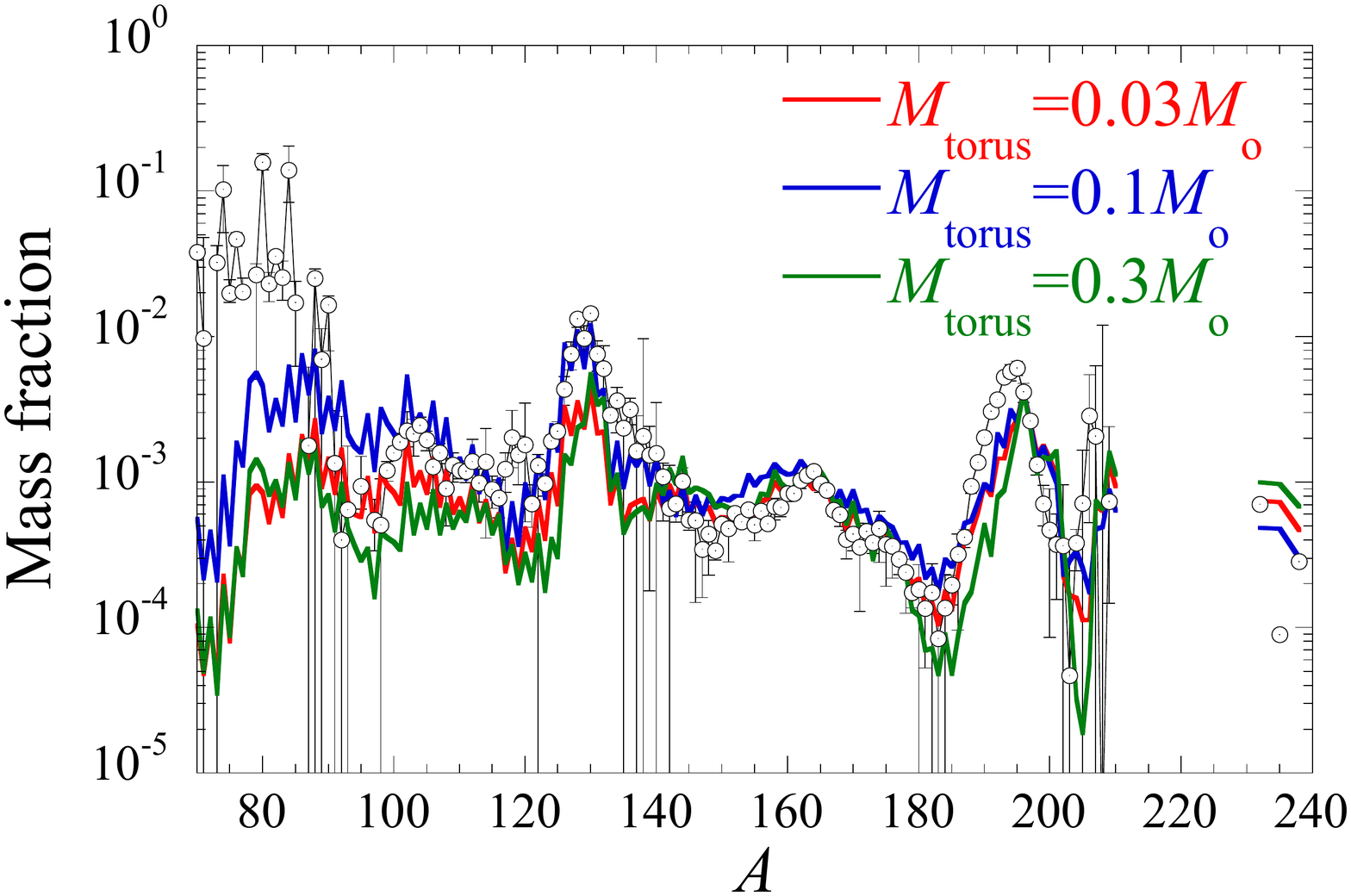}
\vskip -0.7cm
\caption{Abundances versus atomic mass number $A$ in the prompt plus post-merger phase material  of three systems with torus masses $M_{\rm torus}=0.03, 0.1$ and $0.3~M_\odot$. All distributions are normalized to the SoS $A=196$ $r$-abundance (dotted circles). See \cite{Just15} for more details.}
\label{fig_dyn+disk} 
\end{figure}

Simulations of growing sophistication have confirmed that the ejecta from NS mergers are viable strong r-process sites up to the third abundance peak and the actinides  \cite{Just15,Goriely11b,Wanajo14,Roberts11}. The r-nuclide enrichment is predicted to originate from both the dynamical (prompt) material expelled during the NS-NS or NS-BH merger phase and from the outflows generated during the post-merger remnant evolution of the relic BH-torus system. The resulting abundance distributions are found to reproduce very well the SoS distribution (Fig.~\ref{fig_dyn+disk}), as well as various elemental distributions observed in low-metallicity stars \cite{Cowan19}. The physical characteristics and resulting precise r-nuclide composition of the total ejecta show variations depending on the adopted physics (relativistic versus Newtonian mechanics, nuclear input), as exemplified by  \cite{Goriely11b,Wanajo14,Roberts11,Bauswein13,Rosswog14}. However, during the dynamical phase of the merging scenario, the number of free neutrons per seed nucleus can reach so high values (typically few hundreds) that heavy fissioning nuclei are produced, in which case the composition of the ejecta is rather insensitive to the details of the initial abundances and the astrophysical conditions, namely the mass ratio of the two NSs, the quantity of matter ejected, and the EoS \cite{Bauswein13}. This robustness is compatible with the  SoS-like abundance pattern of the rare-earth elements observed in metal-poor stars \cite{Cowan19}. This supports the possible production of these elements by fission recycling in NS merger ejecta. In addition, the ejected mass of $r$-process material, combined with the predicted astrophysical event rate (around 10\,My$^{-1}$ in the Milky Way~\cite{Dominik12}) can account for the majority of $r$-material in our Galaxy~\cite{Bauswein14}. A further piece of evidence that NS mergers are r-nuclide producers indeed comes from the very important 2017 gravitational-wave and electromagnetic observation of the kilonova GW170817  (Section~\ref{prod_r_gcm}) \cite{Abbott17}. 
 
%*****************************************************************
%******************************************************************

\begin{figure}[h]
\centering
\includegraphics[scale=0.5]{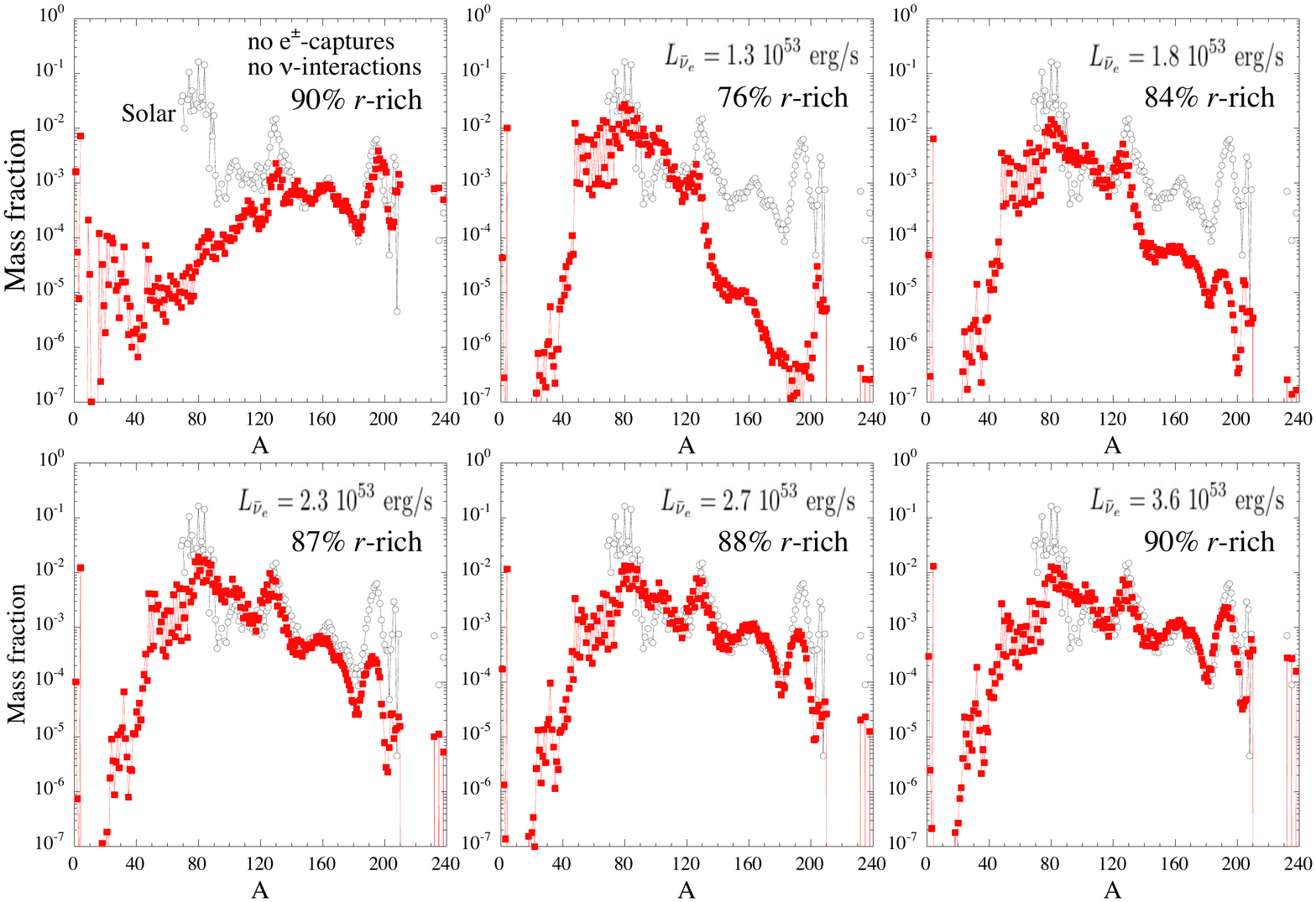}
\vskip -0.5cm
\caption{Impact of the neutrino interactions on the abundance distribution in the ejecta of a 1.35--1.35~$M_\odot$ NS-NS merger model during the dynamical phase obtained in the framework of a parametrized model \cite{Goriely15a}. The upper left panel is obtained without any weak interaction on free nucleons. The others are derived from five different values of the antineutrino luminosity $L_{\bar\nu_e}$, together with $L_{\nu_e}=0.6~10^{53}$~erg/s, $\langle E_{\nu_e}\rangle = 12$~MeV and $\langle E_{\bar\nu_e}\rangle = 16$~MeV. Each panel is labelled with  the mass fraction of $A>69$ $r$-nuclei in the ejected material.}
\label{fig_nut} 
\end{figure}

Despite this success of the NS merger models, many uncertainties remain. One major question concerns the precise role of the interaction of neutrinos with the surrounding matter. The 3D modeling of the energy-dependent neutrino transport including neutrino-flavor oscillations in relativistic very dense and highly magnetized largely aspherical and rapidly time-varying environments raises problems of the highest complexity. Relativistic NS-NS merger simulations \cite{Wanajo14,Sekiguchi15,Ardevol19} including the effect of $\nu_e$ ($\bar{\nu}_e$) captures on neutrons (protons) demonstrate that neutrino reactions can significantly affect the merger dynamics, with direct consequences in particular on the amount of synthesized low-mass ($A<140$) r-nuclides produced in the dynamical ejecta along with the heavier species. However, the validity of the adopted time independence of the
(anti)neutrino luminosities has to be questioned, and its influence on the detailed r-nuclide production has to be answered. Asymmetries in the neutrino fluxes between the polar and equatorial directions has also to be modeled. In addition, it remains to be clarified if the effects of neutrino interactions depend or not on the adopted high-density EoSs and on the precise characteristics of the binary systems that lead to a delayed collapse of the merger remnant. The construction of parametric models allowing a broad examination of the above-mentioned questions may be profitable \cite{Goriely15a}. Figure~\ref{fig_nut} provides an illustration of the results obtained with the help of such a model.
 
%**********************************************************
%*********************************************************

%*****************************************************
\subsubsection{Evolution of the r-nuclide galactic content}
\label{prod_r_gcm}
%***************************************************

A variety of models  with different levels of sophistication have been proposed in order to follow the evolution of the galactic content of r-nuclides. One of the most recent simulations \cite{Haynes19} adopts a chemodynamical approach (coupling hydrodynamics and ``chemistry'') that leads to the conclusion that NS mergers alone cannot account for the observational data for the r-process nuclide Eu, and that the contribution from collapsars is required. A study based on the kilonova that accompanied GW170817 leads to the same conclusion \cite{Siegel19b}. The possible dominance of a collapsar contribution to the r-nuclide content of globular clusters (old galactic sub-systems) is also advocated by \cite{Zevin19}, even if the role of NS mergers cannot be fully excluded. In contrast, another model \cite{Ojima17} assumes the formation of the halo of our Galaxy through the clustering of sub-halos with varying star formation histories, and in which the occurrence of NS mergers is computed stochastically. It concludes that NS mergers alone are capable of accounting for the observations. The predominance of neutron-star mergers in the production of the SoS r-nuclides is also advocated recently by \cite{Bartos19}. 
 
This diversity of conclusions just illustrates that many more or less severe uncertainties remain in the modeling of the r-process from various sites, and additionally from the highly complex models for the evolution of the nuclidic content of our Galaxy. Further efforts are clearly needed in order to put all the astrophysical and nuclear physics aspects of the r-process on reasonably safe grounds. 

%***************************************************
\subsection{An overview of the p-process}
\label{prod_p}
%**************************************************
%*************************************
\subsubsection{Some generalities}
\label{pro_p_general}
%**************************************************

 The SoS content of the neutron-deficient isotopes, referred to as p-nuclides, of the elements heavier than Fe has been discussed in Section~\ref{prod_srp_general} and is displayed in Fig.~\ref{fig_sos_srp}. Roughly speaking, the abundance distribution of the p-nuclides follows the general trend of the s-nuclides, their abundances decreasing with increasing mass number, but at a level of about 1 percent to 1 per-mil of the abundances of the s-isobars. If this same abundance level applies everywhere in the Galaxy, it is quite clear that p-nuclide abundances cannot be observed outside the SoS. Some isotopic anomalies observed in the bulk material, inclusions and various presolar grains of meteorites (see Section~\ref{obs_sos}) are attributable to the p-process. Well documented cases concern anomalies that originate from the decay of the radionuclides \chem{92}{Nb} and \chem{146}{Sm}, a very special and still puzzling type of Xenon referred to as Xe-HL characterized by correlated excesses of the r- ({\it i.e.} Heavy) and p- ({\it i.e.} Light) isotopes of Xe, and various isotopic anomalies in the element Mo of special p-process significance in view of the relatively high SoS abundances of its two p-isotopes \chem{92}{Mo} and \chem{94}{Mo} (see Fig.~\ref{fig_sos_srp} and \cite{Arnould03} for a review).   

The main p-nuclide production results from the destruction of pre-existing s- or r-nuclides by different combinations of ($\gamma$,n), ($\gamma$,p) or ($\gamma,\alpha$) reactions, and more marginally of
(p,$\gamma$) captures (see the review by \cite{Arnould03}, and Fig.~\ref{fig_nucastro}). Some $\beta$-decays, electron captures or (n,$\gamma$) reactions can possibly complete the nuclear flow. These reactions may lead directly to the production of a p-nuclide. In most cases, however, they are synthesized through an unstable progenitor that transforms into
the stable p-isobar by a (chain of) $\beta$-decay(s). Non-thermal reactions in the ISM have been proposed on occasion to explain the synthesis of specific p-nuclides. As discussed further in Sect.~\ref{prod_p_moru}, this concerns in particular the very rare nuclide \chem{180}{Ta^m} ({\it e.g.} \cite{Kusakabe18}). On the other hand, production of the rare p-nuclides \chem{138}{La} and \chem{180}{Ta} by neutral- and charged-current neutrino interactions with the material of different supernova shells (referred to as the $\nu$-process) has also been envisioned (see \cite{Langanke19} for an overview).
 
 %****************************************************
 \subsubsection{Specific p-process sites}
 \label{prod_p_site}
 %**************************************************
 
We just limit ourselves here to a rather brief discussion of possible p-process sites. The reader is referred to \cite{Arnould03} for a more extended review. The beginning of the 1970's has marked a breakthrough in the identification of plausible p-process sites with the suggestion that the neutron-deficient nuclides can be produced thermonuclearly in the deep O-Ne-rich layers of massive stars either in their pre-supernova or supernova phases \cite{Arnould71,Arnould76}. 
\vskip0.2truecm

{\bf Thermonuclear production of p-nuclides in pre-SN conditions and their survival at the SN stage}. Apart from uncertainties concerning radiative nucleon and $\alpha$-particle capture rates and their inverse photodisintegrations, many more or less severe astrophysics problems hamper the reliability of the predicted
pre-SN p-nuclide production. Among others let us cite the production of the
s-nuclides (Section~\ref{prod_s}) which serve as seeds for the p-nuclides. Another serious and
most difficult question concerns the proper description of the mechanisms of transport of
material (``convection'') in the stellar layers of relevance to the pre-SN p-process. This
description affects drastically the abundance predictions, which is an unfortunate situation
in view of the very unsatisfactory state of affairs concerning the convection algorithms
adopted in classical one-dimensional model stars. The question of the survival has been diversely answered, depending of the stellar models used. Some lead to the conclusion of a substantial destruction because of the high p-nuclide concentration in deep layers that become too hot at the time of the explosion. In contrast some survival of the pre-SN p-nuclides is predicted in some calculations, its extent depending quite drastically, however, upon the stellar mass ({\it e.g.} \cite{Rauscher02}). It has to be stressed that all these predictions rely on simple models for one-dimensional non-rotating non-magnetized stars. Quite clearly, more reliable p-nuclide abundance distributions in pre-SN enriched layers and their survival to explosion have to await multidimensional modelings which are the focus of recent efforts showing severe limitations of one-dimensional simulations of the pre-SN O-Ne-rich layers \cite{Mocak18}. 
 
Another site where pre-explosively produced p-nuclides could survive a supernova explosion has been discussed by \cite{Rayet93} (see also \cite{Arnould03}). It concerns the PCSN explosion of very massive stars of Cat.~5. In this scenario, the identity of the p-nuclides that have the best chance to survive the explosion depend on the stellar mass and on the precise outcome of the explosion, particularly regarding the mass of the (BH) remnant (see below). 

 %----------------------------------------------------------------
\begin{figure}[tb] 
\center{\includegraphics[scale=0.3,angle=-90]{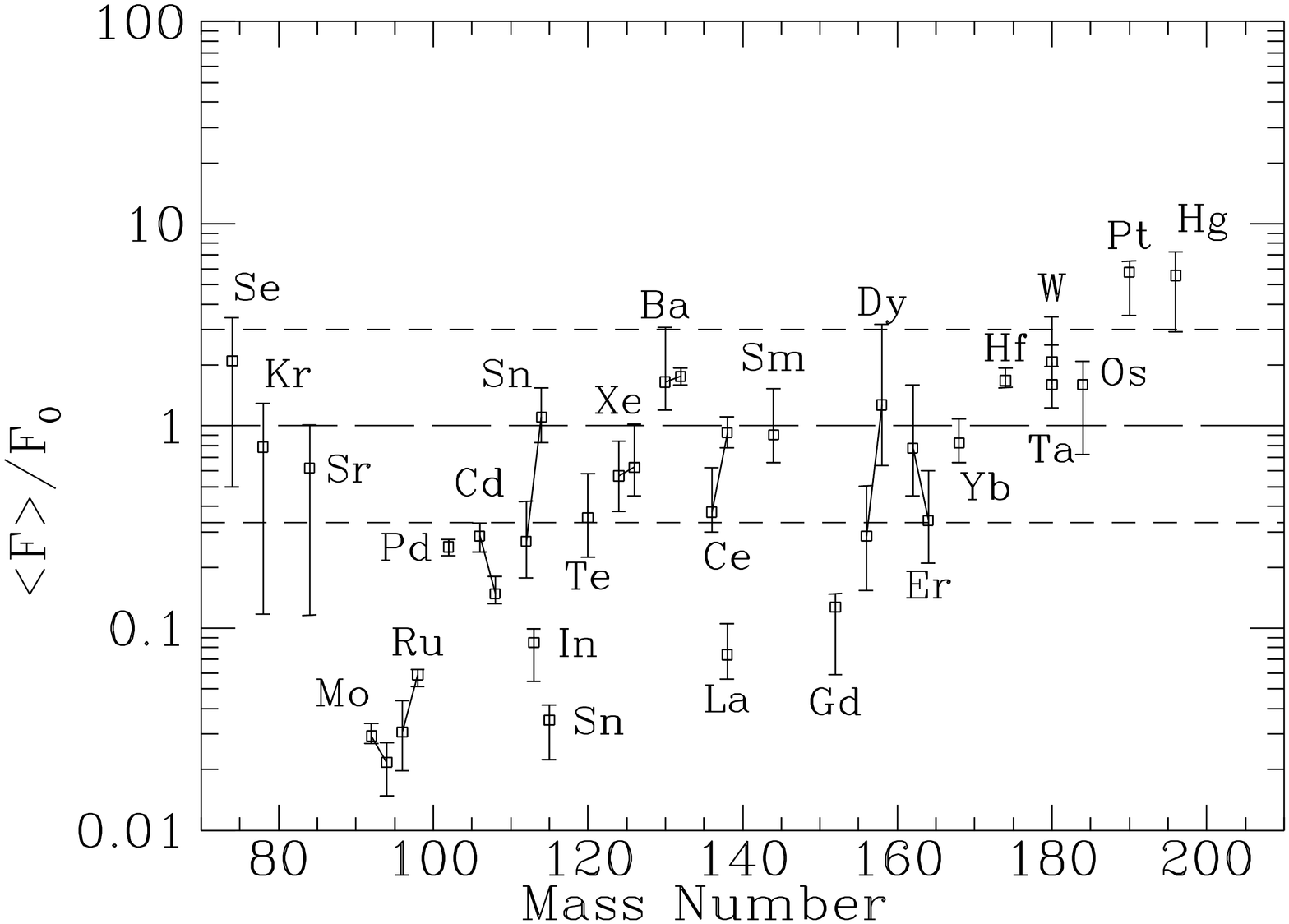}}
\caption{Normalized p-nuclide overproductions $\langle F_i \rangle (M)/F_0(M)$
predicted for the individual CCSN explosions of solar metallicity stars with masses in the range
from 13 to $25 M_\odot$. The mean
overproduction factor $\langle F_i \rangle (M)$ for a star with mass $M$ is defined as the
total mass of the p-nuclide $i$ in the O-Ne producing layers (PPLs) divided by the corresponding mass if the
PPLs had a solar composition. The normalizing factor $F_0(M)$ is the mean overproduction
factor averaged over the 35 p-nuclides, and equals 100. For the sake of simplicity, the ranges obtained for the different considered masses are just schematized by
vertical bars. Open squares indicate the values of the normalized overproductions obtained by
integration over the initial mass function proposed by \cite{Kroupa93} (see \cite{Rayet95} for details). Solid lines join different p-isotopes of
the same element. The displayed \chem{180}{Ta} data correspond to the sum of the \chem{180}{Ta^g} and \chem{180}{Ta^m} abundances. Roughly speaking, \chem{180}{Ta^m} (the only Ta form to be present in the SoS) represents about half of the total \chem{180}{Ta} calculated yields (see Section~\ref{prod_p_moru}).  \newline
{\it Source}: Reprinted figure with permission from \cite{Rayet95}.}
\label{fig_pCCSN} 
\end{figure}
%*********************************************

\vskip0.2truecm

{\bf Thermonuclear production of the p-nuclides in CCSNe}. The most actively pursued avenue of research on the p-process concerns without any doubt CCSN explosions. This topic has been extensively reviewed by \cite{Arnould03}, and we limit ourselves here to a rather brief overview of the main aspects of this scenario. The first full p-process network calculations involving nucleon and $\alpha$-particle captures, as well as their reverse photodisintegrations in the framework of one-dimensional non-rotating models for stars in the approximate 13 to 15 $\leq M \leq 25 M_\odot$ mass range with solar metallicity date back to \cite{Rayet95}. Typical p-nuclide yields derived from such calculations are displayed in Fig.~\ref{fig_pCCSN}. All the normalized overproductions as defined there would be equal to unity if the derived abundance patterns were solar.    

From Fig.~\ref{fig_pCCSN}, it is concluded that about $60\%$ of the  produced p-nuclei fit the SoS composition
within a factor of 3. Some discrepancies are also apparent. They concern In particular the famed underproduction of Mo and Ru p-isotopes. This problem is discussed in
Section~\ref{prod_p_moru}. The nuclides \chem{113}{In}, \chem{115}{Sn}, \chem{138}{La} and
\chem{152}{Gd} are also underproduced, just as in the early
calculations of \cite{Arnould71,Arnould76}. No other clear
source has been identified yet for \chem{113}{In}. In contrast, some \chem{114}{Sn}, \chem{115}{Sn} and a substantial amount of the solar \chem{152}{Gd} and of \chem{180}{W} may owe their origin to the s-process \cite{Goriely99}. This process also
contributes to \chem{164}{Er}. From the level of p-process production of this nuclide shown
in Fig.~\ref{fig_pCCSN}, \chem{164}{Er} may well have a combined s- and p-origin
without having to fear any insuperable overproduction with respect to the neighboring
p-nuclides. The rare odd-odd nuclide \chem{180}{Ta^m} is predicted to emerge naturally and in quantities consistent with its solar abundance. This interesting case is discussed further in
Section~\ref{prod_p_moru}.  

In order for CCSNe to be potentially important contributors to the galactic p-nuclides, it
remains to make sure that they do not produce too much of other species, and in particular
oxygen, the abundance of which is classically attributed to these explosions. This question
has been examined in some detail by \cite{Rayet95}, who conclude that the p-nuclides are
globally underproduced in solar metallicity 25 $M_\odot$ stars by a factor of about
$4 \pm 2$ relative to oxygen when all the abundances are normalized to the bulk solar
values. The mentioned uncertainty relates to changes in the
\reac{12}{C}{\alpha}{\gamma}{16}{O} rate and in the explosion energy. The problem of the oxygen overproduction may be eased in lower
mass CCSN explosions, while it is worsened with decreasing metallicity \cite{Rayet95,Arnould03}. It may
thus be that the p-process enrichment of the Galaxy has been slower than the oxygen
enrichment. There is at present no observational test of this prediction. It may also be
eased due to a higher than nominal production of s-nuclides, which might also help solving
the Mo-Ru underproduction problem made apparent by Fig.~\ref{fig_pCCSN}. 
At this point, it must also be kept in mind that many other sources of uncertainties affect the p-nuclide predicted abundances. They are not only of nuclear physics nature (see \cite{Arnould03} for a review), but probably, and most importantly, reside in the stellar models. As usual, and broadly speaking, a variety of multi-dimensional phenomena, including rotation, are quite likely to enter the picture. Their impact on the CCSN p-process cannot be reliably evaluated at this point. 

Massive stars of the Wolf-Rayet (WR) type also explode by the CCSN mechanisms after episodes of very strong mass losses leading to H-depleted SNIb/c supernovae (see Section~\ref{nucleo_stars}). No specific p-process calculations have been conducted for exploding WR stars. In view of the fact that several of the exploding He star remnants left over in the course of the WR evolution have masses in the range of the He stars associated to CCSNe, one may consider as a first-order approximation that the p-nuclide yields from a SNIb/c explosion of a He star remnant of given mass are not drastically different from those from a CCSN explosion of a He
star of the same mass. However, the equality of
the WR He star remnant and of the He star masses does not guarantee the identity of the
structure and chemistry of the pre-supernova configuration, and consequently of the
explosion and of its yields. This is especially so considering the high sensitivity of the p-process flows to the
exact thermodynamic history of p-process producing layers (PPLs) in massive stars.  

The possible role of other massive stars has also been explored. This concerns more
specifically those exploding as PCSNe. The
only detailed p-process calculations based on this unique scenario have been conducted by
\cite{Rayet93} on grounds of detailed models for a solar metallicity mass losing star with an initial mass of 140 $M_\odot$ evolving from the main sequence all the way to the PCSN. The amount of ejected  p-nuclides comprises a pre-explosion component and an explosion contribution, and depends in particular on the amount of material locked-up in a BH forming during the explosion, the mass of which cannot be properly predicted from the considered explosion model. Clearly, the PCSN simulations still face many uncertainties, and multi-dimensional effects have to be included before putting the models on reasonably safe grounds.  
 
\vskip0.2truecm

{\bf Thermonuclear production of the p-nuclides in SNIa}. As discussed in Section~\ref{nucleo_stars}, a ``canonical'' view of the SNIa phenomenon calls for the total disruption of the CO-WD
member of a binary star which accretes material from its companion at a suitable rate to the
point where its mass reaches a value close to the Chandrasekhar limit of about
$\approx 1.4 M_\odot$. The WD with this mass is
subjected to a hydrodynamical burning triggered by the thermonuclear runaway of carbon, and a
supernova explosion ensues. The modeling of this runaway faces enormous difficulties, the very nature of the burning ({\it e.g.} deflagration, detonation, or detonation followed by deflagration, referred to as ``delayed detonation'') and its details are far from being well established yet. Multi-dimensional simulations are indispensable to capture the very intricate physics of the phenomenon.

\begin{figure}[tb]
\center{\includegraphics[scale=0.35,angle=0]{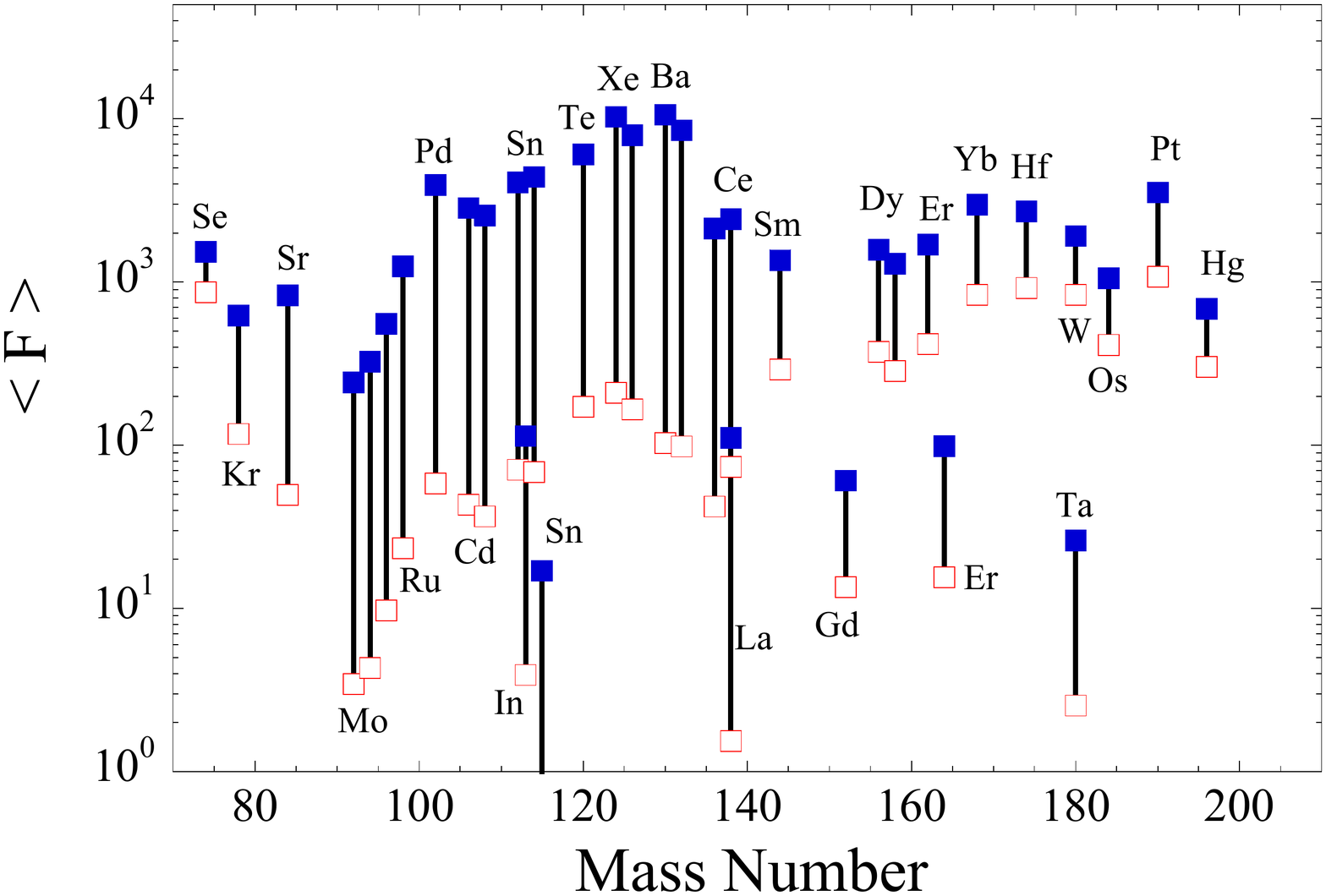}}
\vskip-0.4truecm
\caption{Abundances of the p-nuclides in the SN Ia deflagration model W7 of \cite{Nomoto84}.
The overproduction factors refer to the $0.12 M_\odot$ of the PPLs of the model. They are calculated with solar s-seeds (red solid
squares), or with a seed distribution which is representative of the s-process in solar metallicity AGB stars \cite{Goriely00} (blue open squares). In this case, the s-nuclides with mass numbers
$A \gsimeq 90$ have abundances that can exceed the solar ones by factors as high as $10^3$ to
$10^4$. This s-process enhancement translates directly into an increase in the p-nuclide
overproduction factors.}
\label{fig_p_sn1a}
\end{figure}

The p-process nucleosynthesis possibly accompanying the deflagration or delayed detonation
regimes has been mainly studied in 1D simulations. The corresponding yield
predictions are consequently highly uncertain (see \cite{Arnould03} for a review). Calculated abundances are presented in Fig.~\ref{fig_p_sn1a} for a deflagration model with two representative sets of s-process seeds, one being just solar, the other one being adopted from the production of s-nuclides in thermally-pulsing AGB stars. Similar calculations have also been conducted for a delayed detonation model in a 1D approximation \cite{Arnould03}, and have been extended to 2D models \cite{Travaglio15}.  

By and large, the inspection of Fig.~\ref{fig_p_sn1a} reveals that the calculated deflagration SNIa p-yields are quite similar to those resulting from massive star explosions (Fig.~\ref{fig_pCCSN}) or from delayed detonation scenarios. In particular, the same set of underproduced species
\chem{92,94}{Mo}, \chem{96,98}{Ru}, \chem{113}{In}, \chem{115}{Sn}, \chem{138}{La} is predicted, with the remarkable exception of \chem{180}{Ta^m} which is found to be underproduced in the SNIa case. It is also noticed that the p-nuclide yields, and in particular the level of Mo and Ru p-isotopes underproduction, depends sensitively on s-seed abundance distribution (see Section~\ref{prod_s_site} for some additional discussion of these results).

The predicted SNIa p-nuclide yields suffer from large uncertainties affecting the adopted explosion models as well as the s-seed distributions, detailed information on the composition of the material that is pre-explosively transferred to the WD being missing. Additional uncertainties are of nuclear physics nature, as discussed by \cite{Arnould03}. In such conditions, any attempt to derive conclusions on the contribution of SNIa to the solar p-nuclides \cite{Travaglio15,Travaglio18} is premature and highly unreliable.

%--------------------------------------------------------------------
%---------------------------------------------------------------------
%
\vskip0.2truecm

{\bf Thermonuclear production  of the p-nuclides in sub-Chandrasekhar WD explosions}. CO WDs with sub-Chandrasekhar mass $M < 1.4 M_\odot$ (CO-SCWD) may explode, leading possibly to peculiar sub-luminous SNIa events as a result of the accumulation of He-rich material from a companion. As reviewed by {\it e.g.} \cite{Jose15} (see also \cite{Polin18}), this explosion might result from a detonation wave moving outward to the surface combined with a compression wave propagating to the center where it induces carbon burning. Many uncertainties remain regarding this scenario which has to be put on safer grounds by truly three dimensional simulations. 

\begin{figure}
\center{\includegraphics[scale=0.35,angle=0]{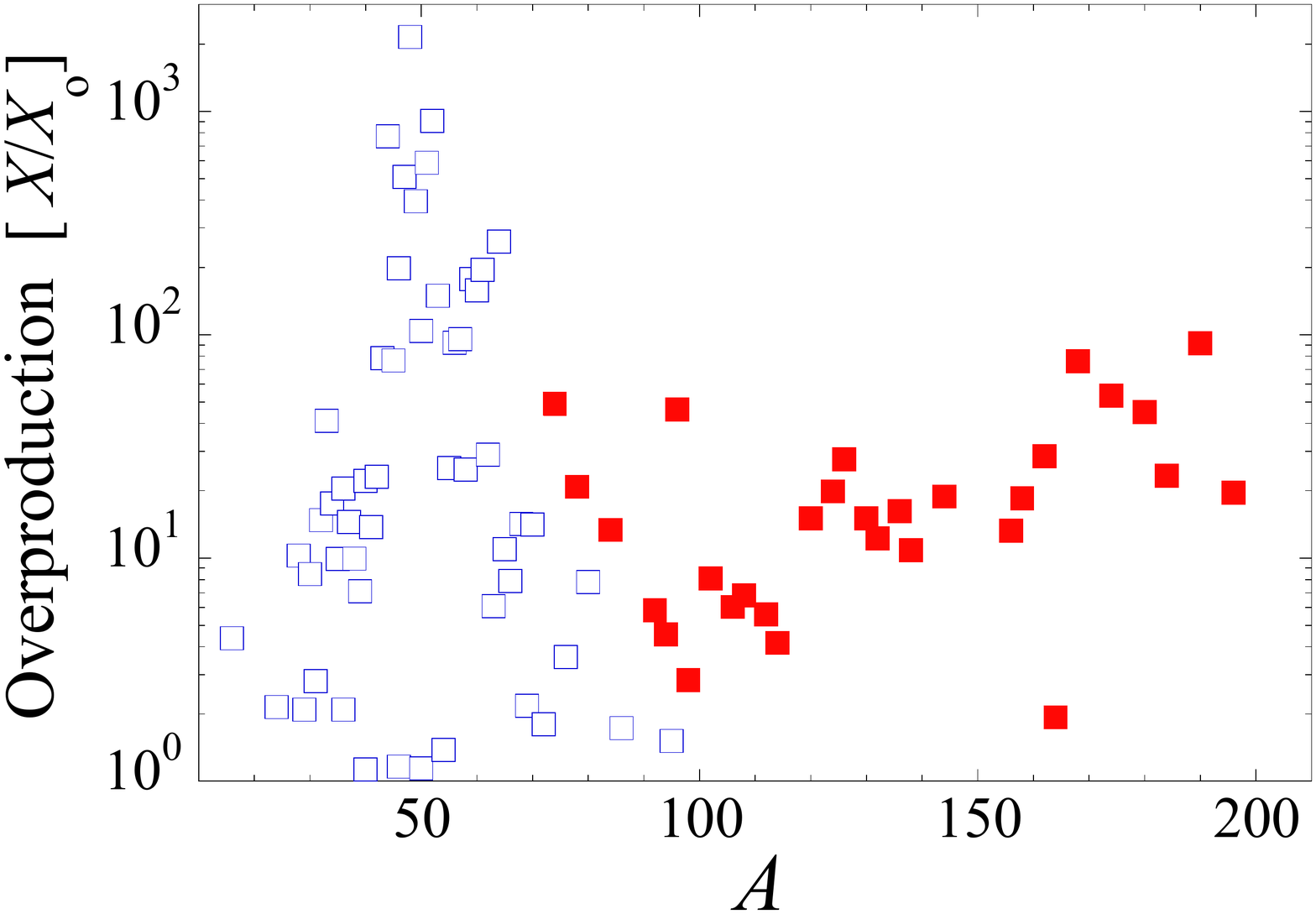}}
\vskip-0.3truecm
\caption{Composition of the ejecta following the 3D simulated explosion of a CO-SCWD. Full symbols denote the p-nuclides. More details can be found in \cite{Goriely05}.}
\label{fig_psubchan}
\end{figure}

The possibility for a p-process to develop during the explosion of CO-SCWDs has been explored by \cite{Goriely02} based on one-dimensional explosion models. A special flow pattern is identified, coined ``proton-poor neutron-boosted rp-process'' or ``pn-process". At early times, an r-process type of flow
develops on the neutron-rich side of the valley of nuclear stability. At somewhat later
times, the material is pushed back to the neutron-deficient side rather
close to the valley of $\beta$-stability. As time passes, an
$\alpha$p-process (Section~\ref{burn_rp}) and a pn-process develop. More details can be found in \cite{Goriely02}. The 1D calculations have been extended to 3D simulations by \cite{Goriely05}. 
Figure~\ref{fig_psubchan} illustrates the results. Roughly speaking, the p-abundance distribution in the disrupted core is very similar to the one found in traditional SNIa (Fig.~\ref{fig_p_sn1a}), with a clear deficient production of \chem{92,94}{Mo} and \chem{96,98}{Ru}, \chem{138}{La} and \chem{180}{Ta} with respect to the other p-nuclides, as well as an overall underproduction of the p-nuclides by a factor of about 10 with respect to \chem{56}{Ni}. The p-nuclide underabundance problems of both the core and the envelope can be cured by considering an enhanced initial abundance of s-nuclides. The astrophysical plausibility of this enhancement remains, however, to be scrutinized. 
 
%--------------------------------------------------------------------
%---------------------------------------------------------------------
 
%******************************************
\subsubsection{A short comment on p-process troublemakers}
\label{prod_p_moru}
%***************************************

\noindent {\it The case of the Mo and Ru p-isotopes}. Much has been written and many speculations have been published on the possible origin of the thermonuclear underproduction of the light isotopes of Mo and Ru in supernova explosions reported above. From a closer look at the problem, it appears that uncertainties in the nuclear reactions involved in the p-process flow leading to Mo and Ru from heavy seeds is unlikely to be the main source of the difficulty. The problem lies more probably in the adopted seed abundances. An exploratory work by \cite{Costa00} shows that indeed the underproduction of the Mo and Ru p-isotopes would be substantially reduced if the neutron producing reaction \reac{22}{Ne}{\alpha}{n}{25}{Mg} largely responsible for the production of the p-process s-seeds in massive stars
were increased by a factor of 5-10 over the adopted value proposed by NACRE \cite{Angulo99}, which leaves the reaction rate well within the large range of the uncertainties reported there. Subsequent studies of this reaction have proposed largely reduced uncertainties ({\it e.g.} \cite{Longland12}). However, the proposed rates are based on various assumptions that remain to be validated \cite{Aliotta16}. It may be fair to state that the final say is still awaiting for improved experimental information.

The possibility that the $\nu$p process can cure the thermonuclear underproduction of the p-isotopes of Mo and Ru has also been examined \cite{Frohlich06}, but still strongly depends on the remaining poorly determined electron fraction of the neutrino-driven wind. A recent calculation concludes negatively \cite{Zhang18}.

\vskip0.2truecm

\noindent {\it The case of \chem{138}{La} and \chem{180}{Ta}}. In spite of the fact that the odd-odd p-nuclides \chem{138}{La} and isomeric  
\chem{180}{Ta^m} are among the rarest SoS species (Fig.~\ref{fig_SoS_A}), their nucleosynthesis origin raises problems. 
The \chem{138}{La} production results from a subtle balance between its
main production by \reac{139}{La}{\gamma}{n}{138}{La} and its main destruction by
\reac{138}{La}{\gamma}{n}{137}{La}. Recent measurements constraining the radiative capture cross sections and photodisintegration rates of relevance for the \chem{138}{La} synthesis \cite{Kheswa15} confirm the 
\chem{138}{La} underproduction in CCSN explosions.

In view of these difficulties, other routes of production have been explored. In particular, the possible role of the 
$\nu$-process has been examined \cite{Goriely01b}. It is concluded that this process may have some efficiency indeed in the \chem{138}{La} synthesis. This conclusion finds some confirmation in the subsequent calculations by \cite{Langanke19,Ko19}, which, however, point to a variety of uncertainties.  A non-thermal GCR production as also been invoked  \cite{Kusakabe18}, but is found not to be efficient enough to explain the SoS \chem{138}{La}.
 
The odd-odd nuclide \chem{180}{Ta} has the remarkable property of having a short-lived
($T_{1/2} = 8.15$ h) $J^\pi = 1^+$ ground state (\chem{180}{Ta^g}) and a very long-lived
($T_{1/2} > 1.2 \times 10^{15}$ y) $J^\pi = 9^-$ isomeric state (\chem{180}{Ta^m}). As a
result, only \chem{180}{Ta^m} is present in the SoS and is the sole naturally occurring isomer in nature with the minute \chem{180}{Ta^m}/\chem{181}{Ta} $\approx 10^{-4}$ abundance. 

The study of the \chem{180}{Ta^m} production in stellar interiors raises the specific problem of 
thermalization of \chem{180}{Ta^g} and \chem{180}{Ta^m}, the solution of which is necessary for evaluating the effective decay lifetime, and thus the probability of survival, of \chem{180}{Ta} in those conditions. The thermalization has been investigated many times over the years (see {\it e.g.} \cite{Arnould07} for some details and references). The equilibration can be
obtained only through a series of mediating levels located at excitation energies $E_x > 1$ MeV for temperatures in excess of about  $T_{9,\rm crit} \gsimeq 0.4$. In these conditions, the stellar effective lifetime is close to the laboratory ground state value. In contrast,  for temperatures $T_9 \lsimeq 0.15$, the population of the excited states is too low for them to play an efficient mediating role, so that the stellar effective lifetime recovers its laboratory value. At intermediate temperatures $0.15 \lsimeq T_9 \lsimeq 0.4$, the non-Maxwellian
populations of the levels involved in the evaluation of the decay rates have to be calculated from kinetic
equations involving the various transition probabilities of relevance \cite{Mohr07}. As estimated by \cite{Arnould03} for the CCSN case, \chem{180}{Ta} equilibrated at temperatures in
excess of $T_{9,\rm crit}$ can survive to a substantial extent (about 40\%) after the freezing of its
production or destruction reactions at temperatures most likely well in excess of this critical temperature. This is comparable to the situation obtained under the extreme assumption that \chem{180}{Ta^g} and \chem{180}{Ta^m} never equilibrate, in which case about half of the total produced \chem{180}{Ta} would be ejected in the ISM. In contrast, the survival of any pre-explosively produced Ta is likely to be insignificant.

The predicted p-process \chem{180}{Ta^m} yields suffer from some nuclear uncertainties on
top of the question of the thermalization discussed above (see \cite{Arnould03} for details). Among them are the ones concerning
the rates of the \reac{181}{Ta}{\gamma}{n}{180}{Ta} and \reac{180}{Ta}{\gamma}{n}{179}{Ta}
photodisintegrations which most directly influence the
\chem{180}{Ta^m} production and destruction. The latter cannot be estimated from experimental
data using the reciprocity theorem, the radiative neutron capture
cross section on the unstable \chem{179}{Ta} being unmeasured. In contrast , the \reac{180}{Ta^m}{n}{\gamma}{181}{Ta} cross section has been measured \cite{Wisshak04}, and the reciprocity theorem can be applied to the reverse photodisintegration. The \reac{181}{Ta}{\gamma}{n}{180}{Ta} cross
section has even been measured directly, the relative production by this reaction of the
ground and isomeric states of \chem{180}{Ta} being obtained experimentally as well \cite{Malatji19}. Of course, the stellar photodisintegration rate of \chem{181}{Ta} can differ significantly from the laboratory value due to the contribution of its thermally excited states. Even so, such measurements strongly constrain the nuclear input, and in particular the $E1$-strength function, and thus help reducing the uncertainties affecting the stellar rate predictions.
 
Figure~\ref{fig_pCCSN} shows that \chem{180}{Ta} can be produced in quantities comparable to those of the neighboring p-nuclides in CCSNe (see also \cite{Rauscher02}), while the situation is predicted to be less rosy in the SNIa or SCWD cases. The ability of CCSNe to be efficient \chem{180}{Ta} producers is confirmed by recent calculations based on the experimental results reported above. Of course, uncertainties in the modeling of all the p-process sites add to the specific uncertainties facing the \chem{180}{Ta} production.

It is noticeable that the possibility of production of \chem{180}{Ta} in supernovae has quite often not been recognized in its appropriate right in the literature. This is why other mechanisms have been invoked, and even sometimes claimed as dominant.  In particular, the s-process production of
\chem{180}{Ta^m} during the thermal pulse phase of low- or intermediate-mass AGB stars has been envisioned. A recent study \cite{Malatji19} based on detailed AGB star models and on experimentally constrained neutron capture rates complemented by HF evaluations leads to the expectation that the contribution of the s-process to the SoS \chem{180}{Ta} is marginal.
 
The $\nu$-process has also been envisioned as a possible \chem{180}{Ta} producer. Its contribution has been re-evaluated recently \cite{Malatji19,Langanke19}, from which it is concluded that this process can indeed be an efficient SoS \chem{180}{Ta} contributor. However, large uncertainties are claimed to be due to the prediction of CCSN neutrino spectra. Lower average $\nu$-energies are found in recent supernova simulations than in earlier ones, which would imply a noticeable reduction in the production of \chem{180}{Ta} as well as of other rare nuclides like \chem{138}{La}. There is some hope to somewhat boost the $\nu$-process efficiency by taking into account the time dependence of the CCSN neutrino emission \cite{Langanke19}.
 
Finally, the GCR interaction with the interstellar material has also been proposed long ago \cite{Audouze70} as a possible \chem{180}{Ta} producer. This mechanism has been revisited recently \cite{Kusakabe18}, with the conclusion that it may be responsible for about 20\% of its SoS abundance. This figure is facing various more or less severe uncertainties, however. 
   
%******************************
\subsubsection{Some words about p-process isotopic anomalies and on the p-radionuclides \chem{92}{Nb^g} and \chem{146}{Sm}}
\label{prod_p_anomalies}
%***************************
Several isotopic anomalies (Section~\ref{obs_sos}) concern the p-nuclides (the reader is referred to \cite{Arnould03} for some details). Broadly speaking, they can be divided into three
categories. The first one involves anomalies attributable
to the decay of radionuclides manufactured by the p-process, and whose lifetimes may be long
enough ($\tau \gsimeq 10^5$ y) for having been in live form in the early SoS before
their eventual in-situ decay in meteoritic solids. This concerns in particular \chem{92}{Nb} and \chem{146}{Sm}, and possibly also \chem{97}{Tc} and  \chem{98}{Tc}. The second category of isotopic anomalies affecting p-nuclides relates largely to presolar grains found in meteorites ({\it e.g.} \cite{Nittler16}). They have been identified as the carriers of Xe-HL referred to above. It is briefly reviewed by \cite{Arnould03} and is not discussed any further here. The third category involves anomalies discovered in specific meteoritic inclusions and in bulk meteoritic samples. They concern notably the p-isotope composition of Sr, Mo, and  Sm. 

The possible origin of the now extinct \chem{92}{Nb^g} (see \cite{Iizuka16} for its abundance at the time of SoS formation) has been discussed in several places, including the review by \cite{Arnould03}, and the more recent series of multi-dimensional SNIa models by \cite{Travaglio14}. The abundances these latter models predict are either in line, or quite lower than earlier evaluations (see {\it e.g.} \cite{Arnould03,Malatji19}). The contribution of the isomer of \chem{92}{Nb} to its decay might even bring its share to the uncertainties in its production level \cite{Mohr16}. 

Considering the uncertainties affecting the \chem{92}{Nb^g} abundance predictions (including the still puzzling p-process origin of its \chem{92}{Mo} neighbor), it is clearly close to impossible to use \chem{92}{Nb} as a p-process chronometer, as sometimes claimed in the literature ({\it e.g.} \cite{Harper96}), as a p-process thermometer \cite{Mohr16}, or even as a constraint for the origin of the SoS p-nuclides, as proposed by \cite{Iizuka16}. 

The study of the $\alpha$-decaying \chem{146}{Sm} as a
potential p-process chronometer has been pioneered by \cite{Audouze72}. This work has
triggered a series of meteoritic, nuclear physics and astrophysics investigations, which have
helped clarify many aspects of the question. This concerns namely the amount of \chem{146}{Sm} that has been injected live into the early SoS ({\it e.g.} \cite{Nyquist94}, and references therein). The
prediction of the \chem{146}{Sm}/\chem{144}{Sm} production ratio has also gained increased reliability through experimental efforts \cite{Somorjai98}. In spite of theses advances, large uncertainties remain, and more or less large discrepancies exist between published values of the production ratio \cite{Arnould03,Travaglio14,Malatji19}. In such conditions, any attempt to build a chronometry based on \chem{146}{Sm} ({\it e.g.} \cite{Harper96}) is a highly risky exercise.

 %****  Sect.8.2   **********
\section{A short overview of nucleo-cosmochronometry}
\label{chronometry}
%***************************

The dating of the Universe and of its various constituents is another tantalizing task in modern science, referred to as ``cosmochronology.'' This field is in fact concerned with different ages, each of which corresponding to an epoch-making event in the past. They are in particular the age of the Universe  $T_{\rm U}$, of the globular clusters $T_{\rm GC}$, of the Galaxy [as (a typical?) one of many galaxies] $T_{\rm G}$, of the galactic disc $T_{\rm disc}$, and of the non-primordial nuclides in the 
disc $T_{\rm nuc}$, with $T_{\rm U} \gsimeq T_{\rm GC} \approx$ ($\gsimeq$) $T_{\rm G} \gsimeq T_{\rm disc} \approx T_{\rm nuc}$. As a consequence, cosmochronology involves not only cosmological models and observations, but also various other astronomical and astrophysical studies, and even invokes some nuclear physics information. Various astronomical methods are used to extract ages from the observations. Each of these methods has advantages ans weaknesses of its own.The reader is referred to {\it e.g.} \cite{Arnould99} for references. Here, we limit ourselves to a short discussion of the chronology of the SoS and of old stars of most direct relevance to astronuclear physics.
 
The discovery of isotopic anomalies attributed to the {\it in situ} decay in some meteoritic material of radionuclides with half-lives in the  approximate $10^5 \lsimeq T_{1/2} \lsimeq 10^8$ y range has broadened the original scope of nucleo-cosmochronology. It likely provides some information on discrete nucleosynthesis  events that presumably contaminated the SoS at times between about $10^5$ and $10^8$ y prior to the
isolation of the solar material from the general galactic material, as well as constraints on the chronology of nebular and planetary events in the early SoS.  This chronometry using short-lived radionuclides is not reviewed here. The reader is referred to {\it e.g.} \cite{Podosek97,Arnould97} for details.

 %*******************************************************************
\subsection{Actinide-based chronometries}
\label{chrono_actinides}
%*******************************************************************

For about 60 years, the long-lived \chem{232}{Th}-\chem{238}{U} and \chem{235}{U}-\chem{238}{U}  
pairs have been classically used to estimate the age of the r-nuclides (assumed to be roughly
equal to the age of the Galaxy) from the present meteoritic content of these nuclides. In contrast to many claims found in the literature over the years, the opinion has been expressed at several occasions that these pairs have just limited chronometric virtues, as reviewed by {\it e.g.} \cite{Arnould07}.

If indeed the bulk SoS composition witnesses the operation of the galactic blender 
(see Fig.~\ref{fig_blender} and Section~\ref{evol_galaxies}), a reliable evaluation of $T_{\rm nuc}$ requires {\it i)} high quality data for the meteoritic 
abundances of the relevant nuclides. Uncertainties in the meteoritic Th and U abundances still amount to 7\% \cite{Palme14}; {\it ii)} the build-up of models for the evolution of nuclides in the Galaxy, primarily in the solar neighborhood, that account for as many astronomical data as possible; {\it iii)} the construction of r-process models that are able to predict the actinide yields with high enough reliability and accuracy. This is an especially severe astrophysics and nuclear physics problem, as we have emphasized in various subsections of Section~\ref{prod_r}.

All these requirements clearly make the chronometric task especially demanding.  
From a detailed discussion, \cite{Yokoi83} conclude, in contradiction to many other claims, that the predicted (\chem{235}{U}/\chem{238}{U})$_0$ and (\chem{232}{Th}/\chem{238}{U})$_0$ ratios at the time $T_\odot$ of isolation of the SoS material from the galactic matter about 4.6 Gy ago are unable to provide chronometric information that cannot be revealed by other methods. In fact, they can at best provide results in agreement with
conclusions derived from other techniques, as they are only very weakly dependent on the galactic age
 $T_G$, at least in the explored range from about 11 to 15 Gy.

One might also confront the predicted actinides production with the observations of r-nuclides in old metal-poor stars. Compared with the SoS case, this chronometry has the advantage of allowing the economy of a galactic evolution model, just as in the case of the isotopic-anomaly based studies. Even so, the difficulties are more substantial than it might appear at a first glance. One of the main problems lies in the necessity to make
the assumption that the r-process is ``universal''. In other words, the observed patterns of
r-nuclide abundances in metal-poor stars have to be considered as exactly solar.  This is indeed
the only way to build a chronology on the metal-poor star content of Th and  U.

Under the assumption of universality, age evaluations have been attempted from observationally-based stellar Th and Eu abundance ratios. From the uncertainties involved, particularly in the calculation of the r-process production of these two elements, it is concluded that no reliable enough chronometry can be derived (see {\it e.g.} \cite{Goriely01}, with application to the star CS 31082-001). An attempt  has also been made to develop a chronology based on observed Pb/Th ratios. The predictions of this ratio are also very uncertain. The situation is even worsened further by the fact that some production of Pb by the s-process cannot be excluded.

All in all, the many uncertainties and difficulties faced by the Th-Eu- or Th-Pb-based chronologies have led \cite{Arnould01} to entitle their paper ``Inquietudes in Nucleocosmochronology''.
 
The Th-chronometry could be put on safer grounds if the Th/U ratios would be known in a variety of stars with a high enough accuracy. In fact, observations of U are quite difficult, so that only seven stars have at present observed U abundances \cite{Holmbeck18,Placco17}. The significant advantage of the use of the Th/U ratio is that these two nuclides are likely to be coevally produced, and that their production ratio is expected to be more accurately predicted than Th/Eu. Even in such relatively favorable circumstances, one would still face the severe question of whether Th and U were produced in exactly the same ratio in presumably the (very) few r-process events that have contaminated the material from which metal-poor stars formed. Even if this ratio would turn out to be the same indeed, its precise value remains to be calculated (see \cite{Arnould90} for an illustration 
of the dramatic impact of a variation in the predicted Th/U ratio on predicted ages).
 
Finally, as stressed in Section~\ref{obs_cr}, the measurement of the actinides content of GCRs is within the reach of present technologies. They have been used to help constraining the origin and age of the GCRs, as well as the actinides production by the r-process. The GCR actinide abundances lead to the conclusion that GCRs are not fresh supernova ejecta. They also provide a way of discriminating between two competing models for their acceleration: the isolated supernova remnant exploding in ordinary, old, ISM, or the super-bubble scenario. More details about these questions can be found in {\it e.g.}  \cite{Goriely01,Arnould07}.

%*************************
\subsection{The \chem{187}{Re} - \chem{187}{Os} chronometry}
\label{chrono_reos}
%***************************
%
First introduced by \cite{Clayton64}, the chronometry using the \chem{187}{Re} - \chem{187}{Os} pair is a remarkable illustration of an interdisciplinary effort devoted to an astrophysics question. It allows to avoid the difficulties related to the r-process modeling. True, \chem{187}{Re} is an r-nuclide. However, \chem{187}{Os} is not produced  directly by the r-process, but indirectly via the \chem{187}{Re} $\beta^-$-decay of $T_{1/2} 
\approx 43$ Gy over the galactic lifetime. This makes it in principle possible to derive a lower bound for 
$T_{\rm nuc}$ from the mother-daughter abundance ratio, provided that the ``cosmogenic"
\chem{187}{Os} component is deduced from its SoS abundance by subtracting its s-process contribution. This chronometry is thus in the first instance reduced to a question concerning the s-process. 

 The development of the Re-Os  chronology requires a variety of information.The precise measurement of the abundances of the concerned nuclides in meteorites is a first important prerequisite. Progress made in the measurement of the abundances of Re and Os  at the time of formation of the SoS leads \cite{Palme14} to quote an uncertainty of 5\% for each element. The uncertainties in the abundances of the Re and Os isotopes of relevance in the chronometry are discussed by \cite{Faestermann98}. 

A second requirement concerns high quality radiative neutron capture cross sections in the $184 \sim 187$ mass range for conditions appropriate to the s-process. Recent efforts in this direction are found in \cite{Shizuma05,Segawa07,Fujii10}. Although the s-process is better understood than the r-process, the calculation of the s-process yields of the Os isotopes is facing specific intricacies \cite{Takahashi98}). The evaluation of the \chem{187}{Os} s-process component from the ratio of its production to that of the s-only nuclide \chem{186}{Os} is indeed not a trivial matter. The difficulty relates to the fact that the \chem{187}{Os} 9.75 keV excited state can contribute significantly to the stellar neutron capture rate because of its thermal population in s-process conditions ($T \gsimeq 10^8$ K). The implied correction factor has been revisited recently by \cite{Shizuma05,Segawa07,Fujii10} on the basis of new experimental and theoretical data. Convergence in the deduced values of the correction is not reached yet. While \cite{Shizuma05} predicts a lowering of the laboratory cross section by about 0.85 at a typical s-process temperature of $4 \times 10^8$~K, a value of about 0.75 is proposed by \cite{Segawa07,Fujii10}. This level of discrepancy has a non-negligible impact on the chronometric predictions. This question clearly deserves further scrutiny, with predictions of cross section uncertainties based in particular on different experimentally constrained models of nuclear level densities, $\gamma$-ray strength functions and optical potentials. It is also important to note that additional problems and uncertainties arise in the evaluation of the s-process \chem{187}{Os}/\chem{186}{Os} production ratio relating to possible s-process branchings in the $184 \leq A \leq 188$ region. \cite{Arnould84,Kaeppeler91}. 

A third requirement concerns the precise knowledge of the stellar \chem{187}{Re} decay rate. This rate depends strongly on its precise ionisation state due to the bound-state $\beta$-decay mechanism 
(Section~\ref{decay_beta}), and consequently on its exact location in a star of given mass and evolution stage. This adds a substantial level of complexity to the problem.

Finally, a major problem is the implementation of the nuclear physics and stellar $\beta$-decay effects mentioned above into a model for the evolution of the Re-Os galactic content, this modeling facing extremely severe problems and uncertainties. This exercise has been attempted by \cite{Yokoi83} on grounds of a reasonable model trying to accommodate for a variety of relevant astronomical constraints provided by observations in  the solar neighborhood. This original work has been updated in several respects by \cite{Takahashi03}, who concludes that the \chem{187}{Re}-\chem{187}{Os} chronology leads to a probable galactic age in the approximate 13 to 18 Gy range. Room for improvement depends largely on astrophysics, and especially on the use of more realistic models for the chemical evolution of the Galaxy, which is far from being an obvious task !

%****************************************** 
\section{Summary}
\label{summary}
%*****************************************

This review is an update and extension of the one published 20 years ago \cite{Arnould99}. Since that time, impressive progress has been made in many fields that relate to astronuclear physics, and some real breakthroughs have been recorded.

%*******************************************************
\begin{enumerate}
%*********************************************************

\item{On the observational side, and as reviewed in the various chapters of Section~\ref{observations},  multi-messenger astronomy has developed more than ever, with often direct implications for astronuclear physics.}

\begin{enumerate}
%1111111111111111111111111111111111111
\item{The electromagnetic view of the components of the Universe has improved dramatically at all wavelengths. Among the very many advances, let us note the remarkable progress made in our knowledge of the microwave background radiation thanks to the advent of several top-quality space-borne observations. At the other end of the wavelength spectrum, x-ray observations have provided invaluable information on the EoS of neutron stars (Section~\ref{eos}), while the detection of $\gamma$-ray lines has in particular provided direct evidence that nucleosynthesis is in operation in explosive events (Section~\ref{nucleo_stars}).  Myriads of high-quality spectra for stars with different ages and locations are now at hand, from which abundances are determined. In this respect, it is worth emphasizing that abundances are not $\it{observed}$, but are 
$\it{derived}$ from the analysis of spectra making use of models the simplifications of which do not capture the whole very complex physics of the stellar atmospheres. The development of multi-dimensional models certainly allows substantial progress to be made in abundance predictions, but their application still remains limited to a quite small fraction of the observational data;}

\item{Neutrino astronomy has made giant steps forward. Particularly noticeable in this field is the resolution of the famed ``solar neutrino problem''. Many spectacular ``neutrino telescope'' are now in operation or in development in order to search for signals from astronomical sources other than the famous supernova SN1987A 
(Section~\ref{obs_neutrino});}

\item{Long-sought gravitational waves have at last been detected \cite{Abbott17}, shedding light on very important aspects of objects of direct relevance to relativistic astrophysics, like the merger of compact stars, and on accompanying nucleosynthetic processes (Section~\ref{prod_r_mergers});}

\item{The composition of GCRs and stellar/solar energetic particles is better known than ever, providing constrains on the GCR sources, acceleration mechanisms, paths within the Galaxy and associated nuclear transmutations (Sections~\ref{obs_cr} and \ref{nucleo_gcrs});}

\item{The accuracy of the SoS composition derived from meteoritic measurements and from the re-analysis of the solar spectra has improved. The study of extra-solar material in the form of grains has continued, with improved knowledge of the specificities of their composition.}
%111111111111111111111111111111111111

\end{enumerate}

%22222222222222222222222222222222222

\item{On the modeling side, progress has been made based on new observations, and even more so on the spectacular increase in computer capabilities.}

\begin{enumerate}

\item{As briefly discussed in Section~\ref{nucleo_bb}, Big Bang models have been put on a much safer ground as a result of the now available high-quality microwave background radiation observations. In particular, the baryonic density of the Universe, which was the last free parameter in BBN nucleosynthesis calculations, is now often claimed to be known with a precision as high as 1\%. We consider that it may be advisable, however, to avoid a too high level of over-confidence in this respect. The analysis of the relevant observations is a highly complex task indeed, and is in particular not free from some basic assumptions which remain to be fully validated. A famed problem that the BB nucleosynthesis predictions cannot be made compatible with the``primordial'' \chem{7}{Li} has been, and still is, the focus of an impressive number of calculations and more or less exotic speculations. The puzzle is still with us today, and does not seem to have a nuclear physics origin. It may be worth reminding at this point (see Section~\ref{nucleo_bb}) that primordial abundances cannot be derived directly from observation, but have to be extrapolated from the measured spectra of the oldest possible stars the surfaces of which are hypothesized not to have been affected by any other source than the BB itself;}

\item{Over the years, more and more simulations of stellar structure and quiescent evolution have appeared in the literature, spanning wider and wider ranges of stellar masses and initial compositions 
(Section~\ref{nucleo_stars}). Some additional sophistication has been introduced in the models, namely with the more or less approximate inclusion of the effects of mass loss, rotation and, to a limited extent, of magnetic fields. However, much remains to be done, especially in the treatment of turbulent transport of matter in the stellar interiors. Clearly, multi-dimensional simulations are badly needed. They have started to be constructed, but remain largely limited to the late stages of quiescent evolution (carbon burning and later phases) which are short enough to be tractable in a three-dimensional approach. The evolution of binary stars has also focussed some attention, but much remains to be done in the field in order to gain a robust picture of the effects of the additional complications raised by mass loss/transfer;}

\item{Huge efforts have been devoted to the simulation of the explosive phases terminating the quiescent evolution of stars of different masses and compositions. Various mechanisms concerning single or binary stars have been invoked, as briefly reviewed in Section~\ref{nucleo_stars}. They refer mainly to neutrinos as possible explosion triggering agents in CCSNe events, and to hydrodynamical burnings of the deflagration or detonation types. Pair-creation, pulsational/relativistic instabilities may also act in very massive or supermassive stars. Even if nuclear physics enters as an important ingredient in the explosion process, especially neutrino physics in the case of CCSNe, the largest uncertainties lie in the detailed multi-dimensional magneto-hydrodynamics simulations, possibly involving more or less rapid rotation and the effect of binarity. One certainly does not have to put under the rug the fact that the explosion simulations are based on more or less simplified one-dimensional pre-explosion models;} 

\end{enumerate}
%222222222222222222222222222222222222

%333333333333333333333333333333333333
 
\item{Nuclear physics enters a very broad variety of astrophysics questions, including the predictions of the nucleosynthesis accompanying the Big Bang (Sectiion~\ref{nucleo_bb}), the modeling of the evolution of stars and the accompanying nucleosynthesis (Section~\ref{nucleo_stars}) of direct bearing on the stellar surface abundances, the composition of the interstellar medium gas and grain components, of the GCRs and stellar/solar energetic particles (Section~\ref{obs_cr}), and ultimately of the SoS (Section~\ref{obs_sos}). This diversity of applications makes the nuclear physics needs even more demanding as the astrophysics conditions are more often than not largely far away from the conditions encountered in the nuclear physics laboratory. In such conditions, it is clear that huge experimental efforts are necessary, but far from being sufficient. The nuclear physics needs are reviewed in the different chapters of Section~\ref{nucdata}, with emphasis on the theoretical achievements;}

%******************************************
\begin{enumerate}

\item{The experimental efforts to measure the masses and structure properties of nuclei have been quite impressive over the last years, with the exploration of regions located further and further away from the valley of nuclear stability. The astrophysical needs are not fully covered, however, as nuclei all the way to the drip lines enter several nucleosynthesis processes (see especially Sections~\ref{prod_r} and \ref{prod_p}). Global microscopic, or at least semi-microscopic models nowadays reflect the state-of-the-art in the mass predictions (Section~\ref{masses}); }

\item{Astrophysics brings its share of additional complications as nuclei are often immersed in a plasma at very high temperatures and/or densities. Nuclear excited states can thus be more or less heavily populated at statistical equilibrium with the ground states (Section~\ref{masses_hight}). At the high densities encountered in compact objects, like NSs, or in the core of exploding stars, the construction of an EoS is mandatory (Section~\ref{eos}). Large efforts have been devoted to the elaboration of more and more sophisticated models relying as much as possible on microscopic approaches;}

\item{The study of the classical $\beta^-$- and $\beta^+$-decay modes has to be extended to highly exotic nuclei, a substantial fraction of which not being reachable in the laboratory. In addition, various effects modify more or less drastically the decay probabilities, as briefly reviewed in Section~\ref{decay_beta}. This is the case for electron degeneracy encountered in many stellar plasmas and the contribution to the effective decay rates of nuclear excited states. The probability of the laboratory known bound electron captures may also be deeply reduced by ionization in the stellar interiors and in the GCRs as well as in stellar/solar energetic particles. Ionization also leads to the development of the bound-state $\beta$-decay phenomenon which has the most spectacular effects in very low $Q_{\beta}$ cases. A remarkable example is provided by \chem{187}{Re}, the half-life of which being  possibly reduced by many orders of magnitude in hot stellar interiors 
(see Fig.~\ref{fig_bound}). Beta-delayed particle emissions or fissions also occur in stars, and are of particular importance in the r-process which involves highly neutron-rich nuclei (Section~\ref{decay_delayed});}

\item{Astrophysics also offers the possibility for additional decay mechanisms to operate. The most important one is the capture of continuum electrons in which the nuclei are immersed. These captures have several important astrophysical consequences. They modify the global excess of neutrons per nucleon in the stellar plasma, which has an important impact on the evolution of the stars and on the accompanying nucleosynthesis. They can even trigger their explosion in certain cases, in particular through the electron captures by highly stable and abundant nuclides, like \chem{20}{Ne} (Section~\ref{decay_continuum});}

\item{Beta-decays have been the focus of much experimental and theoretical effort.  Different models have been developed, ranging from the macroscopic approach of the Gross Theory to microscopic shell models (Section~\ref{decay_models}). We note that the macroscopic Gross Theory models with global parameter values perform remarkably well in their accuracy to reproduce experimental data, at least for not too low $Q_\beta$-values. They also have the important advantage of providing with very limited computing efforts all the $\beta$-decay data that are needed in the modeling of the 
r-process, including in a consistent way the contributions of both the allowed and first-forbidden transitions. The microscopic models still suffer from different limitations, particularly regarding their computational burden to make predictions for a very large body of heavy nuclei;}

\item{Neutron-induced, spontaneous, $\beta$-delayed and photo-fission may have an important impact on the r-process nucleosynthesis (Section~\ref{prod_r}). Theoretical predictions have to complement the large experimental effort in order to cope with the astrophysical requirements covering regions of the nuclidic chart not accessible in the laboratory. Large-scale predictions are based on the liquid-drop approximation and on microscopic mean-field models (Section~\ref{decay_fission}). The state-of-affairs remains rather unsatisfactory, with quite large uncertainties in the fission probabilities and in the predictions of the fission fragment distributions, leading to a variety of highly uncertain r-process yields;}

\item{Various weak-interaction processes involving all sorts of (anti-)neutrinos have an important bearing on a variety of phenomena, namely supernova explosions, the cooling or merging of NSs, and the production of some rare nuclides through the $\nu$-process, the effectiveness of which being poorly established, however, due to large uncertainties in the neutrino capture cross sections, as well as in the supernova neutrino spectra (Section~\ref{decay_neutrino}). Neutrino reactions have also been speculated to be at play in the production of some neutron-deficient heavy nuclides through the $\nu$p process (Section~\ref{burn_rp});}

\item{The knowledge of thermonuclear reaction cross sections is an essential chapter of astronuclear physics. A huge amount (up to several thousands) of neutron, proton and $\alpha$-particle captures on targets covering a large portion of the chart of nuclides occur in the interiors of stars, and are essential to their structure, evolution and nucleosynthesis capabilities. The \chem{12}{C} $+$ \chem{12}{C}, \chem{16}{O} $+$ \chem{16}{O}, and, to a lesser extent, \chem{12}{C} $+$ \chem{16}{O} fusions complete the panoply of important thermonuclear reactions. The important chains of charged-particle induced thermonuclear reactions are briefly reviewed in Section~\ref{burn_general}, while neutron captures are dealt with in the sub-sections of 
Section~\ref{prod_heavy}. As made evident in the various chapters of Section~\ref{reac_thermo}, astrophysics is more demanding to nuclear physics than any other field of science in terms of the huge volume of requested cross sections, but also of the energy regime and region of the nuclidic chart to be explored, forcing the exploration of this ``world of almost no event'' and/or the ``world of exoticism'' (Section~\ref{reac_generalities}). An additional severe intricacy comes from the contribution to the reaction rates of thermally populated excited states, which is clearly out of reach of laboratory investigation. Last but not least, electron screening corrections have to be applied both to experimental data and to the calculated stellar reaction rates between bare nuclei (Section~\ref{reac_generalities}). Decades of experimental and theoretical efforts have been devoted to the study of astrophysically relevant reaction rates, as briefly reviewed in some chapters of Section~\ref{reac_thermo}. Compilations of nuclear data for use in astrophysics are now available (Section~\ref{burn_compil}). At high temperatures (typically in excess of about 10$^9$K), nuclei may be subject to photodisintegrations of the ($\gamma$,n), ($\gamma$,p) or ($\gamma$,$\alpha$) types. Such transformations are important in major burning episodes, in particular Ne burning and subsequent phases (Section~\ref{burn_hesi}), as well as in nucleosynthesis, especially the p- and r-processes (Sections~\ref{prod_r} and \ref{prod_p}). Many experimental and theoretical efforts have been devoted to photodisintegration rates. In view of the difficulties to derive photodisintegration rates through direct approaches, the detailed balance theorem applied to the reverse radiative captures of nucleons or $\alpha$-particles is widely used (Section~\ref{decay_em});} 

\item{GCRs and stellar/solar energetic particles are another important component of the Galaxy, with specific nuclear physics requirements. In contrast to the situation encountered in thermonuclear processes, cross sections need to be known at energies much in excess of the Gamow window and the Coulomb barriers. They have been the focus of much experimental and theoretical efforts, as extensively reviewed recently (Sections~\ref{nucleo_gcrs}, \ref{reac_nonstat}). Clearly, further nuclear physics investigations are needed, especially in order to reduce remaining cross section uncertainties to cope with the high precision of the GCR abundance observations;}

\end{enumerate}
%3333333333333333333333333333333333333

%4444444444444444444444444444444444444
 
\item{An overview of the contributors to the galactic nucleosynthesis is presented. Apart from the Big Bang (Section~\ref{nucleo_bb}) and GCRs (Section~\ref{nucleo_gcrs}), stars are prime agents responsible for changes in their surface compositions, as well as for the evolution of the nuclidic content of galaxies through steady mass losses or through explosive ejecta accompanying supernovae of different types. The ejected material carries the signatures of the quiescent and explosive burnings of H to Si through more or less complex chains of charged-particle induced thermonuclear reactions (Section~\ref{burn_general}) producing the nuclides up to the vicinity of the iron peak. Uncertainties remain in the nucleosynthetic outcome of these processes. Major problems affect the stellar structure simulations, especially the description of supernova explosions. Nuclear physics brings its share of trouble, considering in particular the very special conditions in which nuclear reactions take place in stars;}

\item{Special attention is paid to the synthesis of the nuclides heavier than iron. Neutron capture mechanisms range from the s-process for the production of the stable nuclides located at the bottom of the valley of stability to the r-process responsible for the synthesis of the neutron-rich isobars. The origin of the neutron-deficient isobars observed in the SoS is attributed to the p-process;}

\begin{enumerate}

\item{Central He burning possibly complemented with shell C burning in massive stars and low- and intermediate-mass stars in their AGB phase are classically viewed as the main s-process sites. Massive stars are predicted to produce s-nuclides up to about mass number 100, the neutrons being mainly produced by 
\reac{22}{Ne}{\alpha}{n}{25}{Mg}, while the less massive ones are responsible for the heavier species up to lead, the main neutron source being \reac{13}{C}{\alpha}{n}{16}{O} (Section~\ref{prod_s}). The nucleosynthesis uncertainties arise mainly from the stellar modelings. In particular,  they are at present not able to predict the efficiency of the \chem{13}{C} neutron source with a satisfactory level of confidence. In contrast, the radiative neutron capture data are, with some exceptions, accurate enough. The main remaining nuclear physics uncertainty lies in fact in the stellar rate of the \chem{22}{Ne} burning; }

\item{The r-process remains the least well understood process of heavy nuclide production. The attempts to identify the site(s) where this process could best develop have resulted in a saga over the last decades, made of ups and downs regarding the merits of various proposed models. To make a long story short, it appears generally acknowledged today that NS mergers and/or collapsars are the most promising scenarios (Sections~\ref{prod_r_site} and \ref{prod_r_mergers}). The expected relative contributions of these two sites to the galactic r-nuclide content or to stellar surface abundance patterns remain a matter of hot debate, however (Section~\ref{prod_r_gcm}). Large uncertainties prevent in fact firm conclusions to be drawn. They concern the modeling of the relevant astrophysical sites, as well as the exceptionally vast range of nuclear physics data of importance, involving highly exotic nuclei all the way to the neutron drip line, and even beyond. A special facet of the r-process is the production of actinides. The long-lived \chem{232}{Th}-\chem{238}{U} and \chem{235}{U}-\chem{238}{U} pairs have been classically used to estimate the age of the r-nuclides (assumed to be roughly
equal to the age of the Galaxy) from the present meteoritic content of these nuclides. In contrast to many claims found in the literature over the years, the opinion has been expressed at several occasions that these pairs have just limited chronometric virtues (Section~\ref{chrono_actinides}). Hope in nucleo-cosmochronometry is considered instead to lie in the \chem{187}{Re} - \chem{187}{Os} pair, even if it encounters some serious problems (Section~\ref{chrono_reos});}

\item{In contrast to claims too often made in the literature, the modeling of the p-process is relatively well mastered, even if it is clearly not free from difficulties and uncertainties. The O/Ne-rich layers of massive stars during their late quiescent evolution and explosion are recognized as privileged p-process sites (Section~\ref{prod_p_site}). The proper production of some p-nuclides raises difficulties, however 
(Section~\ref{prod_p_moru}). Uncertainties remain in the calculations of the s-process that provides the seeds for the p-process (see point 5(a) above). In addition, more reliable p-nuclide abundance predictions in pre-SN enriched layers and their survival to explosion have to await multidimensional modelings. As usual, nuclear physics uncertainties add their share to the problem. A brief overview of the situation concerning the p-radionuclides \chem{92}{Nb^g} and \chem{146}{Sm} is presented in Section~\ref{prod_p_anomalies}.}

\end{enumerate}
%4444444444444444444444444444444444444

%**************************************
\end{enumerate}
%**************************************

By way of a very brief conclusion of this review, one may say that a century of remarkable and sometimes really heroic inter- and multi-disciplinary efforts in astronomical observation, modeling of astrophysical sites and laboratory and theoretical nuclear physics have made possible the writing of fascinating chapters of the tale of the atomic nuclei in the cosmos. Quite fortunately, the story has not come to an end yet, and many more scientific adventures in astronuclear physics are still ahead of us !

 \section*{Acknowldegments} S.G. is F.R.S.-FNRS research associate.

\bibliographystyle{elsarticle-num}
\bibliography{biblio_ppnp_resubmit}
\end{document}